\documentclass[preprint]{aastex63}
\usepackage{natbib}
\usepackage{CJKutf8}
\shorttitle{Outflows in FeLoBALQs}
\shortauthors{Choi et al.}
\begin{document}
\begin{CJK}{UTF8}{}
\CJKfamily{mj}

\title{The Physical Properties of Low Redshift FeLoBAL Quasars. I. Spectral Synthesis Analysis of the BAL Outflows using \textit{SimBAL}}

\correspondingauthor{Hyunseop Choi}
\email{hyunseop.choi@ou.edu}

\author[0000-0002-3173-1098]{Hyunseop\ Choi (최현섭)}
\affiliation{Homer L.\ Dodge Department of Physics and Astronomy, The
  University of Oklahoma, 440 W.\ Brooks St., Norman, OK 73019}
\author[0000-0002-3809-0051]{Karen M.\ Leighly}
\affiliation{Homer L.\ Dodge Department of Physics and Astronomy, The
  University of Oklahoma, 440 W.\ Brooks St., Norman, OK 73019}
\author{Donald M.\ Terndrup}
\affiliation{Homer L.\ Dodge Department of Physics and Astronomy, The
  University of Oklahoma, 440 W.\ Brooks St., Norman, OK 73019}
\affiliation{Department of Astronomy, The Ohio State University, 140
  W.\ 18th Ave., Columbus, OH 43210}
\author{Collin Dabbieri}
\affiliation{Department of Physics and Astronomy, Vanderbilt University, Nashville, TN 37235}
\author{Sarah C.\ Gallagher}
\affiliation{Department of Physics \& Astronomy, The University of Western
  Ontario, London, ON, N6A 3K7, Canada}
\affiliation{Canadian Space Agency, 6767 Route de l'A{\'e}roport,
  Saint-Hubert, Quebec, J3Y~BY9}
\affiliation{Institute for Earth and Space Exploration, The
  University of Western Ontario, London, ON, N6A 3K7, Canada}
\affiliation{The Rotman Institute of Philosophy, The University of
  Western Ontario, London, ON, N6A 3K7, Canada}
\author{Gordon T.\ Richards}
\affiliation{Department of Physics, Drexel University, 32 S.\ 32nd St.,
  Philadelphia, PA 19104}
  
\begin{abstract}
We present the first systematic study of 50 low redshift ($0.66 < z < 1.63$) iron low-ionization broad absorption-line quasars (FeLoBALQs) using {\it SimBAL} which represents a more than five-fold increase in the number of FeLoBALQs with detailed absorption line spectral analyses.
We found the outflows have a wide range of ionization parameters, $-4\lesssim\log U\lesssim 1.2$ and densities, $2.8\lesssim\log n\lesssim8\ \rm[cm^{-3}]$.
The objects in our sample showed FeLoBAL gas located at a wide range of distances $0\lesssim\log R\lesssim 4.4$ [pc], although we do not find any evidence for disk winds (with $R\ll0.01$ pc) in our sample.
The outflow strength primarily depends on the outflow velocity with faster outflows found in quasars that are luminous or that have flat or redder spectral energy distributions.
We found that $\sim18\%$ of the FeLoBALQs in the sample have the significantly powerful outflows needed for quasar feedback.
Eight objects showed ``overlapping troughs'' in the spectra and we identified eleven ``loitering outflow'' objects, a new class of FeLoBALQs that are characterized by low outflow velocities and high column density winds located $\log R\lesssim1$ [pc] from the central engine.
The FeLoBALs in loitering outflows objects do not show properties expected for radiatively driven winds and these objects may represent a distinct population among FeLoBALQs.
We discuss how the potential acceleration mechanisms and the origins of the FeLoBAL winds may differ for outflows at different locations in quasars.

\end{abstract}

\section{Introduction}\label{sec:intro}
Iron Low-ionization Broad Absorption Line quasars (FeLoBALQs) are
arguably the most enigmatic of extra-galactic objects.  Their spectra
present a tremendous range of phenomenology that has occasionally 
baffled
experts\footnote{\href{https://www.nytimes.com/1999/08/17/us/rarely-bested-astronomers-are-stumped-by-a-tiny-light.html?smid=url-share}{https://www.nytimes.com/1999/08/17/us/rarely-bested-astronomers-are-stumped-by-a-tiny-light.html}}.   
Although rare and sometimes hard to find, their analysis may prove key to addressing several important questions involving galaxy
evolution and quasar structure and demographics.

Broad absorption line quasars (BALQs) are observed to comprise
10--20\% of optically-selected quasars \citep{hewett03, reichard03, trump06, knigge08,gibson09}.
BALQs are typically
recognized by their \ion{C}{4} absorption, and accompanying
transitions from ions with a single valence electron (\ion{Si}{4},
\ion{N}{5}, \ion{O}{6}) as well as Ly$\alpha$. If these are the only 
lines present, the object is known as a high-ionization BALQ
(HiBALQ). A small fraction also have absorption from
the lower-ionization ions, \ion{Mg}{2} and \ion{Al}{3}.   These are
known as Low-Ionization broad absorption-line quasars (LoBALQs).  Even
more rare are the FeLoBALQs which have absorption from \ion{Fe}{2} and
sometimes other iron-peak elements; observations of \ion{Mn}{2},
\ion{Zn}{2}, \ion{Ni}{2}, \ion{Co}{2}, and \ion{Cr}{2} have 
been reported \citep[e.g.,][]{dekool02f1}.

The precise fraction of quasars corresponding to the different types
of BALQs is not well constrained, principally as a consequence of
selection effects. 
Quasars can be identified by their typically blue colors; the classic
example is the Palomar-Green survey \citep{green86}.  While there
exist relatively blue BAL quasars \citep{hewett03}, on average, BAL
quasars are redder than unabsorbed quasars \citep[e.g.,][]{reichard03,
  gibson09,   krawczyk15}.  Thus while only $\sim$15--20\% of optically selected
quasars are BALQs, a larger fraction has been found among luminous quasars 
\citep{maiolino04, dai08, dai12, bruni19} and  red, infrared-selected,
and radio-selected objects \citep{dai08, urrutia09, fynbo13, krogager15,krogager16,morabito19}.
Observations suggest that a larger fraction of LoBALQs and FeLoBALQs
are missed in optical quasar surveys \citep{urrutia09,   dai12,
  morabito19}.
Part of the problem is that in addition to the heavily absorbed and
reddened spectrum, they can lack strong emission lines and their spectra can be confused
with those of M-type stars and vice versa
\citep[e.g., SDSS~J092853.51$+$570735.3;][]{west11,paris18}.

Traditionally, there are two explanations for these BAL classes.
An equatorial accretion-disk wind may be present in
every quasar, and the BAL type observed depends on the angle of the
line of sight with respect to the accretion disk
\citep[e.g.,][]{krolik98}. Supporting this view 
is the fact that X-ray spectra from a sample of HiBAL quasars are
absorbed but otherwise indistinguishable from normal quasars
\citep[e.g.,][]{gallagher02b,gallagher06}; they also have similar
broadband  spectral energy distributions \citep{gallagher07b}.
Alternatively, BALQs may represent a blow-out phase in the evolution
of galaxies, occurring when a heavily shrouded ultra-luminous infrared galaxy-type (ULIRG) object
shrugs off its cloak of dust and gas \citep[e.g.,][]{farrah12}.  The
fact that LoBALQs and FeLoBALQs are more common among optically red quasars
supports this scenario \citep[e.g.,][]{urrutia09}.  Red quasars are
themselves very interesting objects.  There is evidence that red
quasars evolve differently than blue quasars
\citep[e.g.,][]{glikman18, klindt19}, a result that suggests that
dust-reddened quasars may be an intermediate phase between a
merger-driven starburst in a completely obscured AGN, and a normal,
unreddened quasar \citep[e.g.,][]{glikman17}.
These two origin stories are not mutually exclusive.
In fact, it is plausible that the combination of both explanations may describe BALQ population.

Depending on the physical conditions, the Fe$^+$ ion can contribute 
thousands of absorption lines to the near-UV spectrum.  The wide range of
critical densities and oscillator strengths probed by \ion{Fe}{2}
makes these features richly diagnostic of the physical state of the
absorbing gas \citep[e.g.,][]{lucy14}. Specifically, the relative
strength of the suite of lines shows a strong dependence on density,
ionization parameter, and column density within the outflow.  What
this means is that the physical conditions of the outflows in
FeLoBALQs can be measured with a precision that is arguably unequaled
among outflow phenomena in quasars.   

However, a detailed understanding of FeLoBAL quasars and their origin
has been hampered by the complexity of the spectra.  The combination
of the thousands of absorption lines and velocity dispersion in the
outflows (i.e., the lines are broad) results in significant blending.
Traditional methods of analysis that involve identification and
measurement of individual lines are difficult to use on these
complicated spectra.  As a result, detailed analysis
has been performed on only a handful of objects.  \citet{wampler95}
presented an qualitative analysis of a high-resolution spectrum of
the FeLoBAL quasar Q0059$-$2735 which was discovered as part of
the Large Bright Quasar Survey \citep{morris91}.  The first FeLoBAL
quasars subjected to detailed photoionization analyses were  QSO
2359$-$1241 \citep{arav01, arav08, korista08, bautista10} and three objects
discovered in the FIRST survey \citep{white00}: FIRST
J104459.6$+$365605 \citep{dekool01,   everett02a}, FBQS 0840$+$3633
\citep{dekool02f1}, FIRST J121442.3+280329 \citep{dekool02f2}.  These
masterful and difficult analyses yielded the first well-constrained
distances to the outflows (from 1--700 pc).  The Sloan Digital Sky
Survey \citep[SDSS;][]{blanton17} yielded many more FeLoBALQs \citep[e.g.,][]{hall02}, including
some very unusual and interesting objects.
\citet{hall03} discussed formation scenarios for the narrow \ion{Ca}{2} observed in the
overlapping-trough object SDSS~J030000.56$+$004828.0.  Other FeLoBALQs
that have been analyzed in varying degree of detail include 
SDSS~J0838+2955 \citep{moe09}, SDSS~J0318$-$0600 \citep{dunn10,  
  bautista10}, AKARI~J1757$+$5907 \citep{aoki11},
SDSS~J112526.12$+$002901.3 \citep{shi16}, and PG~1411$+$442
\citep{hamann19b}. 

The analysis of complex spectra of broad absorption-line quasars has become possible with the introduction of the novel
spectral-synthesis modeling software {\it  SimBAL} \citep{leighly18a}.
Because {\it SimBAL} uses forward modeling, line blending can be
accounted for.  In addition, because {\it SimBAL} models the whole
spectrum, it uses the information conveyed by the lines that are {\it
  not} present in the observed spectrum.  The analysis of the $z=2.26$ quasar
SDSS~J135246.37+423923.5 and the discovery of a remarkably powerful outflow
in that  object \citep{choi20} demonstrated an effective application
of the {\it SimBAL} methodology on an overlapping trough
FeLoBALQ.

This paper is the first in a series of four papers.
In Paper II, \citet{leighly_prep}, we discuss the rest-optical spectral properties of a subsample of $z<1$ FeLoBALQs.
Paper III, \citet{choi_prep_paper3}, combines the FeLoBAL properties discussed in this paper and the emission-line analysis results from Paper II \citet{leighly_prep}.
Finally, in Paper IV \citep{leighly_prep_paper4}, we discuss broad-band optical/IR properties of FeLoBALQs and the potential implication for the evolution scenarios for low redshift FeLoBAL quasars.

In this paper we present analysis of the outflows in 50 low-redshift
($0.66 < z < 1.63$) FeLoBAL quasars.  This work increases the number
of FeLoBALQs with detailed analyses by a factor of five.  In addition,
the uniform analysis means that
the properties of the objects can be easily compared.
These objects were 
drawn from two samples (\S~\ref{sec:data}). The first sample includes low-redshift
FeLoBALQs that were observed by {\it Spitzer} in order to constrain
their far-infrared spectral energy distributions \citep{farrah12}. The
second sample includes 30 objects with sufficiently low redshift that high 
quality spectra that the H$\beta$ / [\ion{O}{3}] region of
the spectrum could be analyzed; the results are reported in the 
companion paper, \citet{leighly_prep}. \S~\ref{sec:modeling} gives a brief recap of
the {\it SimBAL} software, and a description of our method for
modeling the continuum using principal component analysis
eigenvectors extracted from SDSS spectra of unabsorbed quasars (also
Appendix~\ref{app:model_cont}).
\S~\ref{sec:calculations} details how we quantified the various physical and kinematic properties of the outflows using the results from the {\it SimBAL} model-fitting.
\S~\ref{sec:results} presents the {\it SimBAL} model-fitting results and
discusses the distributions of the parameters measured directly by the
model fitting including the ionization parameter, density, column
density, velocity, and parameters extracted from the models including
the location of the outflow and its kinetic luminosity.
\S~\ref{sec:analy} describes the relationships among parameters, properties of the opacity profiles, and several special groups of objects.
\S~\ref{sec:disc} discusses the
implications of this large study for our understanding of the origins, formation, and acceleration of broad absorption line outflows.
\S~\ref{sec:summary} summarizes the results and presents planned future work.

\section{Sample Selection and Data}\label{sec:data}
Our total sample of 53, $0.66 < z < 1.63$ FeLoBALQs was chosen from two sources.
The data and other parameters listed in Table~\ref{sample}.
We report the SDSS names of the objects in Column 1 and throughout the paper we use the shortened four by four naming scheme (e.g., SDSS~JHHMM$+$DDMM).
Part of the sample was drawn from the 31 objects presented in
\citet{farrah12} and we analyzed data from 28 of these.
For this paper we rejected objects that did not have sufficient signal-to-noise ratio spectra, clear detection of \ion{Fe}{2} absorption features, and no evidence of strong resonance scattering emission from the BAL wind.
Specifically, we excluded 3 objects from the sample analyzed in \citet{farrah12}.
SDSS~J0911$+$4446 was excluded due to a low signal-to-noise spectrum which shows unusually strong \ion{Fe}{2} emission features potentially coming from resonance scattering from the BAL outflows \citep[e.g.,][]{wang16}.
The spectrum for SDSS~J2215$-$0045 showed BAL features only from \ion{Fe}{3} instead of \ion{Fe}{2}.
Finally,
SDSS~J2336$-$0107 was shown to be a double quasar by \citet{foreman09}, and we could not obtain an adequate spectrum from the SDSS archive for the {\it SimBAL} analysis.

Additionally we included 25 FeLoBALQs from \citet{leighly_prep}.
All of these objects have $z<1$ and sufficiently good signal-to-noise ratios that the H$\beta$ / [\ion{O}{3}] region of the spectrum could be analyzed.
The {\it SimBAL} analysis of the FeLoBALs in the objects included in \citet{leighly_prep} are discussed in this paper; however, we refer to that publication for analysis of the H$\beta$ / [\ion{O}{3}] region and the discussion
of the relationships between the FeLoBAL properties and the quasar optical emission-line properties.
Column 2 of Table \ref{sample} gives the sample origin of the objects.

In some cases, the object was observed by SDSS and then by Baryon Oscillation Spectroscopic Survey \citep[BOSS;][]{dawson13,dawson16}.
The highest signal-to-noise ratio spectrum available from the Sloan Digital Sky Survey \citep[SDSS;][]{blanton17} archive was chosen for analysis.
Multiple spectra are listed when they were averaged for rest-frame optical band analysis when no spectral variability was observed \citep{leighly_prep}.
The $i$ band SDSS magnitudes are listed in Table \ref{sample} (column 4).

As an accurate redshift is essential to analyzing outflow properties, we re-measured the SDSS catalogue redshifts where possible.
In order of preference and as available, redshifts were measured using (1) the low-ionization narrow emission-lines of [\ion{O}{2}] or narrow H$\beta$, (2) the narrow high-ionization [\ion{Ne}{5}] line, or (3) the \ion{Fe}{2} pseudo-continuum template.
Failing those methods, the \citet{hewett10} catalogue redshifts were used.
Best-fitting redshifts and the emission line used for their measurements are listed in Table \ref{sample} (columns 6 and 7)

\startlongtable
\begin{deluxetable*}{LclCCCc}
\tabletypesize{\scriptsize}
\tablecaption{The Sample\label{sample}}
\tablehead{\\
\colhead{SDSS Object Name} & \colhead{Sample\tablenotemark{a}} & 
\colhead{Spectra\tablenotemark{b}} &
\colhead{$m_i$\tablenotemark{c}} &
\colhead{SDSS DR14Q Redshift} & 
\colhead{Redshift Used} & 
\colhead{Redshift Origin\tablenotemark{d}}
}
\startdata
011117.36+142653.6 & F & 5131-55835-0054&17.73 & 1.154 & 1.1551 \pm  0.0002
& [\ion{O}{2}] \\
015813.56-004635.5 & O &  7837-56987-0485&20.44 & 0.896 & 0.8959\pm 0.0002
& narrow H$\beta$ \\
024254.66-072205.6 & F &  0456-51910-0378&18.71 & 1.2166 &
1.2175 \pm  0.0002 & [\ion{O}{2}] \\
025858.17-002827.0 & O &  4242-55476-0028&19.00 & 0.875 &
0.8758^{+0.0009}_{-0.0005}  & [\ion{O}{2}] \\
&  & 9372-58074-0974 \\
030000.57+004828.0 & F &  0410-51877-0623&16.61 & 0.900 &  0.8907 \pm
0.0002 & [\ion{O}{2}] \\
033810.84+005617.6$^*$ & F &  0714-52201-0326&18.35 & 1.6295 &  1.6316 \pm
0.0006 & \citet{hewett10}  \\
080248.18+551328.8 & O &  7281-57007-0616&17.90 & 0.663 &
0.6636^{+0.0001}_{-0.00008} &  [\ion{O}{2}] \\
080957.39+181804.4 & O &  4493-55585-0632&17.44 & 0.970 & 0.970  & SDSS DR14Q \\
081312.61+432640.1 & F & 0547-51959-0242&18.80 & 1.0865 &
1.0894 \pm  0.0008 & [\ion{O}{2}] \\
&  & 0546-52205-0449 \\
&  & 0547-52207-0274 \\
083522.77+424258.3 & F, O & 8280-57061-0366&17.42 & 0.806 & 0.8066\pm 0.0002
&  [\ion{O}{2}] \\
084044.41+363327.8 & F &  8858-57450-0056&16.25 & 1.235 &  1.2372 \pm 0.0001
& FeII \\
091658.43+453441.1 & O & 7517-56772-0266&19.81 & 0.915 & 0.9141\pm 0.0002 &
narrow H$\beta$ \\
091854.48+583339.6 & F &  0484-51907-0598&19.19 & 1.315 &
1.3110  \pm 0.0004 & [NeV] \\
094404.25+500050.3 & O & 7292-56709-0400&19.31 & 0.965 & 0.9656\pm 0.0003 &
[\ion{O}{2}] \\
100605.66+051349.0 & F & 0996-52641-0243&18.67 & 0.9704 &
0.9683 \pm  0.0002 & [\ion{O}{2}] \\
101927.37+022521.4 & F &  0503-51999-0464&18.61 & 1.3643 &
1.3648 \pm  0.0002 & [\ion{O}{2}] \\
102036.10+602339.0 & F &  7087-56637-0979&18.28  & 1.015 &  1.0145 \pm
0.0004 & [\ion{O}{2}] \\
102226.70+354234.8 & O & 4564-55570-0360&19.46 & 0.818 & 0.8207\pm 0.0003 &
[\ion{O}{2}] \\
102358.97+015255.8 & F & 0504-52316-0268&19.12 & 1.0761 &
1.07537 \pm  0.00009 & [\ion{O}{2}] \\
103036.92+312028.8 &  O & 6451-56358-0440&17.65 & 0.864 & 0.8605\pm 0.0002
& [\ion{O}{2}] \\
& & 10465-58144-0458 \\
& & 11383-58485-0722 \\
103903.03+395445.8 & O &  4633-55620-0278&19.66 & 0.864 & 0.8637\pm 0.0002 &
 [\ion{O}{2}] \\
104459.60+365605.1 & O & 8851-57460-0737&16.65 & 0.703 & 0.703 &  SDSS DR14Q   \\
105748.63+610910.8$^*$ & F & 0774-52286-0278&19.49 & 1.2757 &
1.2743 \pm 0.0006 & \citet{hewett10} \\
112526.12+002901.3 & F, O & 3839-55575-0812&17.90 & 0.864 &
0.8636^{+0.0003}_{-0.0002} &  [\ion{O}{2}]  \\
112828.31+011337.9 & F, O & 4730-55630-0172&18.38 & 0.893 & 0.8930\pm 0.0002
&  narrow H$\beta$ \\
112901.71+050617.0 & F &  0837-52642-0400&19.42 & 1.2775 &
1.2814 \pm 0.0008 &  \citet{hewett10} \\
114556.25+110018.4 & F &  1226-52734-0375&18.81 & 0.9330 &
0.9350\pm  0.0006 & [NeV] \\
115436.60+030006.3 & F &  4765-55674-0082&17.70 & &  1.46968 \pm 0.00007 &
FeII \\
115852.86-004301.9 &  F &  0285-51930-0189&19.38 & 0.9833 &
 0.9835 \pm  0.0002 & [\ion{O}{2}] \\
120049.54+632211.8 & F, O &  7106-56663-0915&18.86 & 0.887 & 0.8862\pm
0.0001 &  narrow H$\beta$ \\
120627.62+002335.4 &  F & 0286-51999-0499&18.68 & 1.114 &  1.1369  \pm
0.0008 & \citet{hewett10} \\
120815.03+624046.4 & O &  6974-56442-0404&19.72 & 0.799 &
0.7982^{+0.00004}_{-0.00003} &  narrow H$\beta$ \\
121231.47+251429.1 & O & 5975-56334-0556&19.32 & 0.842 & 0.8427\pm 0.0004 &
[\ion{O}{2}]  \\
121441.42-000137.8 & F & 0287-52023-0514&18.81 & 1.046 &  1.04571^{+0.0002}_{-0.0001} & [\ion{O}{2}] \\
121442.30+280329.1 & O & 6476-56358-0374&17.16 & 0.695 & 0.6945\pm 0.0002 &
[\ion{O}{2}] \\
& & 2229-53823-0557 \\
123549.95+013252.6 &  F & 0520-52288-0001&19.16 & 1.2918  &
1.2902  \pm  0.0002 & [\ion{O}{2}] \\
124014.04+444353.4 & O &  6617-56365-0631&20.10 & 0.964 &
0.9634^{+0.0001}_{-0.0002} &  narrow H$\beta$ \\
132117.24+561724.5 & O &  6828-56430-0710&19.53 & 0.794 & 0.7941\pm 0.0001
&  [\ion{O}{2}] \\
132401.53+032020.5 & F, O & 4761-55633-0136&18.97 & 0.927 & 0.9254\pm 0.0001
&  [\ion{O}{2}] \\
133632.45+083059.9 & O & 1801-54156-0530&18.23 & 0.805 & 0.7988\pm 0.0003 &
\ion{Fe}{2}, broad H$\beta$ \\
135525.24+575312.7 & O & 8199-57428-0180&20.59 & 0.855 & 0.8552\pm 0.0001 &
[\ion{O}{2}] \\
135640.34+452727.2 & O & 6629-56365-0728&19.52 & 0.802 & 0.8025\pm 0.00009 &
 [\ion{O}{2}] \\
142703.62+270940.4 & F &  6018-56067-0412&18.13 & 1.165 & 1.1669 \pm  0.0001
& [\ion{O}{2}] \\
144800.15+404311.7 & O &  5172-56071-0836&16.80 & 0.805 & 0.805 & SDSS DR14Q \\
& & 8498-57105-0724 \\
151708.94+232857.5 & O &  3961-55654-0256&20.30 & 0.810 & 0.8093\pm 0.0001 &
[\ion{O}{2}] \\
152737.17+591210.1 & O &  6799-56478-0458&18.33 & 0.930 & 0.930 & SDSS DR14Q \\
153145.01+485257.2 & O &  6728-56426-0078&19.60 & 0.945 & 0.9445\pm 0.0002 &
 [\ion{O}{2}] \\
155633.77+351757.3 & F &  4965-55721-0548&18.04 &  & 1.501 &
\citet{schulze17} \\
164419.75+530750.4 & O &  8057-57190-0707&17.90 & 0.781 & 0.7813\pm 0.0002 &
 [\ion{O}{2}] \\
173753.97+553604.9$^*$ & F &  0358-51818-0056&19.86 & 1.1017 &
1.1109 \pm  0.0007 & [NeV] \\
210712.77+005439.4 & F &  0985-52431-0522&20.42 & 0.9244 &
0.9266 \pm  0.0002 & [\ion{O}{2}]  \\
213537.44-032054.8 & O &  4385-55752-0286&18.59 & 0.815 &
0.8127^{+0.0002}_{-0.0001} & [\ion{O}{2}]  \\
230730.69+111908.5 & O &  6154-56237-0120&19.51 & 0.878 & 0.8770\pm 0.00009
&  narrow [\ion{O}{2}] \\
\enddata
\tablenotetext{a}{F: Objects drawn from \citet{farrah12}; O: Objects included in \citet{leighly_prep}}
\tablenotetext{b}{Multiple spectra are listed when combined to
  increase signal-to-noise ratio in the optical band.  The first
  spectrum listed set the wavelength sampling and flux level, and
  subsequent spectra were resampled and scaled to match.}
\tablenotetext{c}{The magnitudes are taken from the point source function magnitudes in the AB system from the SDSS DR16, uncorrected for Galactic extinction}
\tablenotetext{d}{The origin of the redshift in Column 5. An emission
  line signals measurement directly from the spectrum. }
\tablenotetext{$*$}{Objects excluded from the analysis (\S~\ref{subsubsec:bal_exclude}).}
\end{deluxetable*}

\section{Spectral Modeling with {\it SimBAL}}\label{sec:modeling}
\subsection{The Spectral Synthesis Software {\it SimBAL}}\label{subsec:simbal}
The traditional method for analyzing BALQ spectra involves line identification of individual BALs and measurement of the ionic column densities for each identified line in order to constrain the physical properties of the absorbing gas \citep[e.g.,][]{bautista10,dunn10,arav13,lucy14}.
This process becomes extremely challenging and ambiguous when the lines are broad and severe line blending causes the BALs to overlap and create wide troughs in the spectrum.
Some FeLoBALQs show thousands of line transitions from \ion{Fe}{2}.
The spectral synthesis code {\it SimBAL} was introduced by \citet{leighly18a} and further developed by \citet{leighly18b} and \citet{choi20}.
We describe the basic properties here but refer to those publications for details.
{\it SimBAL} uses a forward-modeling technique that allows the analysis of heavily absorbed and blended BALQ spectra.
Six physical parameters are required to create an individual synthetic BAL.
The parameters are: the dimensionless ionization parameter $\log U$, the gas density $\log n\ \rm[cm^{-3}]$, a column density parameter $\log N_H-\log U\ \rm[cm^{-2}]$ which represents the thickness of the gas column with respect to the hydrogen ionization front, the velocity offset $v_{off}\rm\ (km\ s^{-1})$, width of the absorption lines $v_{width}\rm\ (km\ s^{-1})$, and a dimensionless covering fraction parameter $\log a$ where larger $\log a$ represents lower partial covering of the emission source.
{\it SimBAL} uses the Markov chain Monte Carlo (MCMC) method, {\tt emcee}\footnote{http://dan.iel.fm/emcee/current/} \citep{emcee}, to compare the synthetic spectrum with the observed spectrum.
From the converged chain of parameter values we construct posterior probability distributions for the fit parameters.
The best-fitting model, the parameters, and their uncertainties are extracted from the posterior probability distributions.

\citet{leighly18a} demonstrated the use of the ``tophat accordion model'' to model the opacity profile of broad absorption lines.
Gaussian opacity profiles often fail to accurately model heavily saturated BAL troughs or fit small scale structures found within the wide troughs.
The tophat accordion model uses a group of velocity-adjacent rectangular bins to divide the opacity profile into smaller velocity segments.
Each tophat bin may be allowed to have independent physical properties or they may be constrained to have a common value.
Therefore, not only do we obtain a better fit to the complex velocity structures of the BAL outflows, but we can also measure the physical properties of the outflowing gas as a function of velocity.
\citet{leighly18b} introduced a two-covering factor model where the absorption lines can have different partial covering parameter $\log a$ values for the continuum and the emission lines.

\citet{choi20} discussed the updated atomic data and the grid of column densities created using {\it Cloudy} version c17.01 \citep{ferland17}.
The main focus of the update was to incorporate more excited-state transitions and more iron-peak elements.
However the line transitions from those excited state ions or rare metal ions do not contribute to the spectrum for low-ionization and low-density BAL outflows.
To save computation time, some of the objects were modeled using the column density grid from the older version of {\it SimBAL} which used the c13.03 version of {\it Cloudy} \citep{ferland13}.

\subsection{Updates to {\it SimBAL}}\label{subsec:simbal_update}
{\it SimBAL} has been updated to include three substantial enhancements that improve the accuracy of outflow parameters: (1) use of spectral eigenvectors for continuum fitting, (2) addition of anomalous reddening \citep{choi20} as an option, and (3) incorporation of wavelength-dependent instrument resolution into the synthetic spectrum generation.
The current version of {\it SimBAL} uses spectral eigenvectors from spectral principal component analysis (SPCA) to model the emission lines.
These eigenvectors enables {\it SimBAL} to employ only a small number of parameters for the emission-line model.
A detailed discussion of the construction of the eigenvectors used in this paper is found in Appendix~\ref{app:model_cont}.

We used a power law for the continuum, and the SMC reddening curve \citep{prevot84} or a general reddening curve \citep{choi20} to reproduce the reddening in the spectra depending on whether the object showed a typical continuum shape or anomalous reddening with a break.

Another update to {\it SimBAL} involves incorporating the instrument resolution into synthetic spectrum \citep{macinnis18}.
The spectrum data file from the SDSS provides information about wavelength dispersion at each pixel.
{\it SimBAL} uses this information to create a resolution matrix and performs a convolution with synthetic spectrum with the matrix.
The process is precisely analogous to the ``Response Matrix Function'' (RMF) matrix used for forward modeling in X-ray spectral fitting.
The shapes of the absorption lines change with the convolution; the principal effect is that the absorption lines with intrinsically narrow widths become noticeably shallower and wider in the spectra once the instrument resolution has been taken into account \citep{macinnis18}.
The other effect is aesthetic; the steps in the step-function opacity (from the tophat accordion model) appear smoothed out.
Without the resolution convolution, {\it SimBAL} would modify the fit parameters to mimic the effect of the instrument resolution,
finding a larger value of the partial covering parameter $\log a$, corresponding to a lower covering fraction, and a larger BAL width.
This added step not only helps {\it SimBAL} to create a more realistic spectrum that more closely matches the data, but also provides more accurate properties of the outflowing gas when the lines are very narrow.

\subsection{{\it SimBAL} Analysis of FeLoBALQ Spectra}\label{subsec:simbal_method}
{\it SimBAL} can model FeLoBALQ spectra with various spectral morphologies using the updates made to {\it SimBAL} (\S~\ref{subsec:simbal_update}) and the software allows the use of user-defined models that are based on various physical models of BAL outflows.
For example, in some objects (13 out of 53) we found evidence for more than one BAL outflow absorber in the spectrum and we included more than one outflow component in their {\it SimBAL} models (\S~\ref{subsec:multiple_BAL}).
Each component was modeled using a set of accordion tophat bins or a Gaussian profile, each with their own independent set of physical parameters.
The multiple outflow components in a given object were found to have overlapping velocity structures or they were observed to be completely separated by velocity (Figure~\ref{fig:multiple_bals_model_1};~\ref{fig:multiple_bals_model_2}).
These multiple BALs were treated as independent outflows so that separate outflows were analysed independently and all figures include the points for these outflows.
These outflows are identified with a letter of the alphabet following the name of the object (Appendix~\ref{app:data_table}).
For some objects, strong narrow emission lines such as [\ion{O}{2}]$\lambda\lambda 3726,3729$ were removed from the spectrum before model fitting.

When using the tophat accordion model for the absorption features, a fixed number of bins need to be specified before the fitting process and a user can decide if they want to allow each bin to have their own independent physical parameters or a single set can be given to the entire ensemble of bins.
We started by fitting the spectra with one ionization parameter ($\log U$) and density ($\log n$) for all bins.
After we retrieved the preliminary model fits, we then tried to fit the objects with more flexible models where we would let the ionization parameter and/or density to vary between bins.
The number of bins for the tophat accordion models was determined based on the total width of the trough as well as whether the trough shows fine velocity structures.
The tophat accordion models for some of the objects with narrow BALs used 3 bins and the objects with the widest troughs used 12 bins.
Considering the spectral resolution of the data ($\sim70\rm\ km\ s^{-1}$) the minimum bin width was $\sim200\rm\ km\ s^{-1}$.
The choice of the total number of bins does not affect the {\it SimBAL} analysis, unless too few or too many bins are used \citep{leighly18a}.

The default partial-covering model in {\it SimBAL}, power-law partial covering parameterized using $\log a$ \citep{leighly18a,leighly18b}, provided robust model fits to the majority of our FeLoBALQs.
Moreover, a {\it SimBAL} model can have two sets of partial-covering parameters each applied to the continuum emission and the emission lines separately to reproduce the difference between the partial covering observed in continuum emission and in emission lines.
However, six objects required a modified partial-covering scheme to obtain robust model fits (\S~\ref{subsubsec:special_covering_model}).
The spectra of these six objects were initially fit with a pair of $\log a$ parameters in order to separately model the partial-covering of the line emission and the continuum emission.
If the results of that fit argued that the line emission is not absorbed by the BAL components, we then proceeded with the modified partial-covering model where the BAL components are not allowed to absorb the line emission.
We performed statistical tests (e.g., $\chi^2$, $F$-test) to confirm our model selection.

Notes on the {\it SimBAL} model fits and specifications (e.g., number of tophat bins) for individual objects are given in Appendix~\ref{app:model_detail}.

\subsubsection{Objects and BALs Excluded from the Analysis}\label{subsubsec:bal_exclude}
Based on preliminary analysis with {\it SimBAL}, we excluded three more objects from the initial sample of 53 objects:
(1) the absorption features observed in the spectra showed properties more consistent with them being an intervening absorber than a quasar-driven outflow, (2) no clear FeLoBAL feature, with observable \ion{Fe}{2} absorption lines, was detected in the spectra.
Following the same reasoning we also excluded several outflow components that were identified with using multi-component outflow models.

Some of the objects had very narrow features and only ground state transitions that were more consistent with metal lines from intervening absorbers.
They could be distinguished from intervening absorbers by partial covering and kinematic properties
\citep[e.g.,][]{hamann11}; see discussion in Appendix~\ref{app:interv_absorp}.
Additionally, the physical properties of a couple of BAL components could not be constrained reliably because only \ion{Mg}{2} (and \ion{Al}{3} in some cases) absorption lines were present in the spectra.
For these reasons, we rejected two objects: SDSS~J1057$+$6109 and in SDSS~J0338$+$0056.
The absorption lines we found in SDSS~J1057$+$6109 has an extreme offset velocity and the very narrow width of the absorption lines that suggested an intervening absorber.
The best-fitting model for SDSS~J0338$+$0056 identified an extreme LoBAL in this object with \ion{Mg}{2} trough located at $\sim2500$ \AA\/ and spanning from $-43,300\rm\ km\ s^{-1}$ to $-26,400\rm\ km\ s^{-1}$, but no absorption from \ion{Fe}{2}.
Also, the assumed \ion{Mg}{2} trough sits on top of the \ion{Fe}{2} emission lines which makes this BAL identification uncertain.
No other absorption lines besides the \ion{Mg}{2} trough was observed and we could not constrain the physical properties of the BAL gas with any certainty.
Therefore we excluded these two object from the analysis.

We excluded one of the two BALs found in SDSS~J1214$-$0001 from further analysis.
We identified two BAL systems in SDSS~J1214$-$0001 \citep{pitchford19}; they were modeled with a Gaussian profile component for a narrow absorption component and a set of tophat bins for the broad trough feature.
However, we could not get a reliable continuum emission model due to the broad tophat component with $v_{width}\sim10800\rm\ km\ s^{-1}$ and $v_{off}\sim-12600\rm\ km\ s^{-1}$ covering the entire \ion{Fe}{2} emission line region ($\sim$2200--2750 \AA\/) with low apparent opacity.
Furthermore, we found the constrained physical properties for this broad BAL component from the model fit to be unreliable.
We kept the narrow BAL component from this object in the BAL sample because it was not affected by the uncertainty in the continuum emission placement.
In addition, we did not find any absorption features in SDSS~J1737$+$5536.
This object was included in the \citet{farrah12} sample but the best-fitting model did not yield any absorption lines.
These three objects were excluded from further discussion, leaving 50 objects in our final sample.

In SDSS~J1324$+$0320 and SDSS~J1531$+$4852, we were able to clearly identify absorption features from metal lines from an intervening absorber.
They were found at extreme offset velocities from the quasar rest-frame ($v_{off}\lesssim-20,000\rm\ km\ s^{-1}$) and had narrow absorption line width ($v_{width}\lesssim50\rm\ km\ s^{-1}$) which strongly suggested that these absorption lines did not originate in a quasar-driven wind.
These absorption features were nonetheless modeled with {\it SimBAL} and included in the best-fitting model plots but are not included in the discussion.

The best-fitting model for SDSS~J1644$+$5307 has two absorption components.
The main \ion{Fe}{2} trough and the most of the BAL features were fit using a 6-bin tophat component but an additional Gaussian component at lower velocity was needed to fit the deep \ion{Mg}{2} trough.
However, the lower-velocity component only appeared in \ion{Mg}{2} and therefore we could not reliably constrain its physical properties.
Therefore this component was excluded from further discussion.
Similarly, the best-fitting models for SDSS~J0916$+$4534 and SDSS~J1531$+$4852 have extra Gaussian components included to fit low opacity absorption features ($I/I_0>0.9$) from the \ion{Mg}{2} transitions located near the main BAL component.
These components are essentially LoBAL absorbers.
They improved the overall model fit but we did not include them in the discussion because the opacity they contributed was insignificant and their physical properties could not be well constrained.

\section{Calculation of Critical Parameters}\label{sec:calculations}
\subsection{\texorpdfstring{Bolometric Luminosity Estimates and Spectral Index $\alpha_{ui}$}{Bolometric Luminosity Estimates and Spectral Index a_ui}}\label{subsec:lbol_aui_measure}
The bolometric luminosities ($L_{Bol}$) of the quasars used throughout the paper were calculated using the bolometric correction factor at 3 $\mathrm{\mu m}$ (BC=8.59) from \citet{gallagher07}.
The flux at rest-frame 3 $\mathrm{\mu m}$ was estimated from fitting the quasar composite SED by \citet{richards06} to the {\it WISE} photometry data and interpolating the flux at 3 $\mathrm{\mu m}$ from the composite SED.
The bolometric luminosity estimates are listed in Table~\ref{tbl:outflow_prop} (column 8)

We defined $\alpha_{ui}$ to be the point to point spectral slope between rest-frame 2000 \AA\/ and 3 $\mathrm{\mu m}$ flux densities:
$$\alpha_{ui}=\frac{\log f_{3\mathrm{\mu m}}-\log f_{2000\mathrm{\AA}}}{\log \lambda_{3\mathrm{\mu m}}-\log \lambda_{2000\mathrm{\AA}}}$$
We used the continuum emission model extracted from the best-fitting {\it SimBAL} model to estimate the observed flux density at rest-frame 2000 \AA\/.
The value of $\alpha_{ui}$ can depend on three quasar properties: reddening, the intrinsic shape of quasar SED, and the strength of the torus emission.
For instance, a quasar that either has a flat SED, is reddened, or has strong hot dust emission will have a flatter, i.e., a larger value of $\alpha_{ui}\sim0$.
For reference, the \citet{richards06} composite quasar SED has $\alpha_{ui}\sim-1.23$.
The values of $\alpha_{ui}$ for the sample are listed in Table~\ref{tbl:outflow_prop} (column 9).

\subsection{Derived Physical Properties and Kinematic Properties of the Outflows}\label{subsubsec:kinematics_measure}
Using the MCMC chains from the best-fitting {\it SimBAL} models, we computed the physical and kinematic properties of the outflows and the associated uncertainties.
Throughout the paper we report and plot median values and 2$\sigma$ (95.45\%) confidence intervals calculated from the posterior distributions as our uncertainty measurements.
Some of the physical properties can be directly extracted from the {\it SimBAL} physical parameters: dimensionless ionization parameter ($\log U$, definition below), gas density ($\log n$ in $\rm[cm^{-3}]$), column density (parameterized as $\log N_H-\log U$ in $\rm[cm^{-2}]$), and dimensionless covering fraction parameter $\log a$ for the inhomogenous partial covering \citep[$\tau=\tau_{max}x^a,\,x\,\in\,(0,1)$;][]{arav05,sabra05}.
We calculated the offset velocities and widths of the BALs, the distance of the BAL gas from the central SMBH $\log R$, the mass outflow rate $\dot M$, and the kinetic luminosity $L_{KE}$.
The covering-fraction-corrected column density of the outflow was calculated by summing $\log U$ and $\log N_H-\log U$ and then correcting the values according to the $\log a$ to account for the power-law partial covering \citep[$\log N_H=(\log N_H\sbond\log U)+\log U-\log(1+10^{\log a})$;][]{arav05,leighly18a,leighly18b,choi20}.

The radius of the outflow (or the distance of the outflow from the central engine) can be calculated from $\log U$ and $\log n$ from the {\it SimBAL} results using the definition of the ionization parameter
$$U=\frac{\phi}{nc}=\frac{Q}{4\pi R^2nc},$$
where $\phi$ is the photoionizing flux in the units of $\rm photons\, s^{-1}\,cm^{-2}$, and $Q$ is the number of photoionizing photons per second emitted from the central engine.
We estimated $Q$ from the SED fits of the photometry data for each object.

The mass outflow rate was calculated using
$$\dot M=8\pi\mu m_p \Omega R N_H v,$$
where the mean molecular weight ($\mu$) is assumed to be 1.4, the global covering fraction is given by $\Omega$, and $R$, $N_H$, and $v$ are calculated from the best-fitting parameters from {\it SimBAL}.
This equation can be derived from taking the time derivative of $M=4\pi\mu m_p \Omega R^2 N_H$, then substituting $dR/dt=v$ \citep[assuming $dN_H/dt=0$;][]{dunn10}.
The value of global covering fraction for FeLoBAL outflows
is uncertain.
The commonly used value $\Omega=0.2$ \citep[e.g.,][]{hewett03} was derived from a fraction of HiBALQs observed in the optically selected sample of quasars; the fraction can be as large as $\sim0.4$ for the luminous infrared-selected quasars \citep{dai08}.
FeLoBALQs are a rarer kind of BALQs; the observed fraction can be as small as $\sim1\%$ in a given quasar sample \citep[e.g.,][]{trump06,dai12},
but it is not clear whether their rarity reflects a small covering fraction, a short lifetime, or a selection bias that makes them difficult to detect.
Detailed discussion on how to best explore the different values of global covering fraction of FeLoBAL outflows can be found in \citet{choi20} where we also performed a multiple global fraction scenario with the idea that a BAL outflowing gas can be seen as different types of BALs depending on the viewing angle.
In this work, we adopt a single global covering fraction $\Omega=0.2$ which yields the mass outflow rate estimates for FeLoBALs that are slightly larger ($\sim0.5$ dex) than the multiple global fraction scenario calculations \citep{choi20}.
The strength of the outflow can be quantified by calculating the kinetic luminosity ($L_{KE}$) of the outflows with the equation $\dot E_k=\dot Mv^2/2$.
The outflow column density, mass outflow rate, and kinetic luminosity for BALs modeled using tophat accordion models were calculated from the sum of the values calculated for each tophat bin.

In order to generate summary statistics for the widths ($v_{width}$) and the offset velocities ($v_{off}$) of the BAL outflows in a consistent manner we adopted a method similar to calculating the balnicity index \citep[BI;][]{weymann91}.
Continuum emission normalized spectra ($I/I_0$) for a single line transition were generated from the best-fitting {\it SimBAL} models and absorption features were defined as regions where the normalized flux $I$ fell below 0.9.
We used the \ion{Mg}{2}$\lambda2796$ line transition which has a higher transition probability ($f_{ik}=0.609$) in the \ion{Mg}{2}$\lambda\lambda 2796,2803$ doublet to generate the $I/I_0$ for all identified BAL outflows.
We estimated the width of the outflow by identifying the start ($v_{max}$) and the end ($v_{min}$) of the absorption features.
The summary offset velocity for each BAL component was estimated by calculating the opacity (column density) weighted velocities.
We used
$$v_{off}=\frac{\int v\times N_H(v)dv}{\int N_H(v)dv}$$
where $N_H$ represents the covering-fraction-corrected hydrogen column density.
The tophat accordion model produces physical parameters of the outflow as a function of velocity with which we calculated the summary offset velocity for an ensemble of tophat bins for a given BAL component.
For Gaussian profiles, this calculation simply yields the velocity at the center of the profile.
By calculating the velocity offsets this way, we avoid overestimating or underestimating the velocities compared to other metrics (e.g., $v_{max}$), especially when the trough has extended low opacity features in one or both velocity directions.
In this paper we retain the signs for the offset velocities as calculated from the quasar rest frame: the outflows have negative offset velocities and the inflows have positive velocities.
For the objects that were modeled with more than a single BAL component, the outflow width and the velocity offset were measured for each component.

\subsection{Opacity Profiles and BAL strengths of Select Transitions}\label{subsubsec:gen_opacity_prof}
We extracted the opacity profiles of several absorption line transitions seen in FeLoBALs from the best-fitting {\it SimBAL} models in addition to the \ion{Mg}{2}$\lambda2796$ used to calculate the kinematic properties (\S~\ref{subsubsec:kinematics_measure}).
The transitions we used are the ground-state \ion{Fe}{2}$\lambda 2383$ with $f_{ik}=0.343$ and the excited-state (0.99 eV above ground) \ion{Fe}{2}*$\lambda 2757$ with $f_{ik}=0.307$.
These are among the stronger transitions from \ion{Fe}{2} multiplet in the near-UV band.
We generated $I/I_0$ models for all BAL components with {\it SimBAL} from the MCMC chains of the best-fitting spectral models.
The wavelength dispersion at each pixel was also taken into account in generating the profiles (\S~\ref{subsec:simbal_update}).
From the models, we visually inspected the shapes of the profiles and measured the kinematic information for each line transition.
The widths for each transition were measured using the same method described in \S~\ref{subsubsec:kinematics_measure}: we used the normalized flux at 0.9 as the boundary of absorption.
In addition, we measured the velocity at the location of the minimum normalized flux for each transition.
We emphasize that $v_{off}$, opacity weighted velocity (\S~\ref{subsubsec:kinematics_measure}), is used throughout the paper as the representative outflow velocity; the velocities at the minima are just used to compare the kinematic properties of the 3 transitions (\S~\ref{subsec:vel_profile}).

We extracted \ion{Ca}{2} K, \ion{He}{1}*$\lambda 3889$, and H$\alpha$ information from the {\it SimBAL} models for each BAL component.
We used the absorption strength (A) defined by \citet{capellupo11} to measure the BAL strengths.
The absorption strength is defined as the fraction of the normalized flux removed by absorption.
This empirical parameter may not accurately reflect the physical properties of the outflowing gas, but it can be used as summary statistics for the apparent BAL strength.
For each BAL component, a single absorption strength value was reported from the averaged model flux within the absorption interval where $I/I_0<0.9$.
To represent the strengths of \ion{Ca}{2} doublet and Balmer series, we chose the transitions with the greatest transition probability which are \ion{Ca}{2} K and H$\alpha$, respectively.
Depending on the redshift of the FeLoBALQs the three line transitions of interest for some of the objects are located outside the bandpass of the SDSS/BOSS spectra used in this work.
Moreover, we only modeled the spectra
to $\sim4700$ \AA\/
which means that the strongest Balmer transition
observed in the given wavelength range was H$\gamma$.

\section{Results}\label{sec:results}
\subsection{Best-Fitting {\it SimBAL} Models}\label{subsec:bestfitmodels}
\begin{figure*}
\epsscale{.97}
\plotone{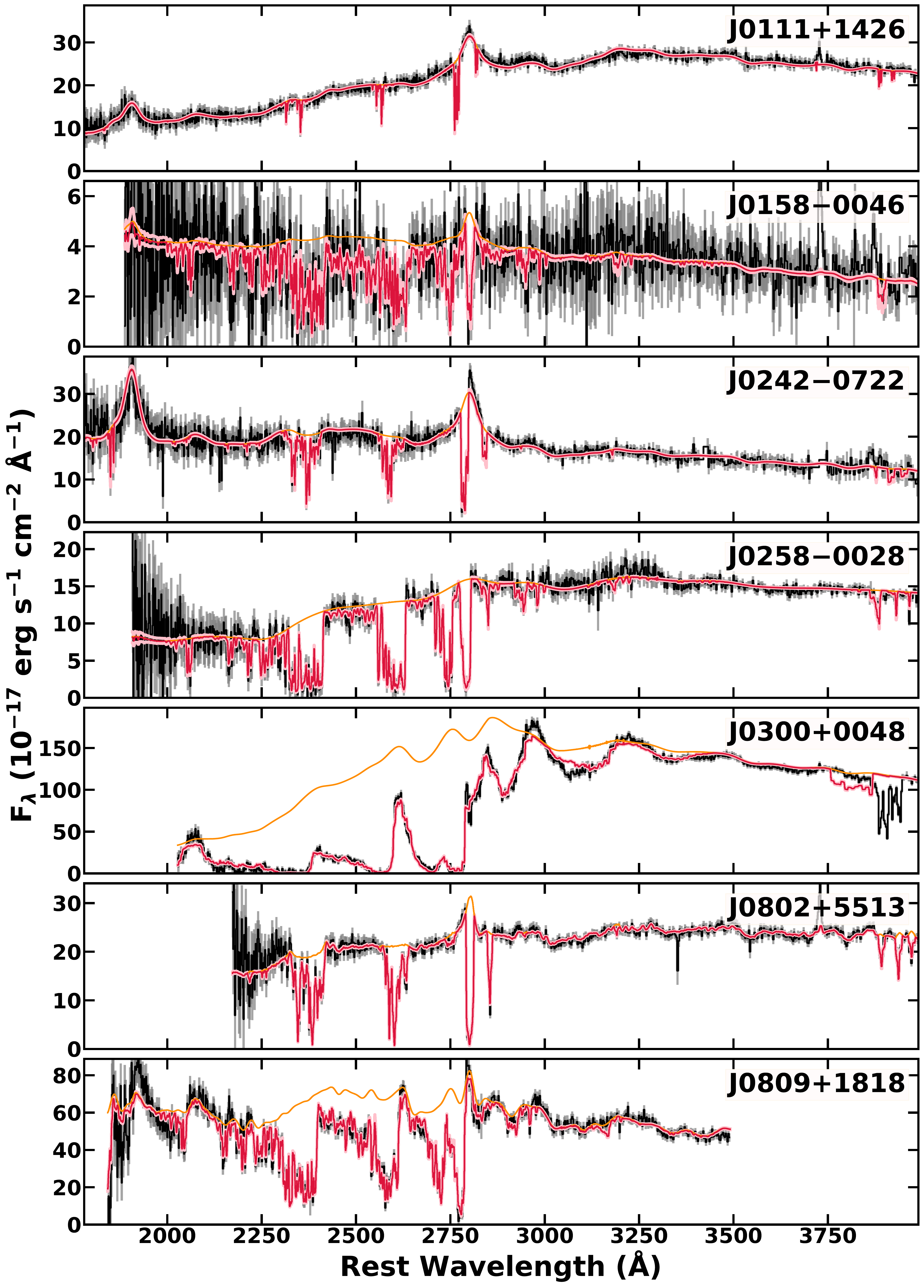}
\caption{The best-fitting {\it SimBAL} models are plotted in red, along with the 2$\sigma$ (95.45\%) confidence models and unabsorbed continuum models in pink and orange, respectively.
The binned data and the associated uncertainties are plotted in black and grey, respectively.
\label{fig:fitfig1}}
\end{figure*}
\addtocounter{figure}{-1}
\begin{figure*}
\plotone{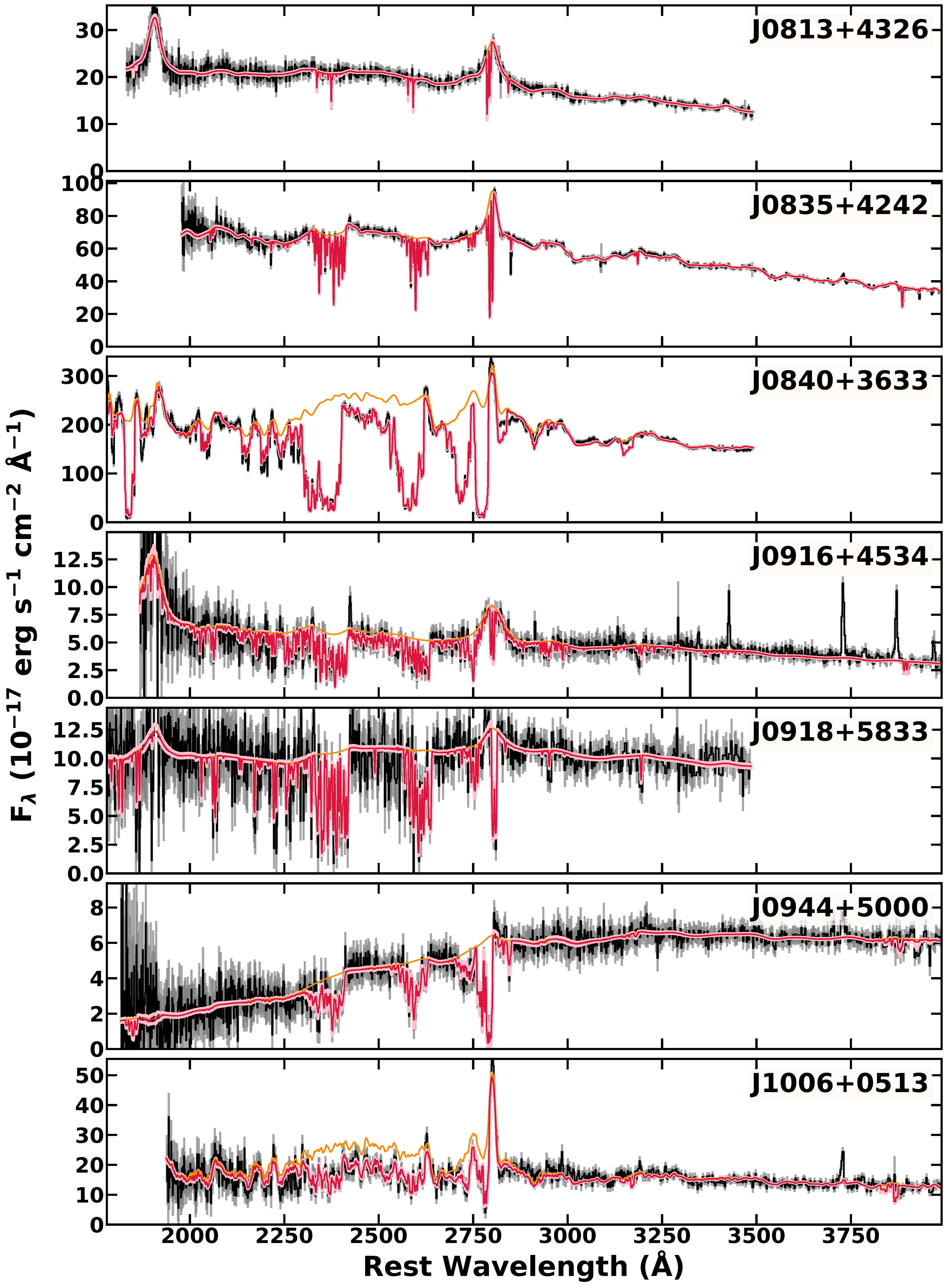}
\caption{(Continued).}
\end{figure*}
\addtocounter{figure}{-1}
\begin{figure*}
\plotone{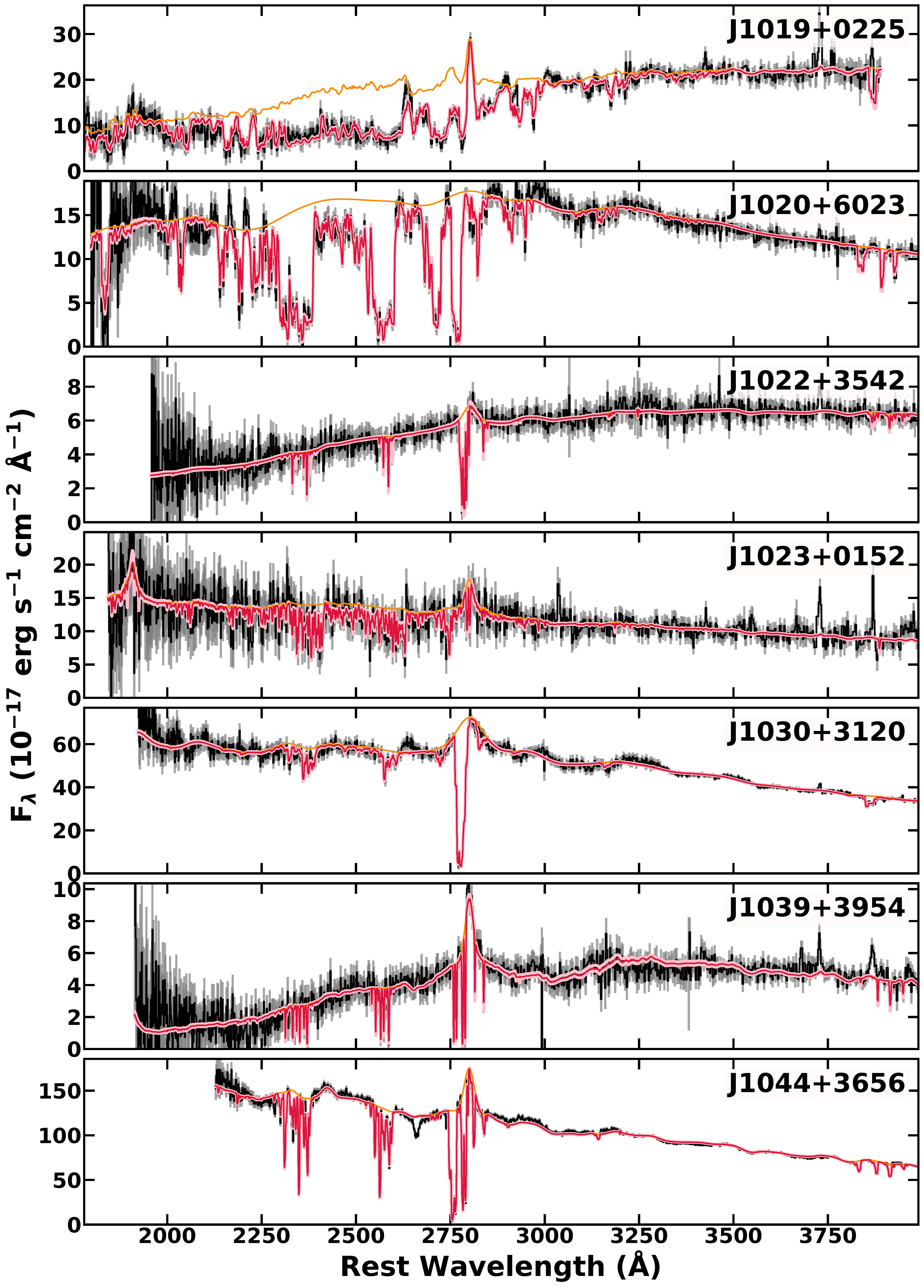}
\caption{(Continued).}
\end{figure*}
\addtocounter{figure}{-1}
\begin{figure*}
\plotone{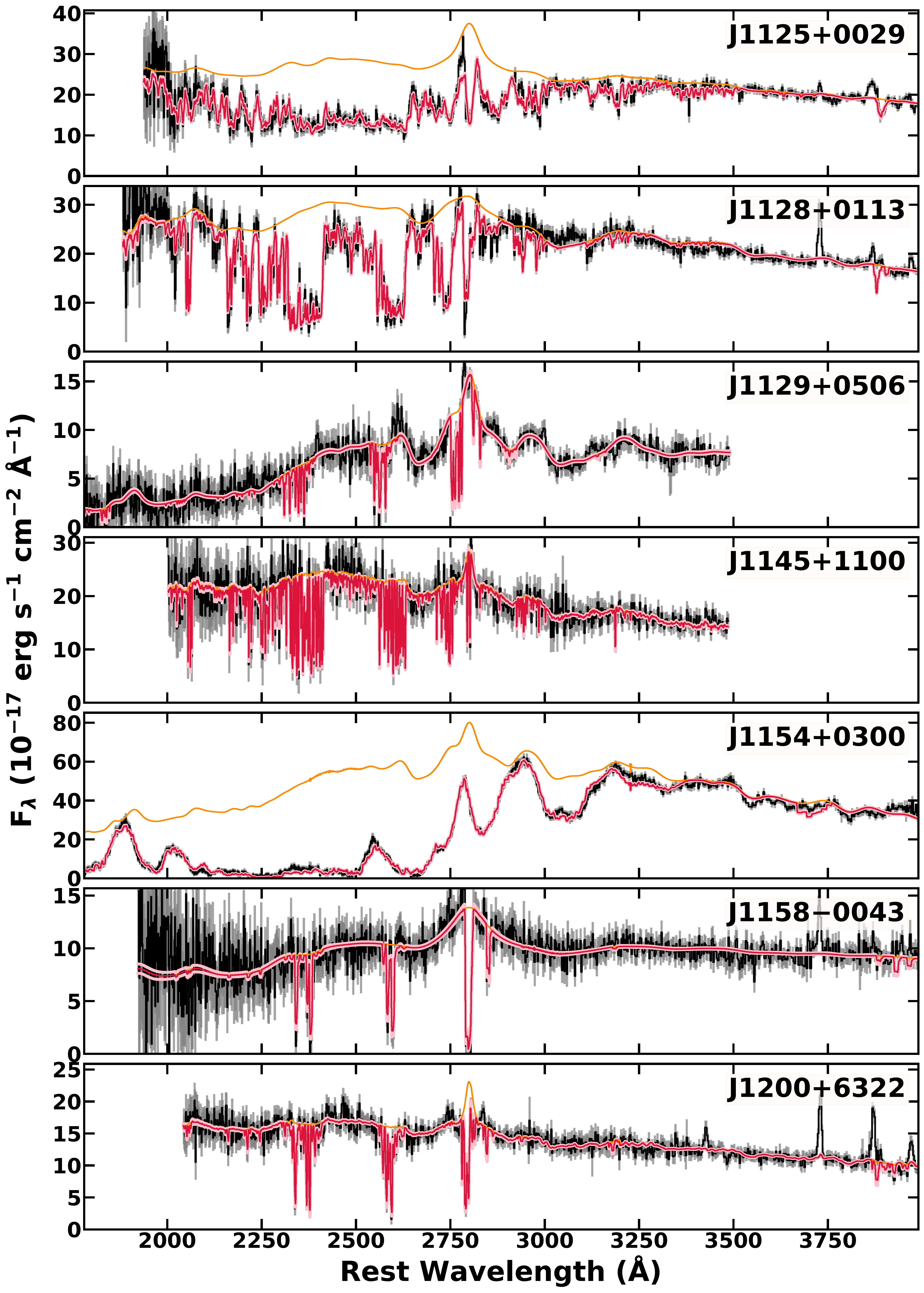}
\caption{(Continued).}
\end{figure*}
\addtocounter{figure}{-1}
\begin{figure*}
\plotone{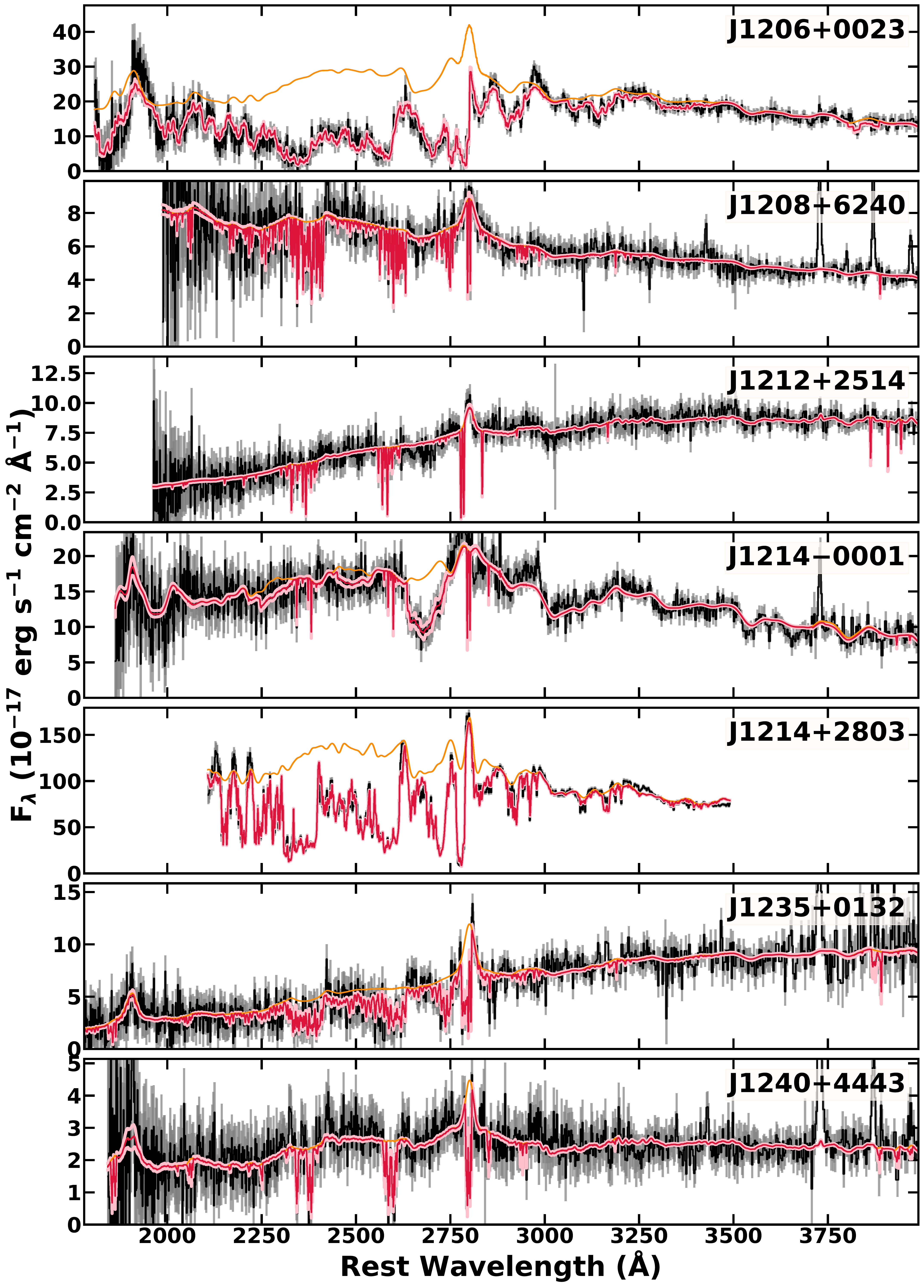}
\caption{(Continued).}
\end{figure*}
\addtocounter{figure}{-1}
\begin{figure*}
\plotone{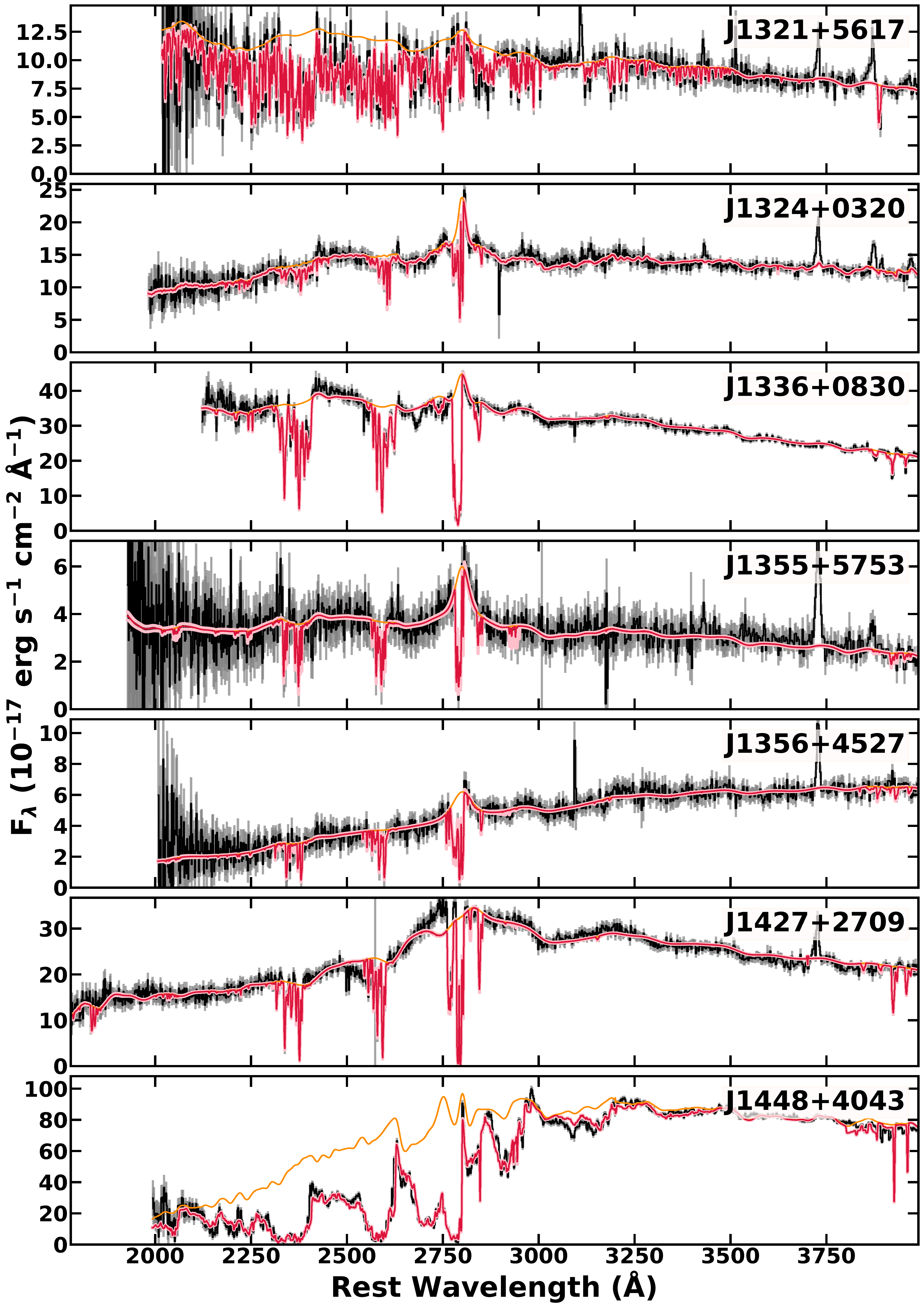}
\caption{(Continued).}
\end{figure*}
\addtocounter{figure}{-1}
\begin{figure*}
\plotone{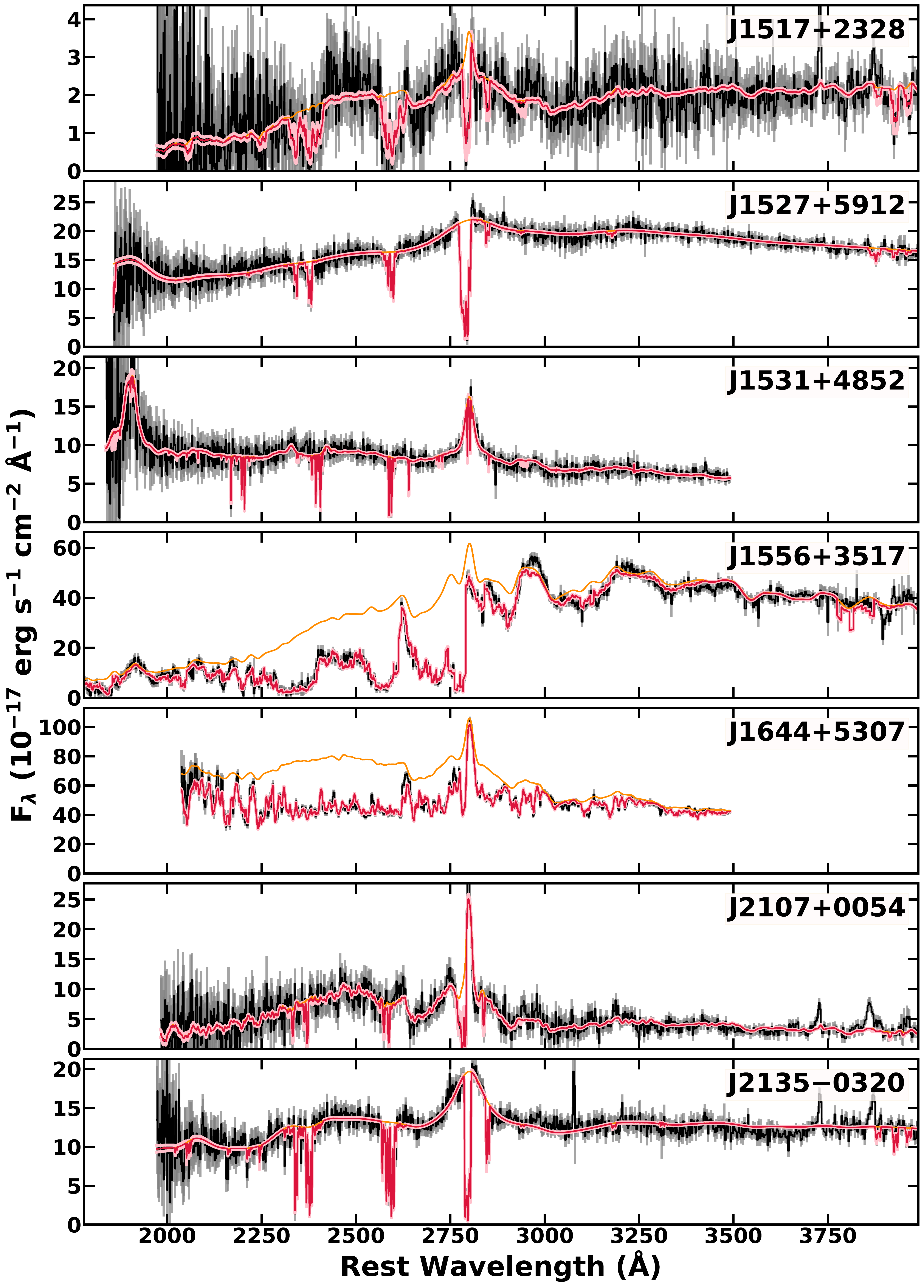}
\caption{(Continued).}
\end{figure*}
\addtocounter{figure}{-1}
\begin{figure*}[ht!]
\plotone{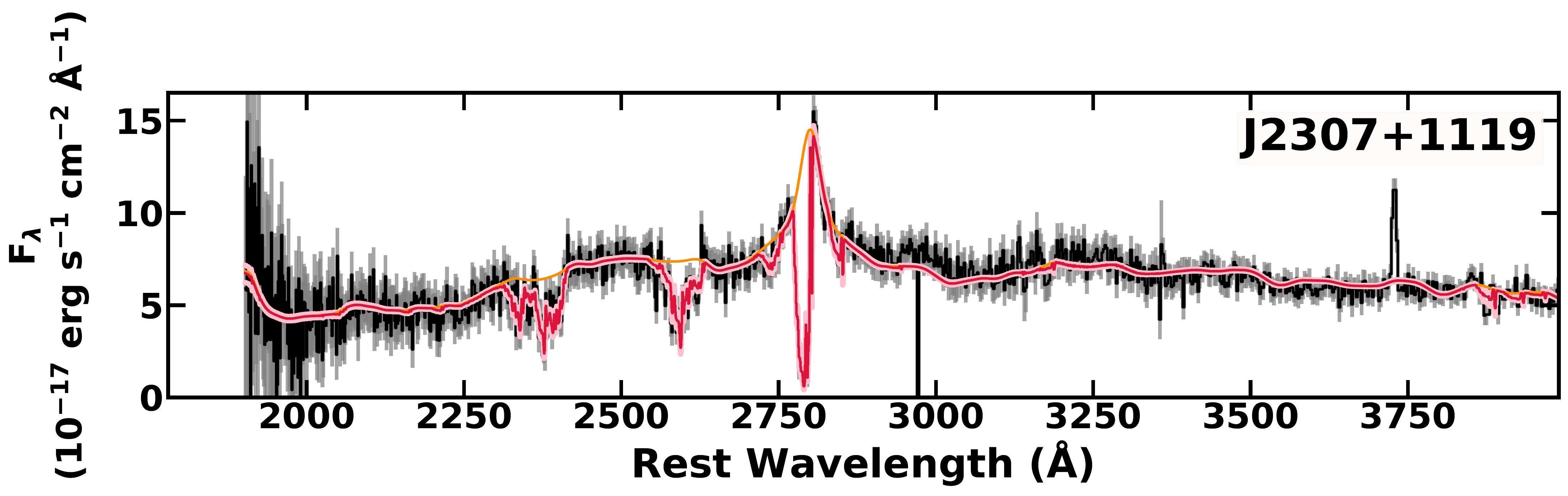}
\caption{(Continued).}
\end{figure*}
We present {\it SimBAL} model fits of the FeLoBALQs.
Figure~\ref{fig:fitfig1} shows the best-fitting models of all 50 objects from the sample.
From these results we derived the physical properties and calculated the outflow properties, with associated uncertainties (Appendix~\ref{app:data_table}).
We identified 60 BAL features and 55 of them were classified as outflows with negative offset velocities.
Eleven objects in the sample were modeled with more than one outflow component where either multiple sets of tophats bins or a combination of tophat bins and Gaussian profiles were used.

Analysis of SDSS~J1352$+$4239 \citep{choi20} showed that the outflow consisted of three components that were distinguished by their distinct physical properties as well as the kinematic properties.
The majority of the multiple-outflow objects we discovered in this sample were found using similar rigorous {\it SimBAL} modeling.
We found that majority of the objects with the accordion tophat models were well fit with a single set of tophat bins with a single ionization parameter and density.
Nonetheless, {\it SimBAL} model fits of some objects revealed a subset of bins that showed significantly different physical properties (e.g., ionization parameter and density).
For those objects, we divided the tophat bins into two or three groups with a single ionization parameter and density for all bins in each group.
In some objects a Gaussian profile was used for the lower-velocity components: SDSS~J0258$-$0028, SDSS~J1125$+$0029, and SDSS~J1448$+$4043 (Figure~\ref{fig:multiple_bals_model_2}).

\subsubsection{{\it SimBAL} Models with Modified Partial Covering}\label{subsubsec:special_covering_model}
\begin{figure}
\epsscale{1.05}
\plotone{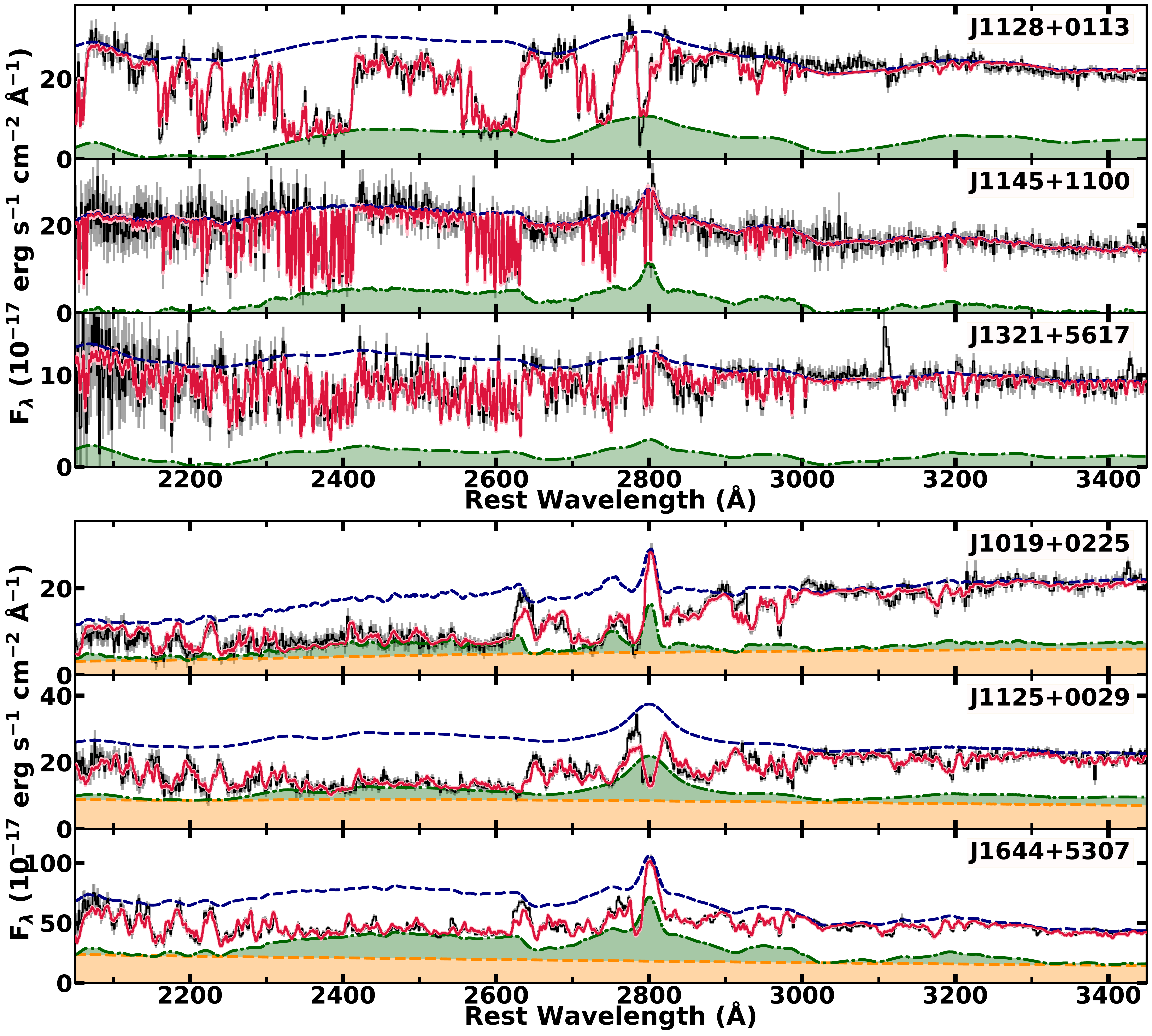}
\caption{The best-fitting {\it SimBAL} models for six objects modeled using the modified partial covering scheme.
The best-fitting models and the unabsorbed continuum emission models are shown in red solid lines and blue dashed lines, respectively.
\textit{Top}: The FeLoBAL gas does not absorb the emission lines (green shaded-region).
\textit{Bottom}: The FeLoBAL gas does not absorb both emission lines and a fraction of the continuum emission (green and orange shaded-regions).
The power-law partial covering ($\log a$) used in \textit{SimBAL} alone could not model the heavy non-black saturation seen in these spectra.
\label{fig:no_line_model}}
\end{figure}

Six objects in the sample required {\it SimBAL} models with the modified partial covering scheme.
The best-fitting models for SDSS~J1128$+$0113, SDSS~J1145$+$1100, and SDSS~J1321$+$5617 included unabsorbed emission line components
and SDSS~J1019$+$0225, SDSS~J1125$+$0029, and SDSS~J1644$+$5307 required a fraction of continuum emission to be unabsorbed by BALs
(Figure~\ref{fig:no_line_model}).
Five out of 6 objects are further classified as ``loitering outflow objects'' (\S~\ref{subsubsec:felobal_classification};~\ref{subsec:loiter}).

In the extreme cases of non-zero flux at the bottoms of the troughs in the spectra of SDSS~J1019$+$0225, SDSS~J1125$+$0029, and SDSS~J1644$+$5307,
an unabsorbed line-emission component alone was not sufficient to model the spectral features, and unabsorbed flux from the continuum emission was necessary to obtain the best-fitting spectral models.
We modified the spectral model for these objects in two ways to account for the large amount of flux seen underneath the troughs.
First, the line emission is not absorbed by the BAL, and second, only a fraction of power-law continuum emission is absorbed by BAL similar to the homogeneous step-function partial covering
\citep[e.g., $I=C_f e^{-\tau}+(1-C_f$), $C_f=$covering fraction;][]{arav05}.
The absorbed part still requires the power-law opacity ($\log a$) to model the significant contribution of weak absorption-lines.
In other words, the BAL winds in these objects have both the inhomogenous partial covering presumably originating from the complex cloud structures within the BAL gas \citep[e.g.,][]{dekool02,leighly18b} and the homogeneous partial covering that originates from the BAL gas absorbing only a fraction of continuum and none of the line emission from the central engine.
With this model setup, we were able to create the overlapping trough features with blended saturated absorption lines and still have a significant amount of flux underneath the troughs.
From the three objects we obtained $C_f\sim0.65-0.71$.

We note that the lower-velocity components in the spectral models for SDSS~J1125$+$0029 and SDSS~J1644$+$5307 were allowed to freely absorb both the continuum and line emission in a standard fashion unlike the higher-velocity components as described above.
That is because the lower-velocity components in these objects are presumed physically separate from the main higher-velocity component and are located at larger distances from the accretion disk and the BLR.
This allowed the lower-velocity components to fit deep \ion{Mg}{2} absorption features seen in the spectra.

\subsubsection{Absorption Lines in FeLoBALs}\label{subsubsec:felobal_anatomy}
\begin{figure}
\epsscale{1.1}
\plotone{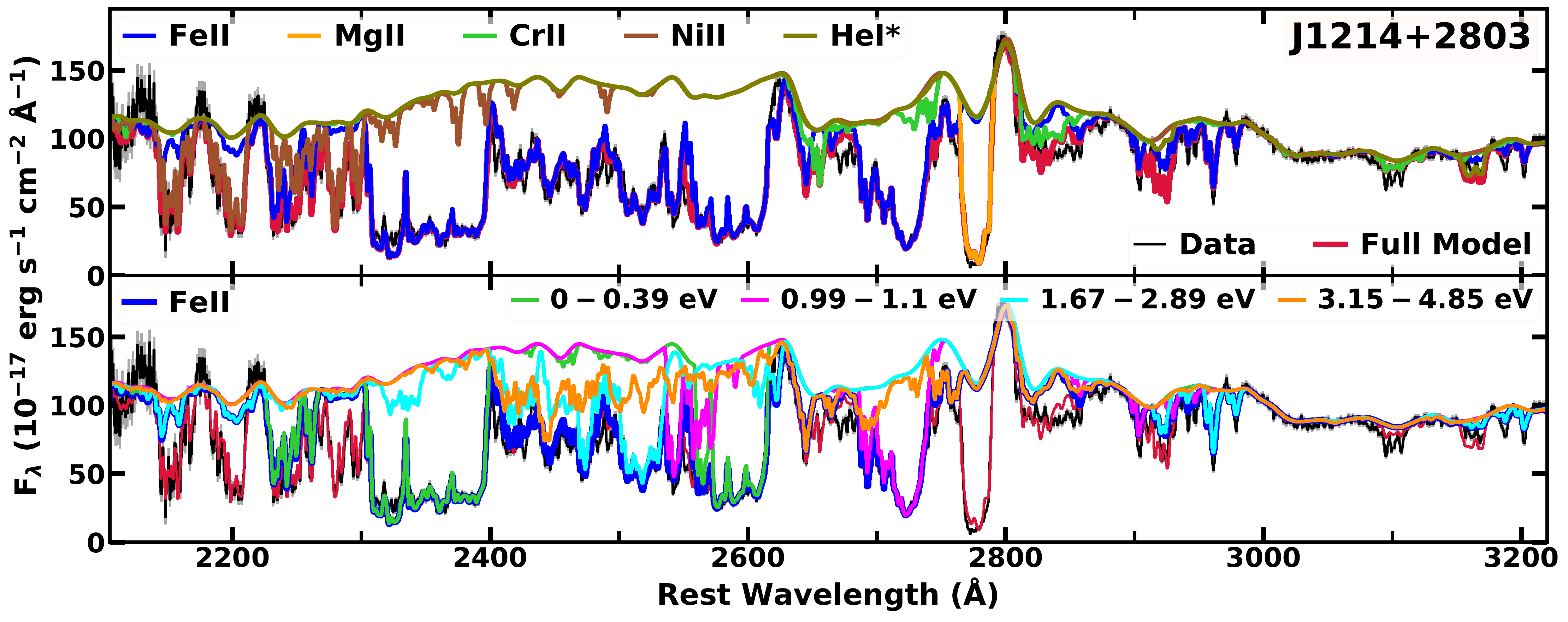}
\caption{Anatomy of the near-UV spectrum of an FeLoBAL quasar.
{\it SimBAL} models for SDSS~J1214$+$2803 showing some of the principal line transitions observed in FeLoBAL troughs is plotted.
{\it Top}: The best-fit model has been divided by the ionic species that are the major contributors of the opacity in the near-UV bandpass for FeLoBALs.
{\it Bottom}: The \ion{Fe}{2} model has been further broken up into 4 different models depending on the lower-level excitation energy.
\label{fig:felobal_anatomy}}
\end{figure}
FeLoBALs are known to show absorption lines from \ion{Fe}{2} as well as various iron-peak elements, such as \ion{Cr}{2}, \ion{Ni}{2}, in the near-UV wavelengths from their high column density gas \citep[e.g.,][]{dekool02f2,choi20}.
Often these transitions are blended and thus isolation of individual line transitions is not possible in many cases, making it extremely challenging to analyze the FeLoBALQ spectra and to constrain the physical properties of the FeLoBAL absorbing gas.
The forward modeling technique used in {\it SimBAL} allows us not only to analyze heavily absorbed spectra with line blending features, but also to study the individual line transitions observed in the spectra by spectral synthesis.

Figure~\ref{fig:felobal_anatomy} highlights some of the major absorption lines observed in FeLoBALs with special {\it SimBAL} models generated using the best-fitting model parameters and user-defined line transition lists.
We used SDSS~J1214$+$2803 as an example because this object has a high-opacity BAL absorbing gas ($\log N_H-\log U\sim23.4\ \rm[cm^{-2}]$).
The FeLoBAL in this object showed most of the prominent absorption lines found in FeLoBALs.
The strength and/or the presence of (or the lack of) absorption lines from certain ionic species provide specific information about the physical conditions of the outflowing gas.
The iron-peak elements are rare (e.g., [Ni/Fe]$\sim-1.2$ for $Z=Z_\odot$) and therefore their absorption lines only appear when the column density is sufficiently high and these lines are often not saturated unlike \ion{Fe}{2} or \ion{Mg}{2} absorption lines.
For example, the absorption feature near $\lambda\sim2200\rm\ \AA$ from \ion{Ni}{2} can be used as an excellent indicator of outflow column density.
They suffer less line blending with \ion{Fe}{2} absorption lines and produce stronger absorption lines compared to other rare iron-peak elements, such as \ion{Cr}{2}.
This particular feature is not observed in FeLoBALQs with low column density BALs (e.g., SDSS~J2307$+$1119, $\log N_H-\log U\sim23.0\ \rm[cm^{-2}]$ in Figure~\ref{fig:fitfig1}).
In addition, absorption line from \ion{He}{1}*$\lambda 3188$, also noted in Figure~\ref{fig:felobal_anatomy}, has been previously known as column density and ionization diagnostics \citep{leighly11}.
For FeLoBALs, the ionization parameter is directly related to the column density of the gas (\S~\ref{subsec:derived_pars}; Figure~\ref{fig:fitpar_aNh}), which means the overall amount of opacity observed in the troughs scales with ionization and this parameter is constrained by not just a subset of line transitions, but by an ensemble of absorption lines.

The bottom panel in Figure~\ref{fig:felobal_anatomy} shows the models of \ion{Fe}{2} absorption lines that have been grouped by the lower-level excitation energy.
\ion{Fe}{2} has a large number of excited state levels and the plethora of absorption lines they produce can be used to constrain the density of the outflowing gas \citep[e.g.,][]{lucy14,choi20}.
Most FeLoBAL quasars show excited state \ion{Fe}{2}* absorption lines with lower-level excitation of $E_{lower-level}\sim1$ eV in their spectra (e.g., SDSS~J0840$+$3633).
Most excited state \ion{Fe}{2} transitions have critical densities $\log n\sim4.5\ \rm[cm^{-3}]$ or greater \citep[e.g.,][]{korista08}.
If the density exceeds this value, then the BAL will show strong absorption lines from \ion{Fe}{2} across various excited energy levels, including high-excitation levels ($E_{lower-level}\sim5$ eV) especially when the density is high.
For instance, SDSS~J1214$+$2803 shown in the figure has a high density outflowing gas ($\log n\sim7.8\ \rm[cm^{-3}]$) and we observe significant opacity from the \ion{Fe}{2}* absorption lines form multiple excited levels.
If an absorbing gas has low enough density and no \ion{Fe}{2} in the excited states, then the FeLoBAL will mostly only show absorption features from the ground state \ion{Fe}{2} (e.g., SDSS~J0802$+$5513, $\log n\sim4.4\ \rm[cm^{-3}]$).
In some BALs with no excited state \ion{Fe}{2}* absorption lines, we were only able to constrain density upper limits ($\log n\ll2.8\ \rm[cm^{-3}]$, column density grid limit).
And thus, these BALs will have lower limits on their distance estimates from the central engines (see \S~\ref{subsubsec:kinematics_measure}).
Also, high ionization condition can populate \ion{Fe}{2} to excited states.

\subsubsection{Classifications of FeLoBALs}\label{subsubsec:felobal_classification}
In the figures throughout the paper, we mark three special types of FeLoBALs identified in the sample: (1) Overlapping troughs, (2) Loitering BALs, and (3) Inflows.
The overlapping trough identification was done based on the spectral morphology.
If an object showed heavily blended \ion{Fe}{2} absorption feature in the spectrum shortward of $\lambda\sim2800\rm\ \AA$, we then classified it as having overlapping trough features (e.g., SDSS~J0300$+$0048 and SDSS~J1154$+$0300).
Second, based on the outflow properties obtained from {\it SimBAL} modeling, we classified ``loitering BALs'' by selecting BALs that have $\log R< 1$ [pc] and $\vert v_{off,\ \mathrm{FeII\ excited}}\vert<2000\rm\ km\ s^{-1}$ (offset velocities calculated from the excited-state \ion{Fe}{2}$^*\lambda 2757$; \S~\ref{subsubsec:gen_opacity_prof}).
Lastly, the inflows ($v_{off}>0\rm\ km\ s^{-1}$) were classified based on the opacity weighted velocities (\S~\ref{subsubsec:kinematics_measure}).
We note that unlike loitering BAL or inflow classifications, the overlapping trough classification is solely based on empirical visual classification and not based on the physical and kinematic properties.
Also, FeLoBAL can have multiple classifications.
For instance, majority of loitering BALs are also classified as overlapping troughs and inflows.
Following the classification scheme we identify: 8 overlapping troughs, 11 loitering BALs, and 5 inflows.
41 out of 60 FeLoBALs do not belong in any of the three special classes and they can be considered typical outflowing FeLoBALs with moderate opacity.
Detailed discussion on overlapping troughs and loitering BALs can be found in \S~\ref{subsec:ot_obj} and \S~\ref{subsec:loiter}, respectively.

Following the theoretical predictions \citep[e.g.,][]{scannapieco04,dimatteo05,hopkins10}, we identified energetic BALs as those that have outflow energy ($L_{KE}$) greater than 0.5\% of the bolometric luminosity ($L_{Bol}$) of the quasar.
Because some of the BAL outflows were constrained with lower limits on the distances of the BAL absorbers from the central engines, not all BALs in our sample have robust constraints on the outflow energy calculations.
Therefore, among the energetic BALs that meet $L_{KE}>0.005L_{Bol}$ condition we only selected the ones with well constrained physical properties (e.g., $\log U$, $\log n$) as powerful BALs.
They are represented in the figures with cyan (or pink) square outlines.

\subsection{Best-Fitting Parameters}\label{subsec:best_fit_pars}

\begin{figure*}
\includegraphics[width=.49\linewidth]{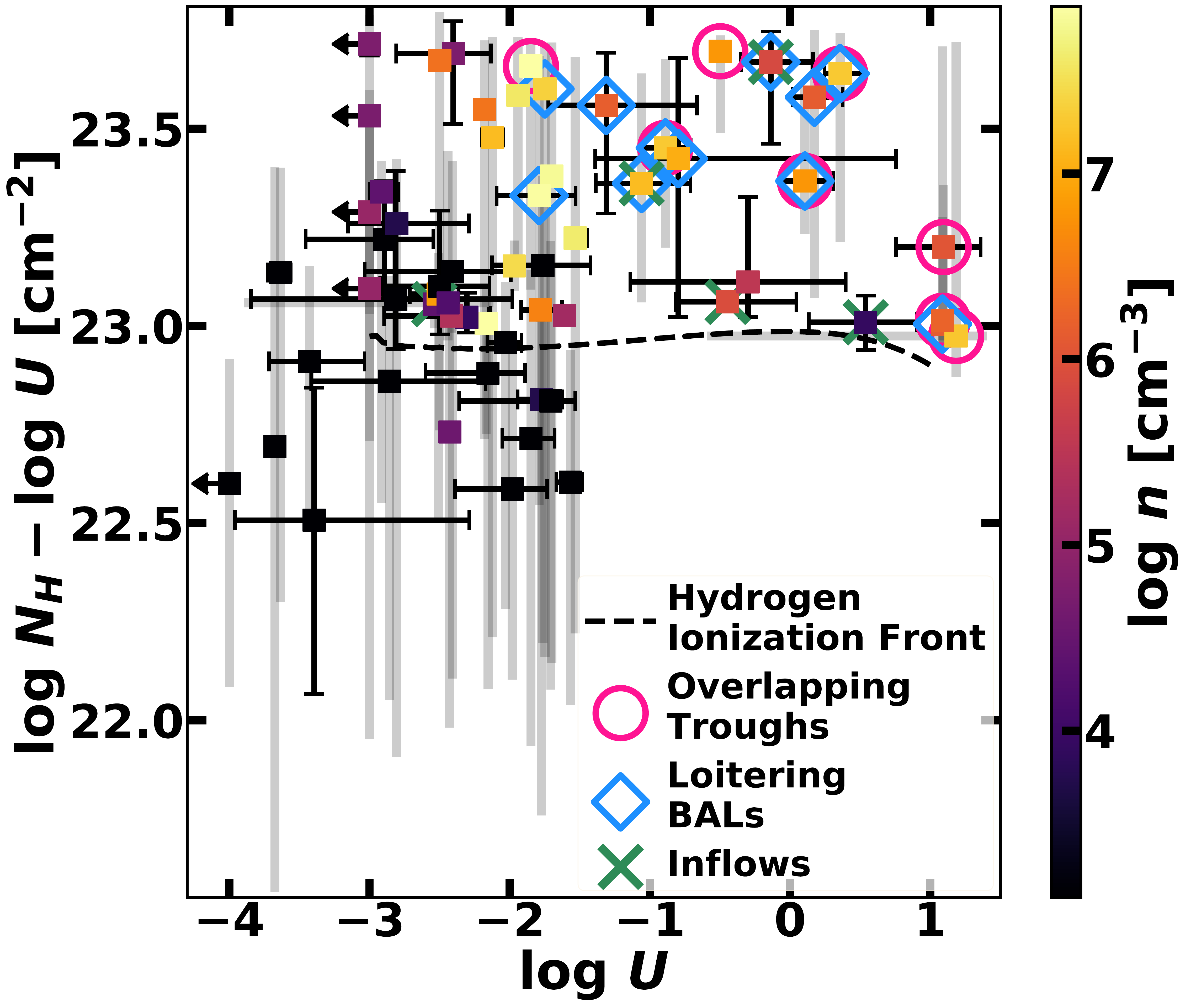}
\includegraphics[width=.485\linewidth]{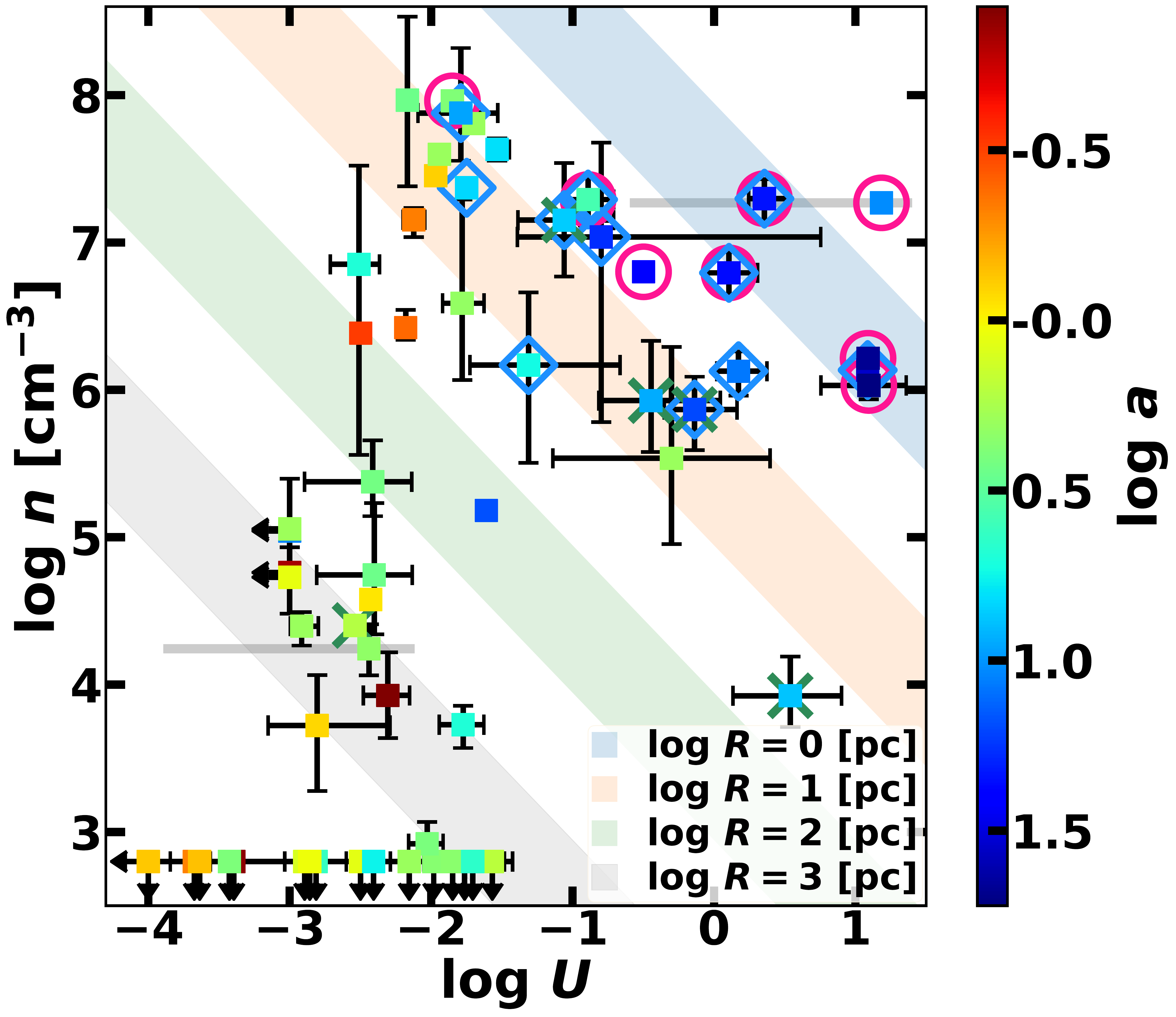}
\caption{The FeLoBAL outflows in our sample show a wide range of physical properties including ionization parameter ($\log U$) and density ($\log n$).
The grey shaded bars represent the range of values among the tophat model bins for each BAL.
For Gaussian opacity profile models or when the tophat bins for a given BAL had a single ionization and/or density parameter, the 2$\sigma$ (95.45\%) uncertainties from the MCMC posterior distributions are plotted as error bars.
Some of the special BALs are marked as follows: red circles for overlapping trough BALs (\S~\ref{subsec:ot_obj}), blue diamonds for loitering BALs (\S~\ref{subsec:loiter}), and green crosses for inflows.
{\it Left panel}:
The dashed line shows the location of the hydrogen ionization front as a function of ionization parameter.
{\it Right panel}:
The locations of the BAL gas from the central SBMH ($\log R$) are marked as shaded areas for reference.
They were calculated assuming the typical ranges of photoionizing photon flux used in our sample of FeLoBALQs ($\log Q\sim55.5-56.5$ [photons s$^{-1}$]).
\label{fig:fitpar_ac_ad}}
\end{figure*}

\subsubsection{Physical Gas Properties of the BALs}\label{subsubsec:best_fit_pars_photo}
Figure~\ref{fig:fitpar_ac_ad} shows the distribution of $\log U$, $\log N_H-\log U$, and $\log n$ of the FeLoBALs in our sample.
The FeLoBALs have a wide range of ionization parameter ($\log U\sim-4$ to 1.2) and density ($\log n\sim2.8-8.0\ \rm[cm^{-3}]$); these values span our computational grid.
Most of the BALs have column density parameter ($\log N_H-\log U$) high enough to encompass the hydrogen ionization front ($\log N_H-\log U\sim23\ \rm[cm^{-2}]$).
This high column density is expected for FeLoBALs
because a column density thick enough to include the hydrogen ionization front is necessary to produce the observed \ion{Fe}{2} absorption \citep[e.g.,][]{lucy14}.
As shown in the left panel of Figure~\ref{fig:fitpar_ac_ad}, we found no FeLoBALs with $\log U>-1.5$ and $\log N_H-\log U<22.9\ \rm[cm^{-2}]$, because insignificant \ion{Fe}{2} is produced in that region of parameter space.

Starting at lower $\log U$ FeLoBALs ($\log U\lesssim-2$), the outflows have lower density ($\log n\lesssim 5\ \rm[cm^{-3}]$) and some of them 
have column densities insufficient to breach the hydrogen ionization front ($\log N_H-\log U\lesssim23\ \rm[cm^{-2}]$).
The physical condition of the gas at low ionization parameter and density does not dramatically change across the hydrogen ionization front in the lower $\log U$ gas and thus the ionic column density of \ion{Fe}{2} gradually increases across the ionization front.
Therefore, BAL gas with lower $\log U$ can populate \ion{Fe}{2} ions even at the slightly lower $\log N_H-\log U$ before the gas column density encompasses the hydrogen ionization front.
Their spectra are least absorbed with no significant line blending or saturation and often the individual absorption lines can be easily identified.
The opacity mainly comes from the ground state \ion{Fe}{2} and \ion{Mg}{2} with weak or no observable opacity from the excited state \ion{Fe}{2} or other rare metal ions (e.g., SDSS~J0835$+$4242, SDSS~J1240$+$4443 in Figure~\ref{fig:fitfig1}).
Although we were able to constrain the densities of most FeLoBALs, we assigned density upper limits for some of the low-$\log U$ FeLoBALs that showed no absorption lines from the excited-state \ion{Fe}{2}.

Most of our FeLoBALs have a moderate ionization parameter of $\log U\sim-2$.
These medium-$\log U$ FeLoBALs have the widest range of densities, spanning the entire range we found in our sample ($2.8\lesssim\log n\lesssim8.0\ \rm[cm^{-3}]$).
These FeLoBALs have the spectral morphology of ``typical'' FeLoBALs with strong absorption lines from \ion{Mg}{2} and ground state \ion{Fe}{2} as well as excited state \ion{Fe}{2} and other iron-peak elements depending on the gas density and column density (e.g., SDSS~J0840$+$3633, SDSS~J1214$+$2803 in Figure~\ref{fig:fitfig1}).
Moreover, standard BAL spectral features (e.g., non-black saturation, line blending) can be easily found in their spectra.
Most of the previously well-studied FeLoBALs belong to moderate-$\log U$ FeLoBALs (e.g., QSO2359$-$1241, \citealt{arav01}; FIRST J104459.6$+$365605, \citealt{dekool01}; FBQS 0840$+$3633, \citealt{dekool02f1}; FIRST J121442.3$+$280329, \citealt{dekool02f2}).

We discovered a number of FeLoBALs with high ionization parameter ($\log U\gtrsim-1$; e.g., SDSS~J0158$-$0046, SDSS~J1154$+$0300 in Figure~\ref{fig:fitfig1}), which were responsible for the large range of ionization parameter we found in our FeLoBAL sample.
High ionization FeLoBALs have higher column density gas that includes a significantly larger number of excited state ions compared with the low $\log U$ FeLoBALs.
Most of the FeLoBALs with high ionization parameter also have special classifications, due to their spectral morphology and physical properties (e.g., overlapping troughs, loitering BALs; \S~\ref{subsubsec:felobal_classification}).
These high opacity FeLoBALs (e.g., overlapping trough BALs) have not been previously analyzed in detail due to the difficulty in analyzing FeLoBALQs with the extreme absorption features that are often seen in these objects.

The FeLoBALs with higher $\log U$ also have higher $\log n$ and larger $\log a$ (less covering; Figure~\ref{fig:fitpar_ac_ad}).
The absence of FeLoBALs with high ionization and low density can be explained by the geometrical constraints expected from the outflows.
The physical thickness of outflowing gas ($\Delta R\sim U/n$, for a fixed $\log N_H-\log U$) cannot be greater than the distance of the gas from the central engine ($\log R$).
The overlapping trough BALs and loitering BALs (marked with red circles and blue diamonds, respectively) have both higher $\log U$ and $\log n$ compared to the rest of the BALs.
The trend between $\log U$ and $\log a$ might be as a selection effect.
In high-$\log U$ outflows, the large amount of opacity from larger number of
excited state ions and rare metal ions would heavily absorb the quasar spectrum.
Therefore, unless the covering fraction is low (high $\log a$), FeLoBALs with high $\log U$ can not be detected easily.

\subsubsection{Kinematic properties of the BALs}\label{subsubsec:best_fit_pars_kine}
\begin{figure*}
\includegraphics[width=.499\linewidth]{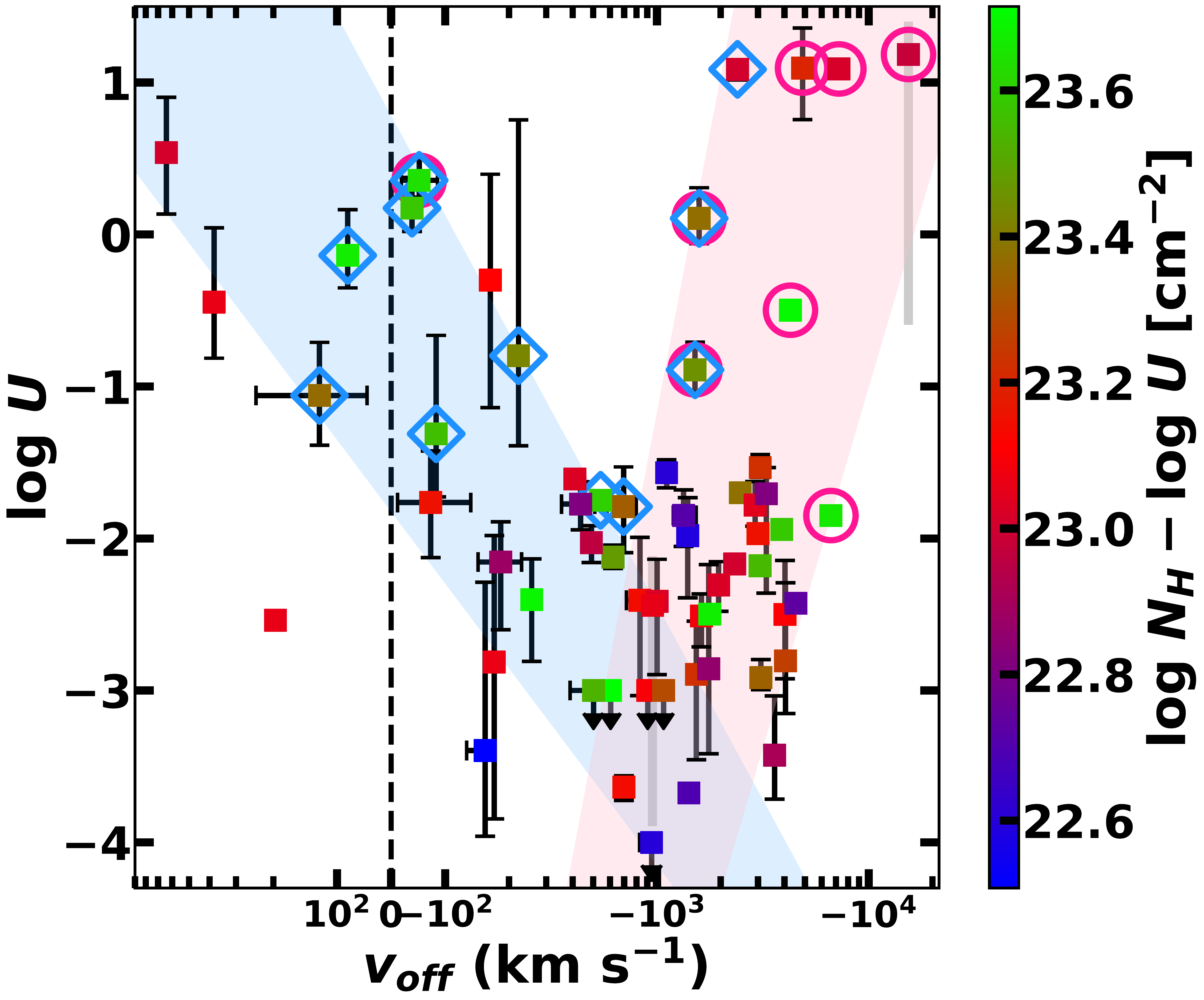}
\includegraphics[width=.475\linewidth]{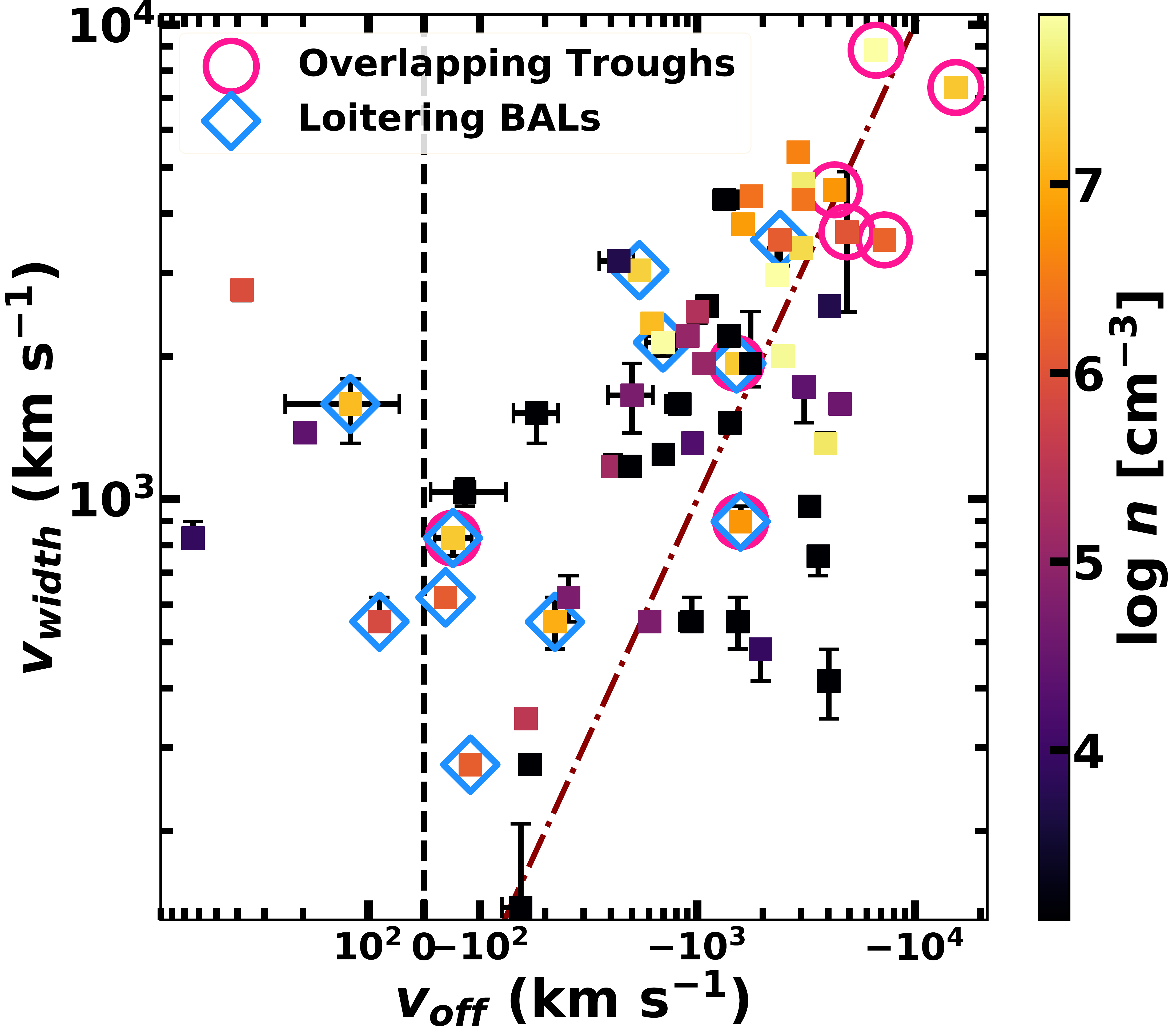}
\caption{{\it Left panel}:
We found no strong correlations between the offset velocities and the physical properties of the BAL gas ($\log U$ and $\log N_H-\log U$) for the whole FeLoBAL sample.
Instead, there is a `v' shape with high $\log U$ BALs present at the lowest and highest velocities (red and blue shades).
{\it Right panel}:
The widths of the BALs scale with the velocity offsets.
The distribution roughly follows a one-to-one ratio (brown dotted-dashed line) which indicates that the most of FeLoBAL outflows are not detached from the emission line at rest.
A linear scale was used in the region $\vert v_{off}\vert<100\rm\ km\ s^{-1}$ and log scale was used elsewhere in the x-axis.
Markers and error bars as in Figure~\ref{fig:fitpar_ac_ad}.
}
\label{fig:fitpar_vel_a_width}
\end{figure*}

Figure~\ref{fig:fitpar_vel_a_width} shows how the kinematics of the outflow (e.g., outflow velocity and width) compare with the physical properties of the gas (i.e., $\log U$, $\log n$, and $\log N_H-\log U$).
Of the 60 BALs in our sample, we found 5 BALs with positive offset velocities (SDSS~J0158$-$0046, SDSS~J0802$+$5513, SDSS~J0916$+$4534, SDSS~J0918$+$5833, and the second BAL component in SDSS~J1125$+$0029).
We have robust redshift measurements for theses objects from the narrow emission lines of [\ion{O}{2}], narrow H$\beta$ component, or high-ionization [\ion{Ne}{5}] line.
The offset velocity of the FeLoBALs does not seem to be correlated with either $\log U$ or $\log N_H-\log U$, and for a given outflow velocity we found a wide range of physical properties.
However, we found that the FeLoBALs with the highest velocities have higher ionization parameters, but the converse is not true.
The subtle v-shaped distribution seen between $v_{off}$ and $\log U$ potentially suggests that there may be more than a single population within FeLoBALQs;
this topic as well as the analysis of the relationship between quasar properties (e.g., Eddington ratio) and outflow (kinematic) properties are explored further using the H$\beta$ / [\ion{O}{3}] properties in \citet{choi_prep_paper3}.
We discuss the potential acceleration mechanisms for the FeLoBAL outflows in \S~\ref{subsec:redden}.

The distribution of the widths and the velocities of the outflows shows a nearly one-to-one relationship (the right panel in Figure~\ref{fig:fitpar_vel_a_width}).
This suggests that most of the outflows in our sample have widths similar to the offset velocities; the outflows are not detached but rather they start from near rest ($\sim0\rm\ km\ s^{-1}$).
Our result is consistent with what has been seen in composite spectra of BALQs that showed troughs beginning from near the peaks of the emission lines \citep{hamann19,rankine20}.
The outflows with the higher velocities have proportionately larger widths and tend to have higher density compared to other outflows in our sample, although there are also FeLoBALs with low velocity and high density.
Most of the overlapping trough BALs (red circles in Figure~\ref{fig:fitpar_vel_a_width}) have the highest outflow velocities with large widths as well as high density.

\subsection{Derived Physical Properties of the Outflows}\label{subsec:derived_pars}
\begin{figure}
\epsscale{.56}
\plotone{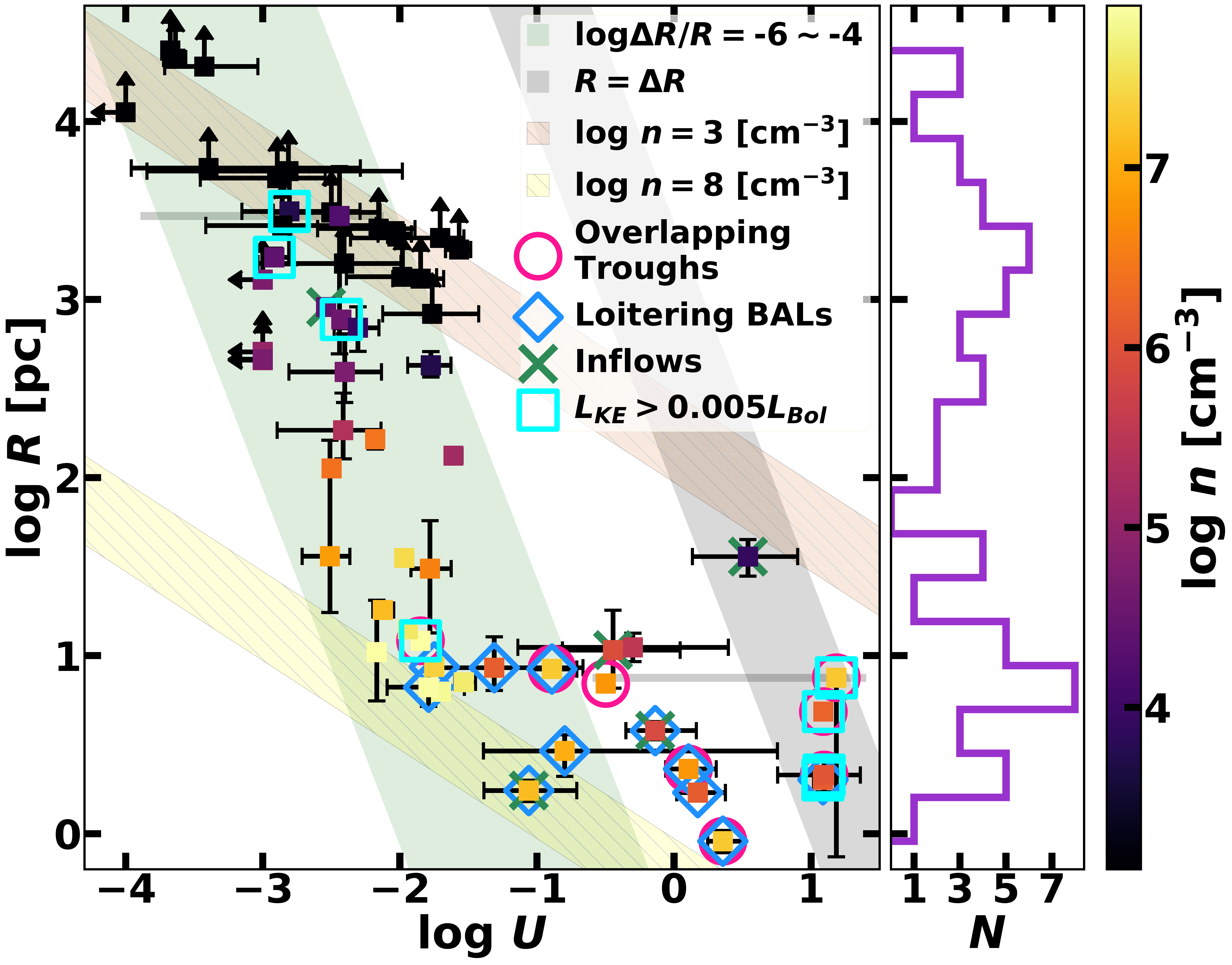}
\caption{The BALs have a wide range of distances from the central engine, ranging from a torus-scale ($\sim$pc) to a host galaxy scale ($\sim$kpc).
The cyan squares mark the powerful outflows that have well-constrained physical parameters ($\log U$ and $\log n$) and kinetic luminosity ($L_{KE}$) greater than 0.5\% of the bolometric luminosity of the quasar ($L_{Bol}$).
The powerful outflows are located in a various distances from the central black hole and they have a variety of physical properties.
The histogram on the y-axis shows the distribution of $\log R$ in our sample.
There is a lack of FeLoBALs near $\log R\sim2$ [pc] (\S~\ref{subsec:loc_origin_felobal}).
A range of BAL cloud volume filling factors (or normalized radial widths, $\log \Delta R/R$) and two $\log n$ values are marked as shaded areas for reference.
Markers and error bars as in Figure~\ref{fig:fitpar_ac_ad}.
\label{fig:fitpar_aR}}
\end{figure}

The large range in ionization parameter and density observed means that the outflows are located throughout the quasar, ranging from near the torus at $\log R\sim0$ [pc] to the host galaxy $\log R\sim3$ [pc].
We recovered an inverse relationship between $\log R$ and the two gas parameters $\log U$ and $\log n$ that can be explained with the definition of $\log U$ ($U\propto1/nR^2$).
Given the range of the total number of photoionizing photons per second ($Q$) from the AGN for the objects in our sample
and the range of densities we have for the BAL clouds, we can identify a region in the $\log R\sbond\log U$ parameter space where we expect the BALs to be located.
If the values of $\log U$ and $\log n$ were distributed evenly for our sample of FeLoBALs, we would find them evenly distributed in the area between the two dashed-shaded strips for $\log n=3$ and 8 $\rm[cm^{-3}]$ in the Figure~\ref{fig:fitpar_aR} (calculated assuming $\log N_H-\log U=23\sim23.5\rm\ [cm^{-2}]$ and $\log Q\sim55.5-56.5$ [photons s$^{-1}$]);
it is clear they are not.
Instead of a uniform distribution we find that
high $\log U$ outflows have higher $\log n$ and the outflows with lower $\log n$ tend to have lower $\log U$ as well.
This behavior is found because most of our FeLoBALs have a small range of volume filling factors (or normalized radial widths, $\log \Delta R/R\sim-5$; $\log\Delta R=\{\log U+(\log N_H-\log U)\}-\log n$) except for the compact BALs that are located $\log R\lesssim1$ [pc]
which have large volume filling factors ($\log \Delta R/R\gtrsim-3$).
We discuss this point and other geometrical constraints in Paper III \citep{choi_prep_paper3}.

We found mass outflow rates of $\dot M_{out}=0.049\sim520\rm \ M_\odot\ yr^{-1}$.
The mass outflow rates span more than four orders of magnitude which reflects the wide range of $v_{off}$ we found in our FeLoBAL outflows.
We also report mass inflow rates of $\dot M_{in}=0.016\sim55\rm \ M_\odot\ yr^{-1}$ from the 5 BALs with $v_{off}>0\ \rm km\ s^{-1}$.
For a couple of BAL components with $v_{off}\sim0\ \rm km\ s^{-1}$ that were modeled using tophats, we calculated both the mass outflow and inflow rates depending on the velocities of the bins.
Except for SDSS~J1125$+$0029 in which there is an additional outflowing BAL, SDSS~J0158$-$0046, SDSS~J0802$+$5513, SDSS~J0916$+$4534, and SDSS~J0918$+$5833 only showed inflow BALs and did not have other outflowing BALs in the spectra.
These inflow BALs are mostly located within $\log R<2$ [pc] (one at $\log R\sim3$ [pc]) and have low offset velocities ($v_{off}\lesssim600\ \rm km\ s^{-1}$).
We note that there is a caveat that mass outflow rates are dependent on the assumed global covering fraction ($\Omega$) of (FeLo)BALs.
We used $\Omega=0.2$ in this work (\S~\ref{subsubsec:kinematics_measure}), and detailed discussion on global covering fraction for FeLoBALs can be found in \citet{choi20}.

\begin{figure}
\epsscale{.56}
\plotone{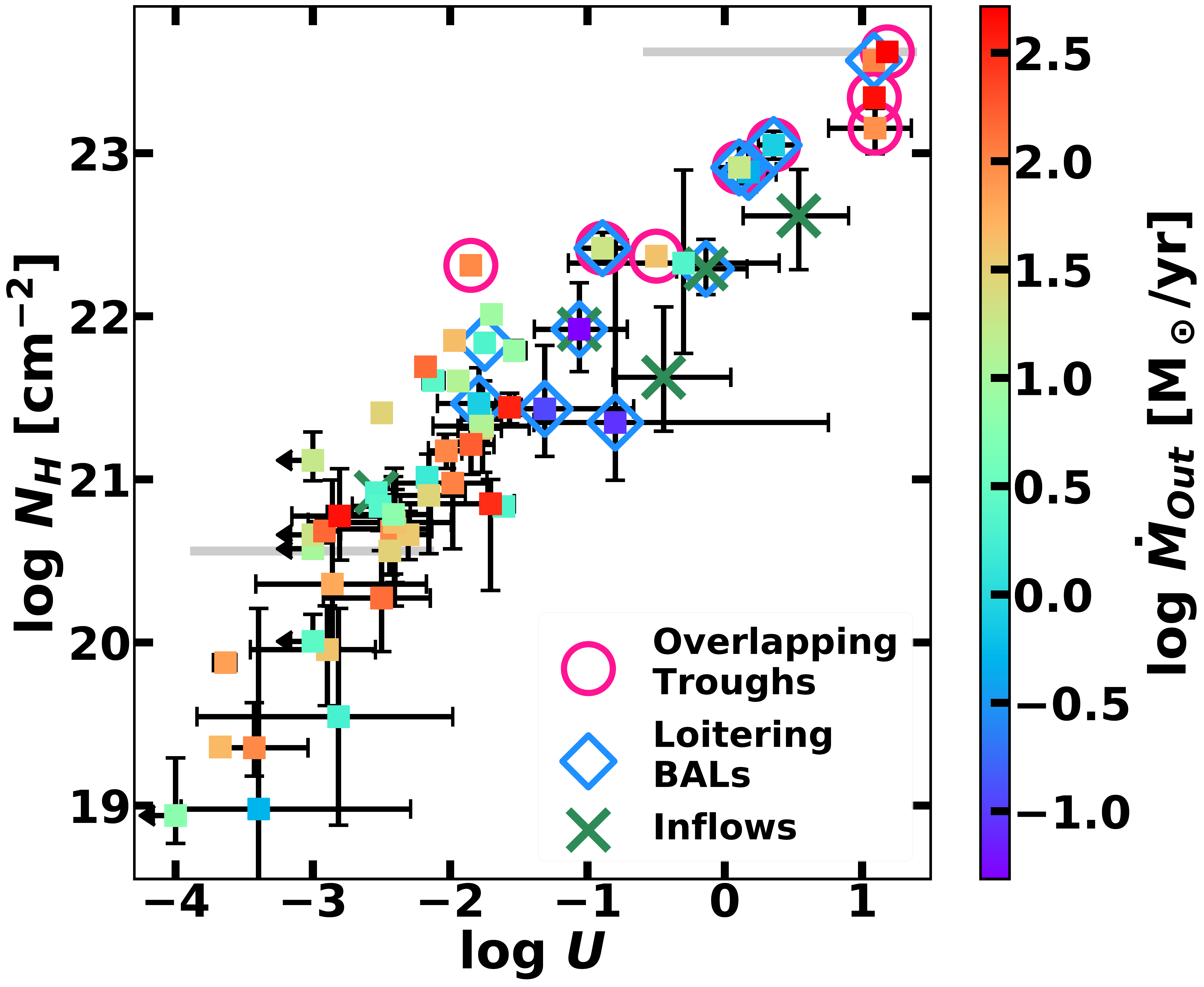}
\caption{The BALs with high $\log U$ have larger covering-fraction-corrected hydrogen column densities.
The strong correlation between the two parameters is expected for FeLoBALs that require the gas column densities to be high enough to reach the hydrogen ionization front ($\log N_H-\log U\sim23\rm\ [cm^{-2}]$).
Markers and error bars as in Figure~\ref{fig:fitpar_ac_ad}.
\label{fig:fitpar_aNh}}
\end{figure}

Figure~\ref{fig:fitpar_aNh} shows that the higher $\log U$ outflows have higher partial-covering-corrected column density.
The correlation may be an artifact of sample selection because the column density of the outflowing gas needs to be high enough to reach the hydrogen ionization front for the FeLoBALs ($\log N_H-\log U\ga23.0\rm\ [cm^{-2}]$, left panel in Figure~\ref{fig:fitpar_ac_ad}; \S~\ref{subsubsec:best_fit_pars_photo}).
Moreover, we see relatively small dynamic range of about 1 dex across $\log N_H-\log U$ whereas $\log U$ ranges from $\sim-4$ to $\sim1.5$.
From the distributions of the parameters alone we can expect $\log N_H$ is highly dependent on $\log U$ so that the gas with higher $\log U$ also has higher $\log N_H$.

The mass outflow rate ($\dot M$) does not seem to be strongly correlated with either ionization parameter or partial-covering-corrected column density.
Outflows with high mass outflow rates ($\dot M_{out}\sim100\rm \ M_\odot yr^{-1}$) were found across the entire $\log N_H$ and $\log U$ range.
This is because the magnitude of the mass outflow rate is mainly determined by the outflow velocity rather than the location or the physical properties of the gas (\S~\ref{subsec:corr}).

\begin{figure}
\epsscale{.56}
\plotone{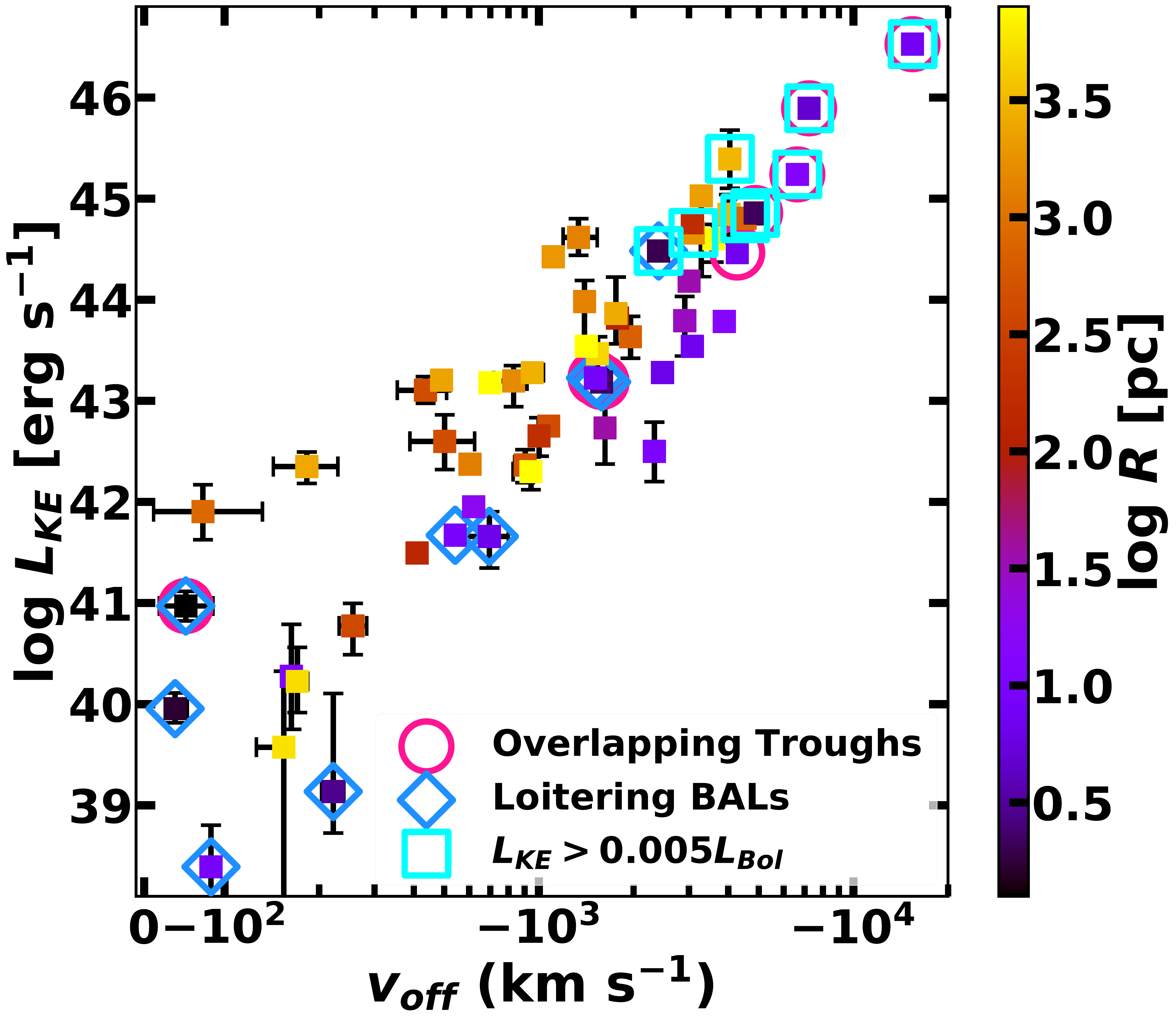}
\caption{The kinetic luminosities ($L_{KE}$) of the FeLoBAL outflows scale with the outflow velocities with the slope of $\sim3$.
This relationship is expected from the definition of $L_{KE}$.
The scatter in the $L_{KE}$ direction can be mainly ascribed to the range of mass outflow rate ($\dot M$) of the outflows as well as gas physical conditions such as covering-fraction-corrected hydrogen column densities.
The colors of the markers represent the locations of the outflows ($\log R$) and for a FeLoBAL outflow at a given outflow velocity, larger $\log R$ will yield larger $\dot M$ and greater $L_{KE}$.
We did not find any particular trend with $\log R$ and the properties plotted in the figure; however, the outflows with the highest velocities in our sample were all found to be within the vicinity of torus $\log R\lesssim1$ [pc].
The 5 BALs with inflows are not plotted.
\label{fig:vel_ke}}
\end{figure}

We found a strong correlation between the outflow strength and the outflow velocity (Figure~\ref{fig:vel_ke}).
Considering that $\dot M\propto v$, $L_{KE}$ of an outflow is proportional to
$v^3$
, which explains the tight correlation between $L_{KE}$ and $v_{off}$ observed in our sample of FeLoBALs.
All the outflows with the highest outflow velocities (8 out of 55 FeLoBALs with $v_{off}\lesssim-2,400\rm\ km\ s^{-1}$) in our sample have enough energy to produce quasar feedback (\S~\ref{subsec:feedback}).
While many of these powerful outflows with high velocities are located near the vicinity of torus $\log R\leq1$ [pc], we did not find robust evidence suggesting a connection between $\log R$ and $L_{KE}$ (\S~\ref{subsec:corr}).
This lack of correlations suggests that the outflow strength or the outflow's role in quasar feedback is mainly determined by the outflow velocity and the physical properties of the gas.
In other words, where the outflows are located at present does not have significant influence on the inferred energetics of the outflow.

\section{Analysis of the Full Sample}\label{sec:analy}
\subsection{Correlations}\label{subsec:corr}

\begin{figure*}
\plotone{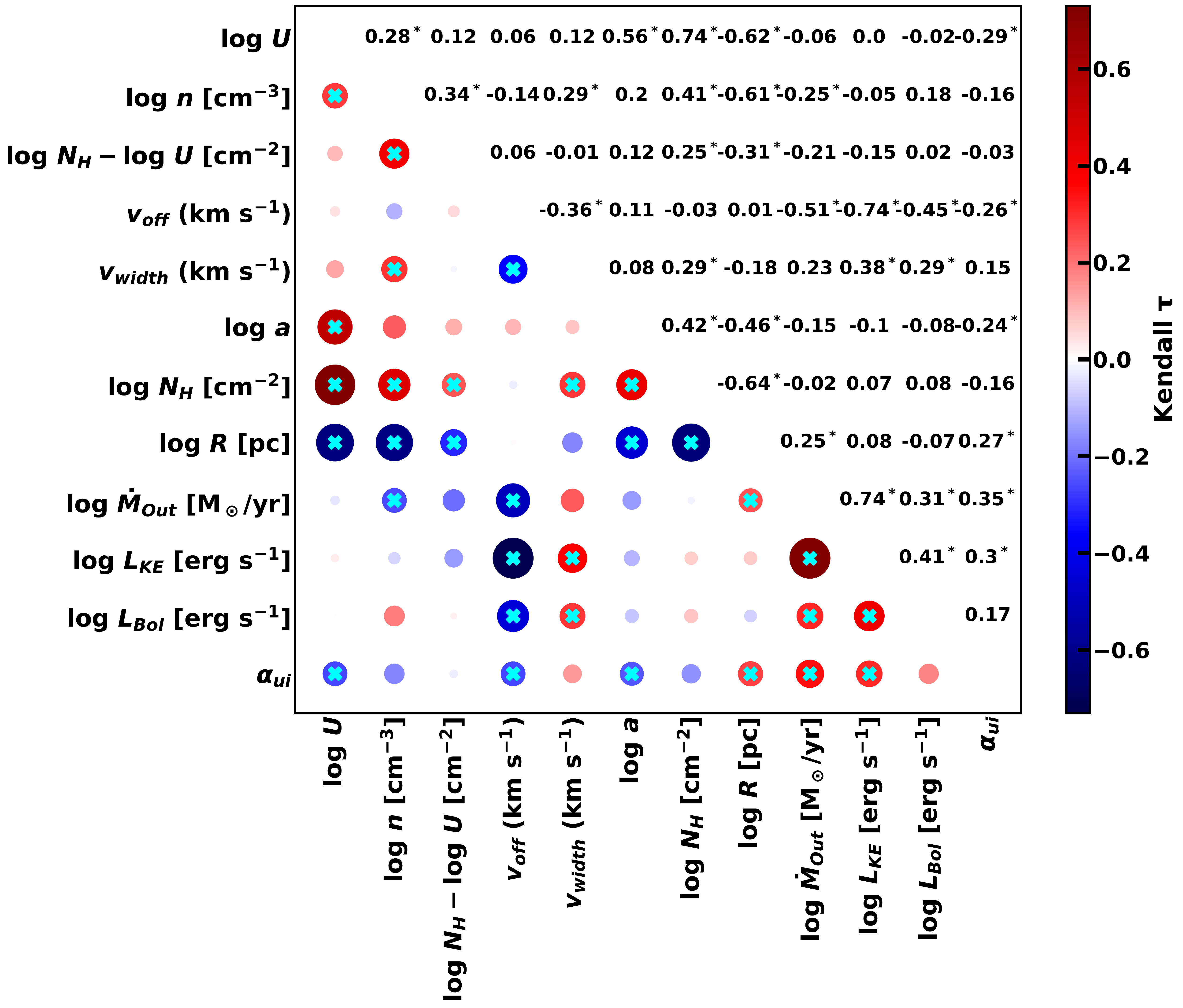}
\caption{The correlation coefficients calculated for all measured BAL property pairs.
The size and the color of the markers reflect the values of Kendall's rank correlation coefficient.
The values with asterisks and cyan crosses represent the pairs that showed statistically significant correlations ($p<0.01$).
The first six parameters ($\log U$, $\log n$, $\log N_H-\log U$, $v_{off}$, $v_{width}$, and $\log a$) are the best-fitting model parameters from {\it SimBAL} and the rest are the derived properties calculated from the model parameters.
\label{fig:corr_matrix}}
\end{figure*}

In order to systematically study the relationship between the physical and kinematic properties of FeLoBALs and the derived wind properties, we calculated the Kendall rank correlation coefficients for all pairs of the measured quantities and examined the ones that showed significant correlations ($p<0.01$).
Because we have BALs with upper or lower limits of $\log n$ or $\log U$, we used {\tt pymccorrelation} by \citet{privon20}\footnote{https://github.com/privong/pymccorrelation/} which provides a python implementation of Kendall's tau rank correlation coefficient calculator for censored data \citep{isobe86}.
Only the values from the BALs from the outflowing gas ($v_{off}<0\rm\ km\ s^{-1}$) were used for correlation analysis with outflow properties ($\dot M_{out}$, $L_{KE}$).

\subsubsection{Outflow Physical Properties}\label{subsubsec:corr_felobal}
Among the fit parameters that were directly constrained from the models,
we found correlations between $\log n$ and $\log U$ and $\log N_H-\log U$ for gas physical properties, and between $v_{off}$ and $v_{width}$ for gas dynamical parameters.
The correlations with $\log n$ suggest that for the FeLoBALs in our sample, the gas with higher density also tends to be more highly ionized and thicker with a higher column density.
The covering fraction parameter $\log a$ was also highly correlated with $\log U$.
The correlations between $\log U$, $\log n$, and $\log a$ were seen on the right panel of Figure~\ref{fig:fitpar_ac_ad} where the BALs with higher $\log U$ were found to be denser with lower partial covering (higher $\log n$ and $\log a$).
As we discussed in \S~\ref{subsec:best_fit_pars}, the correlation between the partial covering parameter $\log a$ and the ionization parameter maybe due to a selection effect: FeLoBALs with high $\log U$ and low $\log a$ (large covering) are unlikely to be detected due to heavy absorption.
We only observe high $\log U$ outflows that have high $\log a$ (i.e., not completely covered).

While we also found that $\log U$ shows correlations with both $\log R$ and $\log N_H$, we interpret these correlations mainly arising from a selection effect (\S~\ref{subsec:derived_pars}).
$\log R$ is calculated using the definition of $\log U$: $R\propto (Q/nU)^{1/2}$ (\S~\ref{subsubsec:kinematics_measure}).
The range of number of photoionizing photons per second emitted from the AGN, $Q$, this sample only spans about a dex for the FeLoBAL quasars, therefore we can expect $\log R$ to be negatively correlated with both $\log U$ and $\log n$, as found.
Similarly, the correlation we observe between $\log U$ and $\log N_H$ (partial-covering-corrected column density) is expected for a FeLoBAL sample because FeLoBALs require the outflow gas to be thick enough to reach the hydrogen ionization front $\log N_H-\log U\sim23.0\rm\ [cm^{-2}]$ (\S~\ref{subsec:derived_pars}).

A negative correlation between velocity offset and velocity width is seen in the right panel of Figure~\ref{fig:fitpar_vel_a_width} (as well as with the correlation coefficient).
The FeLoBAL outflows in our sample show a near one-to-one correlation between the two velocity parameters, which indicates that FeLoBAL features are seldom seen as detached troughs; the absorption starts from rest.
We can also see from the $p$-value and the correlation coefficient that there is a weak positive correlation between $\log n$ and $v_{width}$ but not with $v_{off}$.
Although the statistical test suggests that there exists a correlation between these parameters in our sample of low redshift FeLoBALQs, we cannot confidently conclude whether they represent a true global trend in the parameter space.
We will need to analyze a larger homogeneous sample of FeLoBALQs in order to substantiate this result.

The parameters that measure the strength of the outflow ($\dot M_{out}$, $L_{KE}$) showed strong correlations with BAL velocity offset.
As discussed in \S~\ref{subsec:best_fit_pars}, it is expected for $\dot M_{out}$ and $L_{KE}$ to be strongly dependent on $v_{off}$.
The correlation analysis from our FeLoBALs further supports the relationship between the two properties and also suggests that the dependence of $\dot M_{out}$ and $L_{KE}$ on the dynamics of the gas ($v_{off}$) is so strong that physical properties of the gas (e.g., $\log U$, $\log N_H-\log U$) do not significantly influence the energetics of the BAL outflow or the outflow's potential impact on the host galaxy.
We did not observe any significant correlations between $L_{KE}$ and other 
parameters used to calculate this value ($\log R$ or $\log N_H$), again emphasizing its strong dependence on velocity above all other parameters (Figure~\ref{fig:vel_ke}).

\subsubsection{Quasar Properties}\label{subsubsec:corr_quasar}
In addition to the BAL properties, we also investigated potential connections between the outflows and the properties of the quasars.
The quasar properties (e.g., $L_{Bol}$) were measured using the observation data and {\it SimBAL} modeling (\S~\ref{subsec:lbol_aui_measure})

We found a strong positive correlation between the velocity offset and the bolometric luminosity of the quasar ($L_{Bol}$).
Our result is consistent with previous studies that found that quasars with higher $L_{Bol}$ have outflows with higher velocities \citep[e.g.,][]{laor02,ganguly07,fiore17}.
The outflow velocity is expected to depend on the bolometric luminosity normalized by the Eddington value ($L_{Bol}/L_{Edd}$) for radiatively driven quasar winds.
We explore this correlation in Paper III \citep{choi_prep_paper3} using the objects that have black hole mass measurements from the rest-optical emission-lines \citep{leighly_prep}. 
Because the velocity offset determines the strength of the outflow,
we found a positive correlation between $L_{Bol}$ and $L_{KE}$; more luminous quasars tend to have faster and more energetic outflows.

We found that $\alpha_{ui}$ is strongly correlated with $\log R$ and $v_{outflow}$ ($v_{off}$).
Quasars with flatter SEDs ($\alpha_{ui}\gtrsim-0.5$) have redder colors and have faster and thus more powerful outflows.
The quasars with steeper or bluer SEDs ($\alpha_{ui}\lesssim-0.5$) have FeLoBALs that are located closer to the central engine (smaller $\log R$) that are more ionized (higher $\log U$) and they tend to have smaller outflow velocities.
Most of these compact, low-velocity FeLoBALs are loitering BALs that do not show properties expected for typical quasar-driven winds.
We further discuss the properties of these FeLoBALs in \S~\ref{subsec:loiter}.

Our result seems somewhat similar with what have been found in Extremely Red Quasars (ERQs).
These objects show a higher incidence of outflow signatures in [\ion{O}{3}]$\lambda\lambda 4959,5007$ emission lines than typical blue quasars and a correlation between the outflow speeds and the redness of the SED  has been found \citep{hamann17,perrotta19}.
We note, however, that the FeLoBALQs in our sample are not ERQs and the quasar winds seen in emission lines can exhibit different properties than the BAL winds.
We discuss the implications of the SED color on the acceleration mechanism of the BAL outflows in \S~\ref{subsec:redden}.

In summary, the only physically significant correlations that we found from the FeLoBAL and quasar properties were between $L_{Bol}$ and $v_{off}$, which propagates because of the functional dependence to $\dot M_{out}$ and $L_{KE}$.
We also found correlations with $\alpha_{ui}$ where FeLoBALs found in objects with flatter (redder) spectral slope have larger $\log R$ and $L_{KE}$.

\subsection{Objects with multiple FeLoBAL outflows}\label{subsec:multiple_BAL}
We identified more than one outflow component in 9 objects (excluding the broad \ion{Mg}{2} component in SDSS~J1214$-$0001 and the extra component in SDSS~J1644$-$5307 only seen in \ion{Mg}{2}; see \S~\ref{subsec:bestfitmodels} and Appendix~\ref{app:model_detail}).
While three of the nine objects showed several troughs distinctly separated by velocity, the others required a rigorous modelling with {\it SimBAL} to identify multiple outflows in blended troughs.

\begin{figure*}
\epsscale{.95}
\plotone{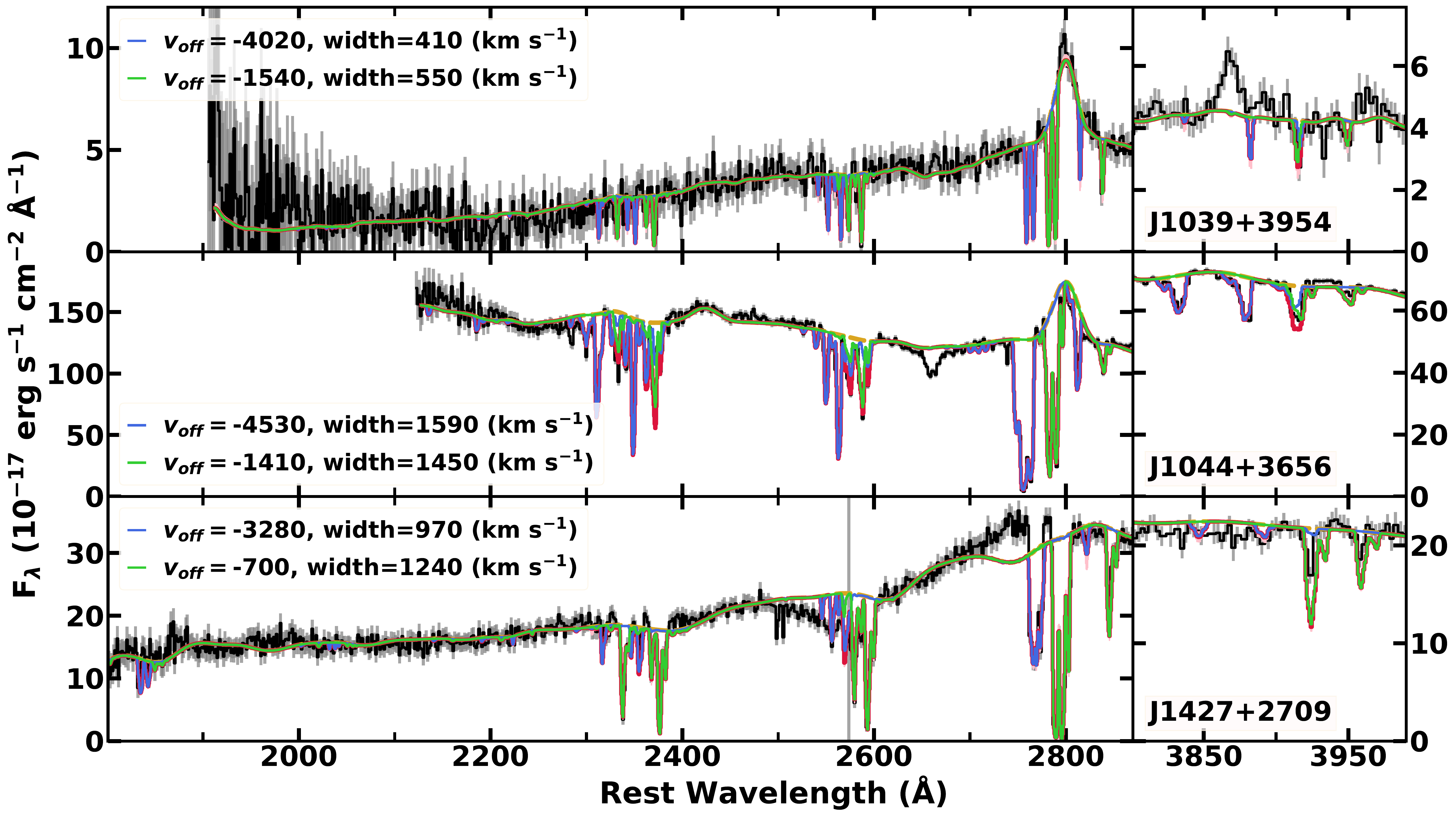}
\caption{Objects that show multiple outflows where each BAL component (indicated with blue and green model curves) is clearly separated (\ion{Mg}{2}$\lambda\lambda 2796,2803$ troughs are not blended, left column).
The data (error) is plotted in black (grey) and the best-fitting model is shown in red with the BAL components are over-plotted in green and blue.
The column on the right shows the wavelength region where \ion{He}{1}*, \ion{Ca}{2} H and K absorption lines are found.
Note the BALs in these objects have low ionization and no \ion{He}{1}* absorption-lines are observed.
\label{fig:multiple_bals_model_1}}
\end{figure*}
\begin{figure*}
\epsscale{.95}
\plotone{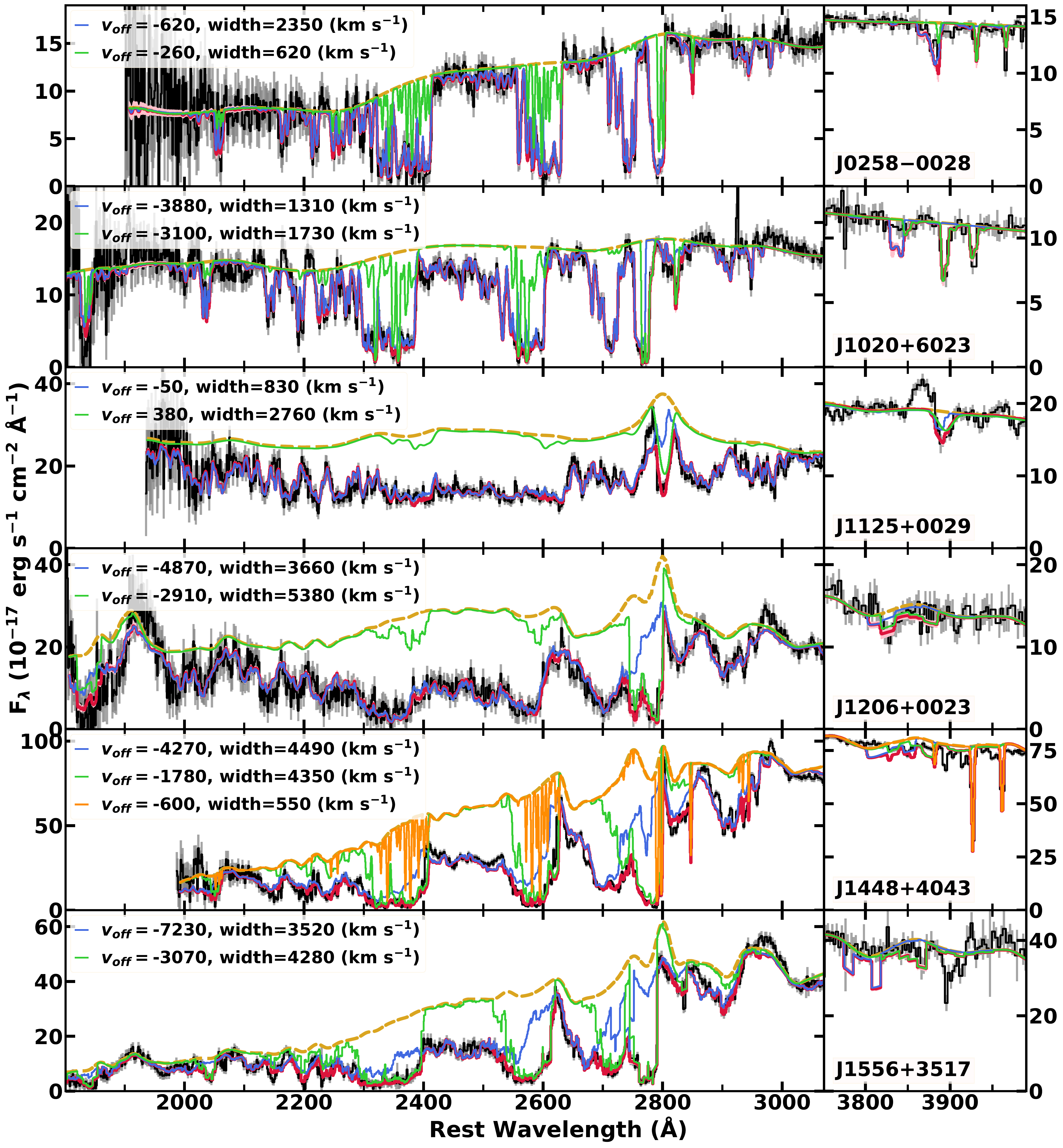}
\caption{
Objects that have multiple blended BAL components.
The differences in gas physical properties can be observed most clearly with \ion{Ca}{2} H and K lines where these absorption lines are only found the lower-velocity components with lower $\log U$ and $\log n$ (right column).
For example, we found three BAL components in SDSS~J1448$+$4043 with the lowest-velocity component showing deep \ion{Mg}{1} absorption line as well as \ion{Ca}{2} H and K absorption lines (plotted in orange).
The data (error) is plotted in black (grey) and the best-fitting model is shown in red with the BAL components are over-plotted in green and blue with the continuum plotted in dashed tan lines.
\label{fig:multiple_bals_model_2}}
\end{figure*}
\begin{figure}
\epsscale{.5}
\plotone{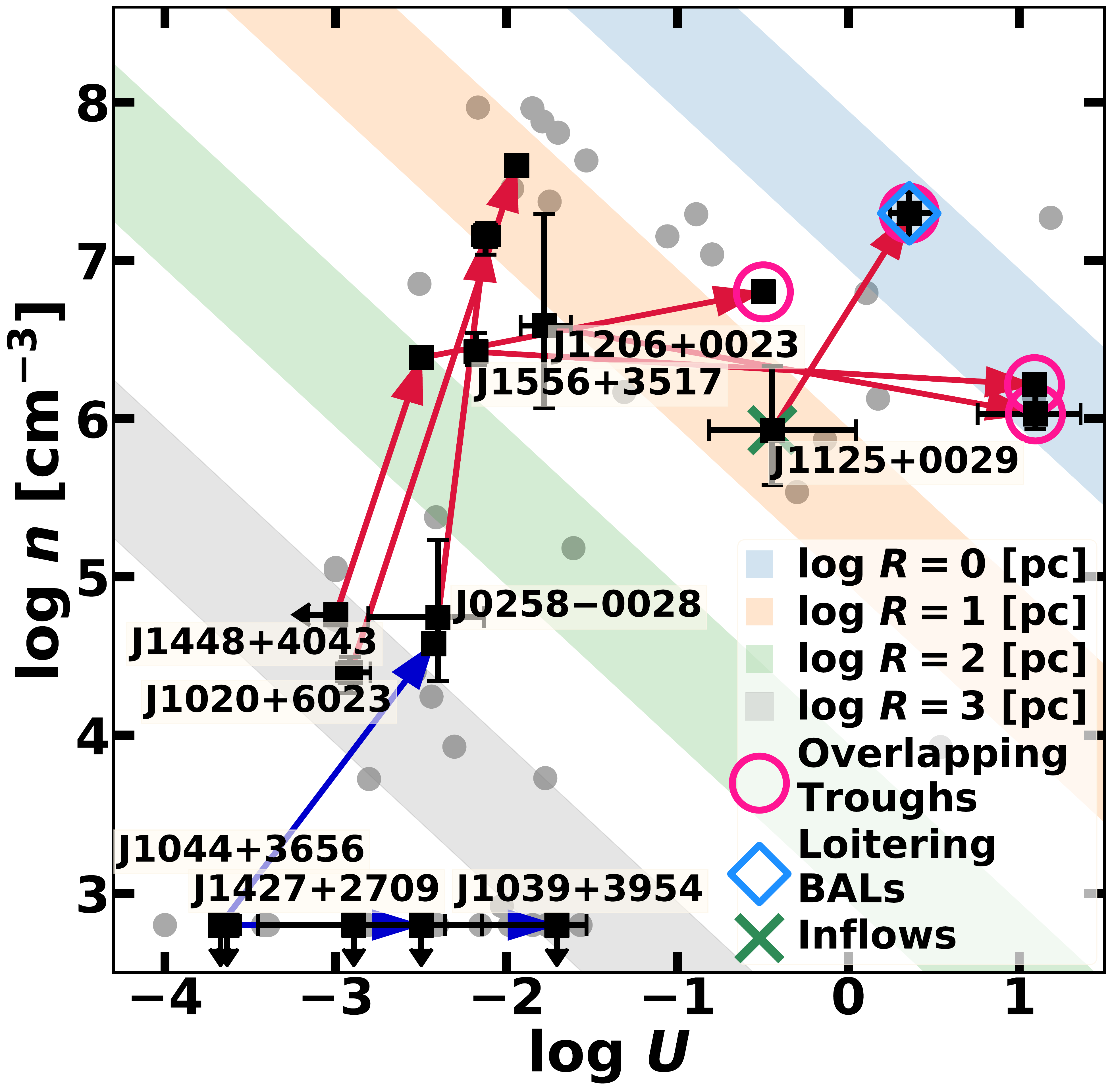}
\caption{The outflows from the objects that have more than one FeLoBAL feature are plotted with squares and the rest of the outflows shown by grey dots.
The arrows start from the lower-velocity component and end at the higher-velocity component in the same object.
Blue and red arrows (labelled by object) represent objects shown in Figure~\ref{fig:multiple_bals_model_1} and \ref{fig:multiple_bals_model_2}, respectively.
The shaded diagonal bars for $\log R$ were calculated using the typical range of photoionizing photon flux found in our sample of FeLoBALQs ($\log Q\sim55.5-56.5$ [photons s$^{-1}$]).
All of the higher-velocity components have higher ionization parameter and the majority of them have higher density compared to the lower-velocity components in the same object.
Thus the higher-velocity components are located closer to the central black hole.
Four of the higher-velocity components with significantly higher $\log U$ are overlapping trough BALs.
The error bars show 2$\sigma$ (95.45\%) uncertainties.
\label{fig:multiple_bals}}
\end{figure}

Figure~\ref{fig:multiple_bals_model_1} and Figure~\ref{fig:multiple_bals_model_2} show the two groups of objects with more than one BAL component.
The three objects in the first group have more than one narrow \ion{Mg}{2} trough that are separated by velocity (e.g., SDSS~J1044$+$3656 in Figure~\ref{fig:multiple_bals_model_1}).
In these objects the differences in gas physical parameters were not extreme and the estimated distances for the BALs only differ by a moderate amount (Figure~\ref{fig:multiple_bals}).
In the second group (Figure~\ref{fig:multiple_bals_model_2}), the features are blended.
{\it SimBAL} analysis identified multiple components and large differences in the gas properties, especially $\log U$, were found (Figure~\ref{fig:multiple_bals}).
The difference in $\log U$ resulted in large differences in $\log R$.
The BAL components with higher $\log U$ include a plethora of absorption line transitions including the rare transitions from the excited state ions, but due to the large $\log a$ (low covering fraction) the depth of these troughs is small.
These particular components played a critical role in creating the overlapping trough features because they provided excited state \ion{Fe}{2} transitions around $\sim2500$\AA\/ (i.e., between the ground-state multiplet features near $\sim2400$\AA\/ and $\sim2600$\AA\/) as well as the absorption lines observed longward of $\sim2800$\AA\/
\citep[\S~\ref{subsec:ot_obj}; see also][]{lucy14}.

The higher-velocity components in the multiple-outflow objects have higher $\log U$ and lower covering fraction compared with the lower-velocity component (e.g., SDSS~J1020$+$6023 in Figure~\ref{fig:multiple_bals_model_2}; Figure~\ref{fig:multiple_bals}).
They produced opacity from both the excited-state and the ground-state \ion{Fe}{2} transitions.
In contrast, the lower-velocity components produced higher opacity overall from the ground-state \ion{Fe}{2} and \ion{Mg}{2} only and were responsible for the most of the strong \ion{Mg}{2} absorption features observed in the spectra.
\citet{choi20} and \citet{leighly18a} both observed a similar trend where they found higher ionization parameter and lower covering fraction for the tophat bins with higher velocities in the best-fitting {\it SimBAL} models.
The distinct gas properties constrain the location to different radii: the higher-velocity component is found much closer to the SMBHs than the lower-velocity component in a given object.

The FeLoBALQs with multiple blended components show some of the most extreme spectral features, such as overlapping troughs  (e.g., SDSS~J1556$+$3517 in Figure~\ref{fig:multiple_bals_model_2}).
However, the BAL component decomposition revealed that the lower-velocity components in these objects actually resemble a typical FeLoBAL.
In other words, without the higher-velocity components these FeLoBALQs would show spectra that are indistinguishable from those of moderately absorbed FeLoBALQs with low to moderate ionization parameter (\S~\ref{subsubsec:best_fit_pars_photo}).
Because of the difference in location we expect the dynamic time scales between the higher-velocity components and the lower-velocity components would be dramatically different, with the higher-velocity components expected to have much shorter time scales.
It is plausible that these higher-velocity components at $\log R\lesssim1$ [pc] may represent a a transient phenomenon.
For example, objects that have overlapping trough features originating from higher-velocity components may show BAL variability in which overlapping trough disappears to reveal spectra that look like typical FeLoBALQs.
Such variability has been seen in a number of FeLoBAL quasars \citep[e.g.,][]{rafiee16}.

There could be many origins of the lower-velocity components for the multi-component objects.
{\it SimBAL} modeling assumes that these multiple components are not physically related (i.e., independent photoionization modeling for each absorber).
Because the BAL components in a given multi-BAL FeLoBALQ are separated by a large radial distance, it is unlikely that they are physically related and formed in the same gas.
One possibility is that the lower-velocity components might represent the remnants from earlier ejection episodes that are located along the line of sight \citep[e.g.,][]{choi20}.
A variability study of these objects could potentially give us more detailed picture of the geometry of the multi-BAL system.
For instance, we may expect systematically different variability pattern between the two absorbers if we assume the two gas clouds are not physically related \citep[e.g.,][]{leighly15}.

Nevertheless, it is possible that the multiple absorbers in a given object may be related in terms of their photoionization processes.
\citet{voit93} proposed structures in outflowing gas cloudlets that can produce BALs with stratified ionization conditions based on the observations of high-ionization BALs (\ion{C}{4}$\lambda\lambda 1548,1550$) and low-ionization BALs (e.g., \ion{Mg}{2}$\lambda\lambda 2796,2803$).
In their models, highly ionized condition can be formed in higher velocity portion of the BAL gas clumps and lower ionization environment is found near the lower velocity end, similar to how we found higher-velocity components with higher ionization.
If the multiple components found in a given object are physically related and their photoionization processes are interdependent, it would be conceivable that we may be overestimating the distances between the lower-velocity BALs and the higher-velocity BALs.
In such scenario, a single absorbing gas may be able to produce multiple components with different ionization states; however, {\it SimBAL} neither has grids nor performs dynamical photoionization modeling needed to reproduce such physical conditions.
Furthermore, the objects in our current sample have too low redshift to observe the \ion{C}{4} lines that we need to investigate such scenario;
therefore we do not have robust observational evidence supporting complex ionization structure in absorbing gas.
We will investigate this question using a sample of higher redshift FeLoBALQs.

A third possibility is that the lower velocity gas may originate from more distant structures in the quasar system.
At the distances calculated from the gas properties (10s--1000s pc), molecular clouds in the host galaxy could be the source for the material illuminated by the quasar \citep[e.g.,][]{fg12b}.

\subsection{Opacity profiles of the outflows}\label{subsec:vel_profile}
\subsubsection{\texorpdfstring{\ion{Mg}{2} and \ion{Fe}{2} Absorption Lines}
{MgII and FeII Absorption Lines}}\label{subsubsec:mg_fe_profile}
Comparing the opacity profiles between line transitions from the ions that have different properties can potentially tell us about the structure of the BAL winds.
For instance, \citet{voit93} proposed a schematic picture of BAL clouds based on the differences in the opacity profiles between the high-ionization lines (e.g., \ion{C}{4}) and the low-ionization lines (e.g., \ion{Mg}{2}).
They suggested that the BALs originate from a turbulent absorbing region and absorbing gas clouds may have structures with dense cores with inhomogeneous photoionization conditions.

\begin{figure*}
\includegraphics[width=.499\linewidth]{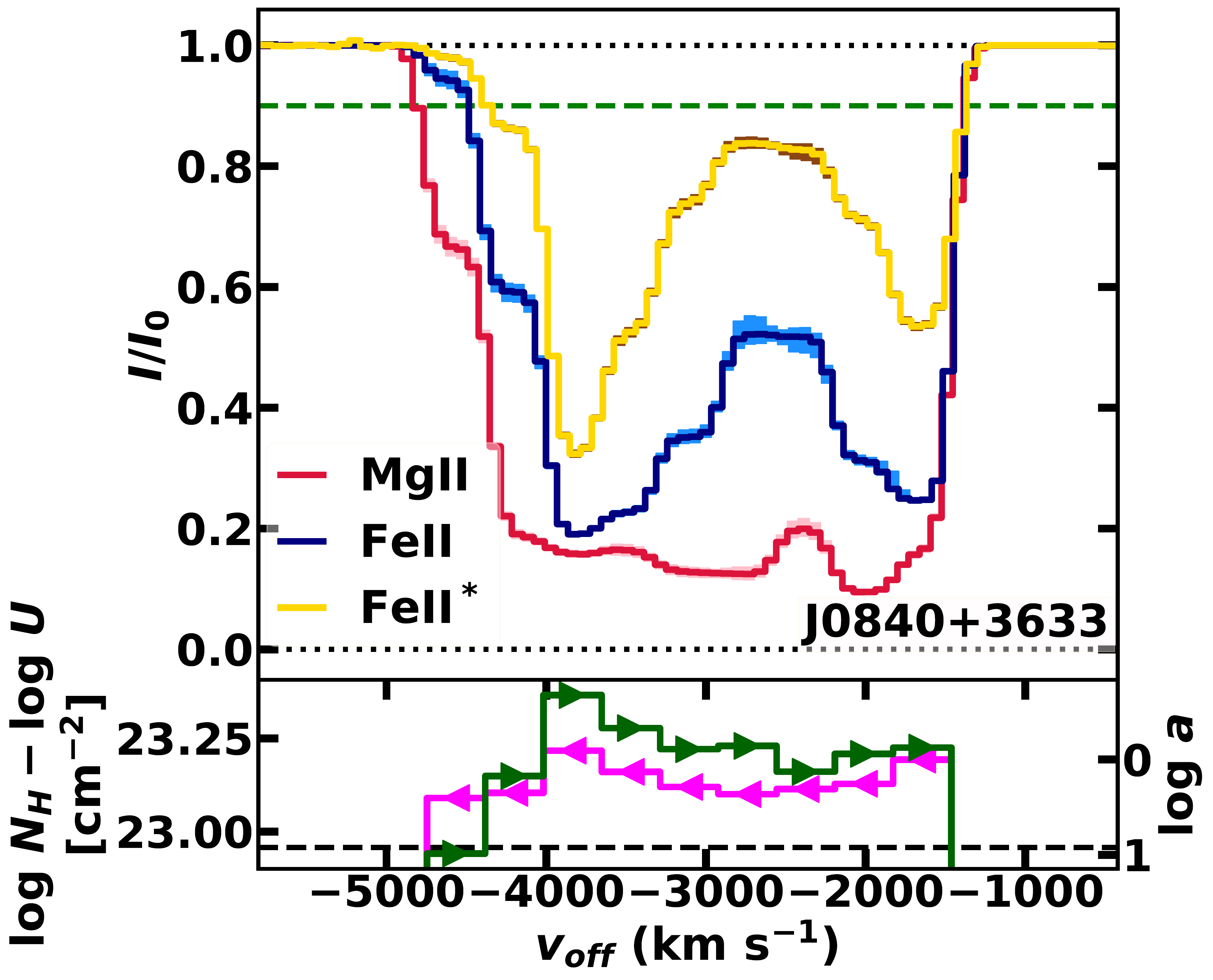}
\includegraphics[width=.48\linewidth]{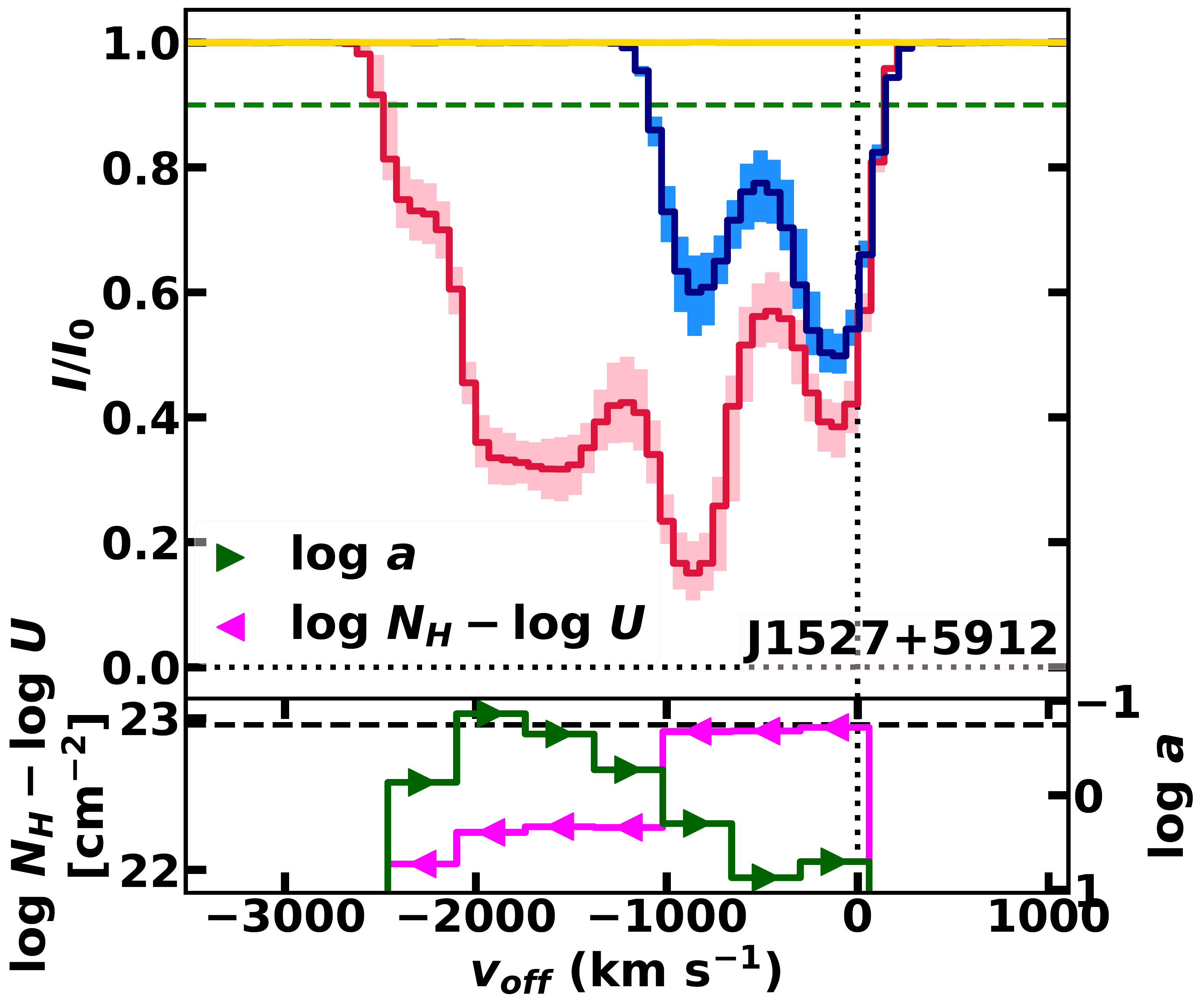}
\caption{The top panels show normalized spectrum ($I/I_0$) models for three line transitions (\ion{Mg}{2}$\lambda 2796$, \ion{Fe}{2}$\lambda 2383$, and \ion{Fe}{2}$^*\lambda 2757$) from SDSS~J0840$+$3633 and SDSS~J1527$+$5912.
The lighter shaded regions around each model represent 2$\sigma$ (95.45\%) uncertainties.
In the bottom panels, the column density parameter ($\log N_H-\log U$) and the covering fraction parameter ($\log a$) as a function of velocity are plotted in pink and green, respectively.
The complete figure with $I/I_0$ models from all 60 BALs is available in the online journal.
\label{fig:ii0_example}}
\end{figure*}

We found evidence for gas structure in the profiles in our sample.
Although most of the FeLoBALs were modeled with a single ionization parameter and a single density (\S~\ref{subsec:bestfitmodels}), we found dramatic changes in column density and covering fraction across the BAL velocity profile.
Two of the $I/I_0$ models for SDSS~J0840$+$3633 and SDSS~J1527$+$5912 are shown in Figure~\ref{fig:ii0_example}.
These models show remarkable differences between the profiles of the \ion{Fe}{2} and \ion{Mg}{2} lines that is caused by a change in both $\log N_H-\log U$ and $\log a$ across the trough.
In particular, the excited state \ion{Fe}{2} opacity profile for SDSS~J0840$+$3633 has a prominent double-peak structure with lower apparent opacity in the center that is not obvious in the opacity profile of the \ion{Mg}{2}.
The $I/I_0$ model for SDSS~J1527$+$5912 shows the presence of a high-column density core at the lower velocity end of the profile; therefore, we observe a narrower width for \ion{Fe}{2}.
This BAL has low density ($\log n\sim4.7\rm\ [cm^{-3}]$) and the excited state \ion{Fe}{2} transition is not observed.

\begin{figure}
\epsscale{.56}
\plotone{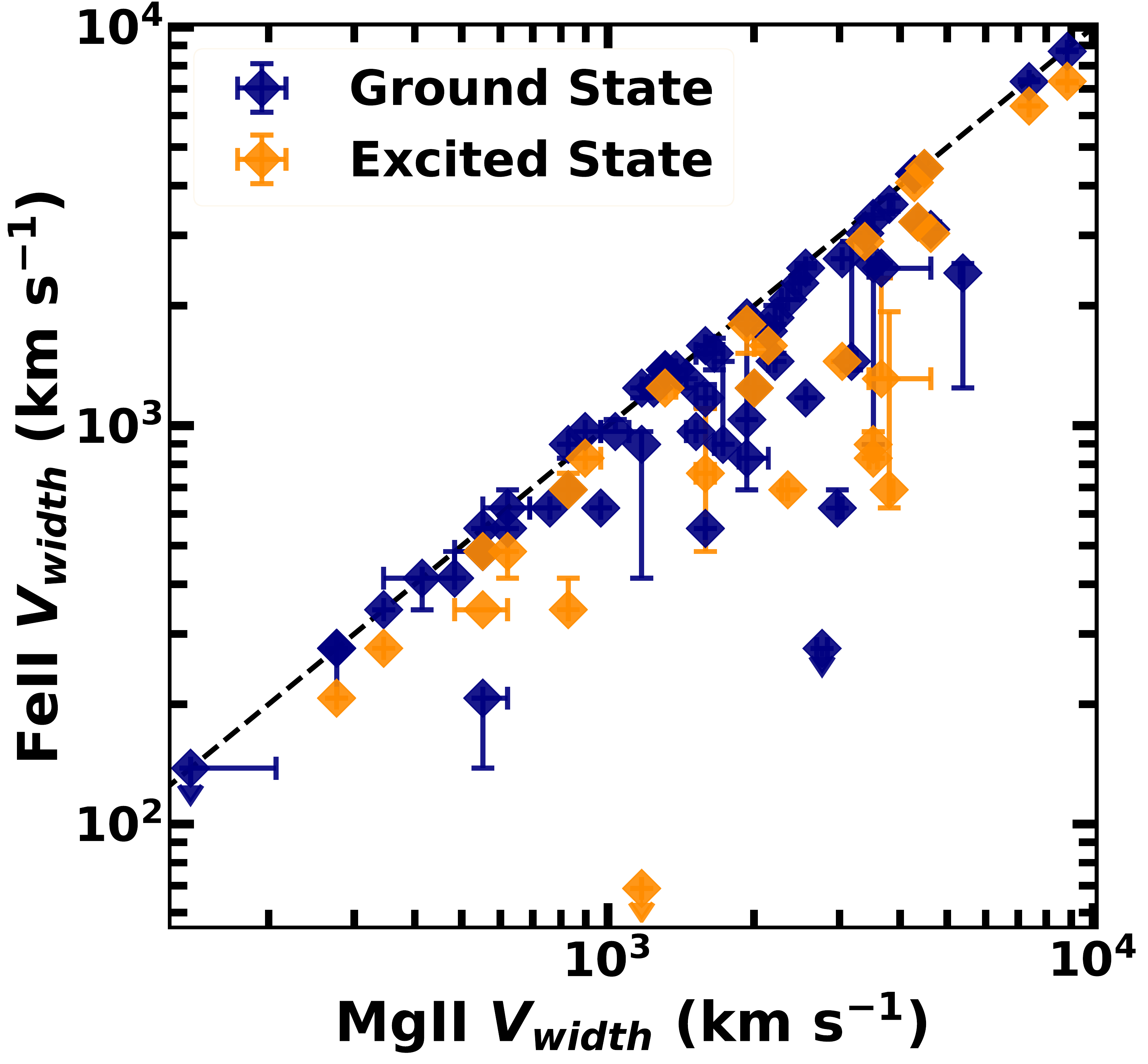}
\caption{The width measurements from different line transitions are plotted with 2$\sigma$ (95.45\%) error bars.
The widths measured with the excited state \ion{Fe}{2} are noticeably smaller than the widths measured from the \ion{Mg}{2} or the ground state \ion{Fe}{2} transitions.
Objects with the excited state \ion{Fe}{2} transition are plotted with orange diamonds.
\label{fig:multiple_widths}}
\end{figure}

Figure~\ref{fig:multiple_widths} shows that the BAL widths measured from \ion{Fe}{2} transitions are generally smaller or similar to those measured from \ion{Mg}{2}.
The widths of the excited state \ion{Fe}{2} transitions were significantly smaller than the widths of the ground state \ion{Fe}{2} and \ion{Mg}{2}.
Since a higher column density is required to accumulate significant opacity in excited state \ion{Fe}{2}*, this means that the trough column density is not constant with velocity.
In other words, the differences in widths can be ascribed to the intrinsic inhomogeneous physical structure of BAL gas, such as change in $\log N_H-\log U$ across the BAL troughs.
For example, the model for SDSS~J1527$+$5912 show a significantly wider profile in \ion{Mg}{2} than in the ground state \ion{Fe}{2} because the bins at higher velocities show a drop in $\log N_H-\log U$ (below the hydrogen ionization front at $\log N_H-\log U\sim23\ \mathrm{[cm^{-2}]}$) and thus they do not produce opacity from \ion{Fe}{2}.
BALs identified with substantially larger \ion{Mg}{2} widths than the \ion{Fe}{2} widths showed extended low $\log N_H-\log U$ structure where the bins at the lowest and highest ends in velocity only produced \ion{Mg}{2} opacity.
Similarly the opacity from the excited state \ion{Fe}{2} only appeared in the bins with high $\log N_H-\log U$ and oftentimes these high opacity concentrations or ``cores'' were only found in a small subset of bins for a given BAL component.

In addition to the change in column density, the covering fraction also varies with velocity and the combination of the two can change the relative widths of these lines.
For saturated lines, the line depths and shapes are mostly controlled by the partial covering.
{\it SimBAL} takes both of the effects into consideration when using tophat accordion models to fit the spectra.
For example, in Figure~\ref{fig:ii0_example}, both $\log N_H-\log U$ and $\log a$ change significantly across the trough and as a result we observe a sawtooth-shaped line profile in excited state \ion{Fe}{2} in the $I/I_0$ model for SDSS~J0840$+$3633.

We measured the offset velocities for all three transitions separately (\S~\ref{subsubsec:gen_opacity_prof}) using the $I/I_0$ models and they
showed no systematic difference from the summary outflow velocities we calculated from the distributions of intrinsic opacity for each BAL (\S~\ref{subsubsec:kinematics_measure}).
This result shows that our definition of summary outflow velocities is not biased against any particular transition and that in general where the intrinsic opacity is high, the maximum of the apparent opacity is also high despite partial covering heavily influencing the apparent line and opacity profiles.

\citet{voit93} found that the low-ionization lines such as \ion{Mg}{2} and \ion{Al}{3} are typically only located at the low-velocity ends of the BAL troughs with narrow line profiles whereas high-ionization lines \ion{C}{4} appear in a wider velocity range extending to much higher velocity showing much broader line profiles.
They concluded that nonmonotonic acceleration or deceleration of outflow gas scenarios can explain such velocity structure.
Although we found physical properties change across the BAL troughs and thus the opacity profiles for different transitions may show completely different shapes for a given BAL component, we did not find robust systematic trends between the profiles of \ion{Mg}{2}, \ion{Fe}{2}, and \ion{Fe}{2}*.
Future work with a sample of high redshift FeLoBALQs will allow us to study the absorption line profiles of \ion{C}{4} using SDSS spectra and we will be able to examine how their line profiles differ from the low-ionization lines and to investigate the potential ionization structure in BAL absorbing clouds.

\begin{figure}
\epsscale{.98}
\plotone{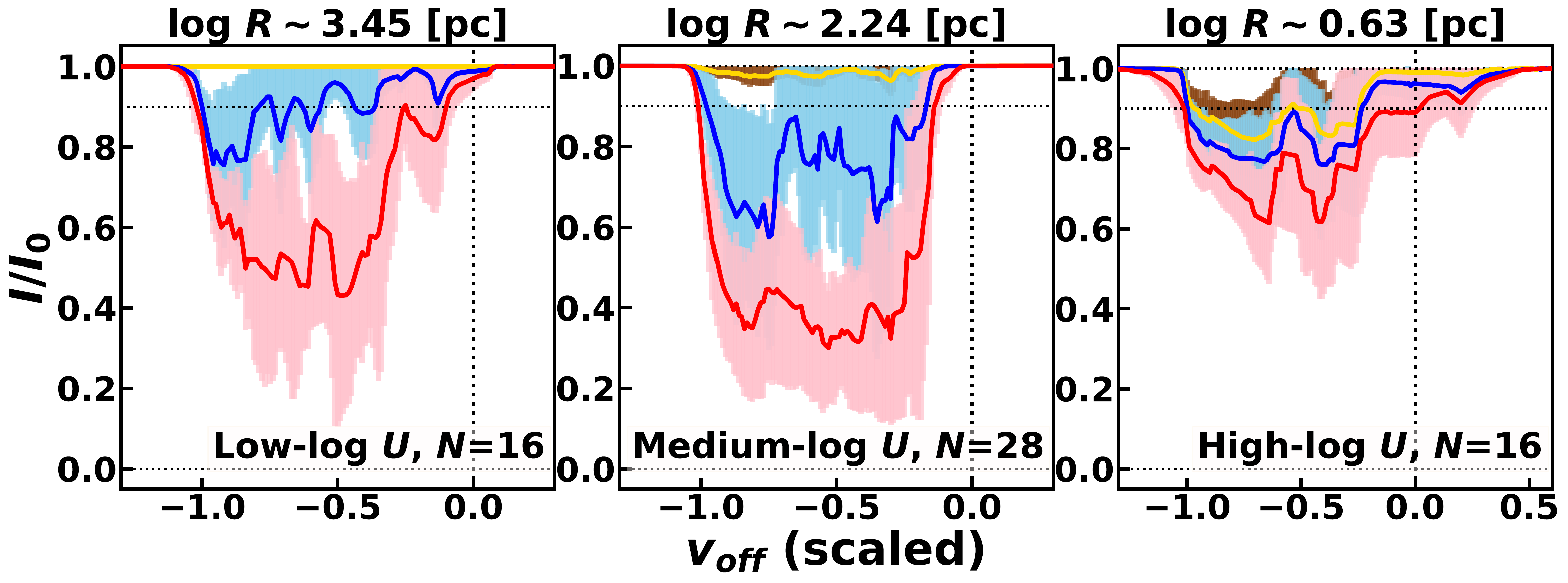}
\caption{The three composite $I/I_0$ model spectra are plotted.
The left panel shows the median composite $I/I_0$ model spectrum for the low-$\log U$ BALs and the middle and the right panel shows the same model for the medium-$\log U$ BALs and high-$\log U$ BALs, respectively.
The median $\log R$ for each group is noted above each panel.
The numbers of BAL components used to generate each composite are also noted.
The \ion{Mg}{2}$\lambda 2796$, \ion{Fe}{2}$\lambda 2383$, and \ion{Fe}{2}$^*\lambda 2757$ transition models are plotted in red, blue, and yellow, respectively.
The pink, light blue, and brown shaded areas represent the median absolute deviation for the \ion{Mg}{2}$\lambda 2796$, \ion{Fe}{2}$\lambda 2383$, and \ion{Fe}{2}$^*\lambda 2757$ transitions, respectively.
The normalized $v_{off}$ was used on the x-axis where 0 represents $v_{off}=0\ \rm km\ s^{-1}$ and $-1$ (or 1 for $v_{off}>0$) represents the maximum BAL velocity measured for each $I/I_0$ model.
\label{fig:ii0_composite}}
\end{figure}
Trends in the absorption-line structure hold promise for illuminating the nature of the acceleration mechanisms or gas cloud structure as a function of radial position in the quasar.
To explore this possibility, we created composite $I/I_0$ profiles for subsamples with similar ionization and radial location for the outflowing gas.
Specifically, we grouped the BALs into three groups using $\log U$ values: 16 low-opacity BALs with low $\log U$ that have lower limits on $\log R$ estimates (due to upper limits on $\log n$ constraints); 28 intermediate-opacity BALs with $\log U<-1.5$ that have well-constrained $\log R$ (and $\log n$); 16 high-opacity BALs with $\log U>-1.5$.
The median $\log U$ for the three groups are $-2.66$, $-2.24$, and $-0.02$.
Because there is a strong correlation between $\log R$ and $\log U$ (\S~\ref{subsec:derived_pars}, \S~\ref{subsec:corr}), these groups also showed clear differences in $\log R$ properties: the median $\log R$ for the low-$\log U$, intermediate-$\log U$, and high-$\log U$ BAL group are 3.45, 2.24 and 0.63 [pc], respectively.
The three groups represent (1) distant, low ionization BALs with kiloparsec-scale winds, (2) intermediate-scale medium ionization BALs, and (3) compact high ionization BALs that are located within $\sim10$ pc from the central engine with size scales comparable to the torus scale.
The $I/I_0$ models for each BAL component were normalized with respect to their maximum offset velocity where in the new normalized $v_{off}$ axis $-1$ (or 1 for $v_{off}>0$) represents the maximum velocity measured from the $I/I_0$ models \citep[a similar parameterization can be found in][]{borguet10}.
We then median combined the $I/I_0$ models and calculated the median absolute deviation.
Figure~\ref{fig:ii0_composite} shows the composite $I/I_0$ model spectra from the three BAL groups.

The composite $I/I_0$ models showed large differences in the line depths of \ion{Mg}{2}$\lambda 2796$ and the excited-state \ion{Fe}{2}$^*\lambda 2757$.
Noting that $\log U$ is correlated with $\log n$ and $\log a$,
we expect the BALs in the low $\log R$  BAL group with high $\log U$ to also have higher $\log n$ and $\log a$ (less covering) as observed.
The high $\log a$ (less covering) makes all the line transitions appear shallow, which is why the line depth of \ion{Mg}{2} is the smallest in the composite for the small-$\log R$ group.
On the other hand, the higher value of $\log n$ can populate the excited state \ion{Fe}{2} ions in the gas and create strong excited-state \ion{Fe}{2} absorption lines,
so we see the deepest excited state \ion{Fe}{2} line profile compared to the other composites.
The composite for the low $\log U$ group with distant winds shows no opacity from the excited state \ion{Fe}{2}.
While the differences in the line depths seen in the composites for different $\log R$ can be unambiguously explained by the differences in the physical properties, we did not find robust evidence for a systematic difference in gas structures or opacity profiles between the composite $I/I_0$ models.
We may expect to find a more definitive answer with composite $I/I_0$ models from a larger sample; however, the large dispersion (median absolute deviation) seen in Figure~\ref{fig:ii0_composite} suggests that the line profiles depend more strongly on the individual physical conditions of each FeLoBALs than any global trend.
We also expect larger differences between the low-ionization lines considered here, and high-ionization lines such as \ion{C}{4}$\lambda\lambda 1549,1551$.

\subsubsection{\texorpdfstring{\ion{Ca}{2}, \ion{He}{1}*, and Balmer Absorption Lines}
{CaII, HeI*, and Balmer Absorption Lines}}\label{subsubsec:he_h_profile}
\ion{Ca}{2} H and K, \ion{He}{1}*$\lambda 3889$, and Balmer absorption lines found in the rest-optical wavelengths can provide crucial information about the physical properties of the outflowing gas.
\citet{leighly14} analyzed \ion{Ca}{2}, \ion{Na}{1}, and \ion{He}{1}* absorption lines observed in the optical spectrum of the nearby Seyfert 1 object Mrk 231 to discover evidence for an interaction between a quasar outflow and surrounding the ISM.
\citet{leighly11} discussed the advantages of using \ion{He}{1}*$\lambda 3889$ absorption line to study BAL winds that have high column densities.
The density of the BAL gas needs to be high ($\log n\gtrsim7$) in order to produce an observable amount of opacity from Balmer transitions \citep[e.g.,][]{leighly11}.
Therefore, Balmer absorption lines can be used as diagnostics to detect compact BAL winds (small $\log R$) with high densities.

We found that many of our FeLoBALQs show \ion{Ca}{2}, \ion{He}{1}*, and Balmer absorption lines in the spectra.
The best-fit models provided excellent fit both for the main \ion{Fe}{2} and \ion{Mg}{2} troughs and for the rest-optical absorption lines as well (Figure~\ref{fig:fitfig1};~\ref{fig:multiple_bals_model_1};~\ref{fig:multiple_bals_model_2}).
In contrast, previous BAL studies only found a small number (FeLo)BALQs with \ion{Ca}{2} BAL features \citep[e.g.,][]{boksenberg77,arav01,hall03}.
Moreover, Balmer absorption lines have been observed in only a small number of FeLoBALQs \citep[e.g.,][]{hall07,shi16,schulze18}.
That is maybe because there has been no systematic study of FeLoBALQs prior to this work.
We investigated how these rest-optical absorption lines are related to the FeLoBAL winds and what information about outflows we can gain from analyzing these transitions.

\begin{figure}
\epsscale{.97}
\plotone{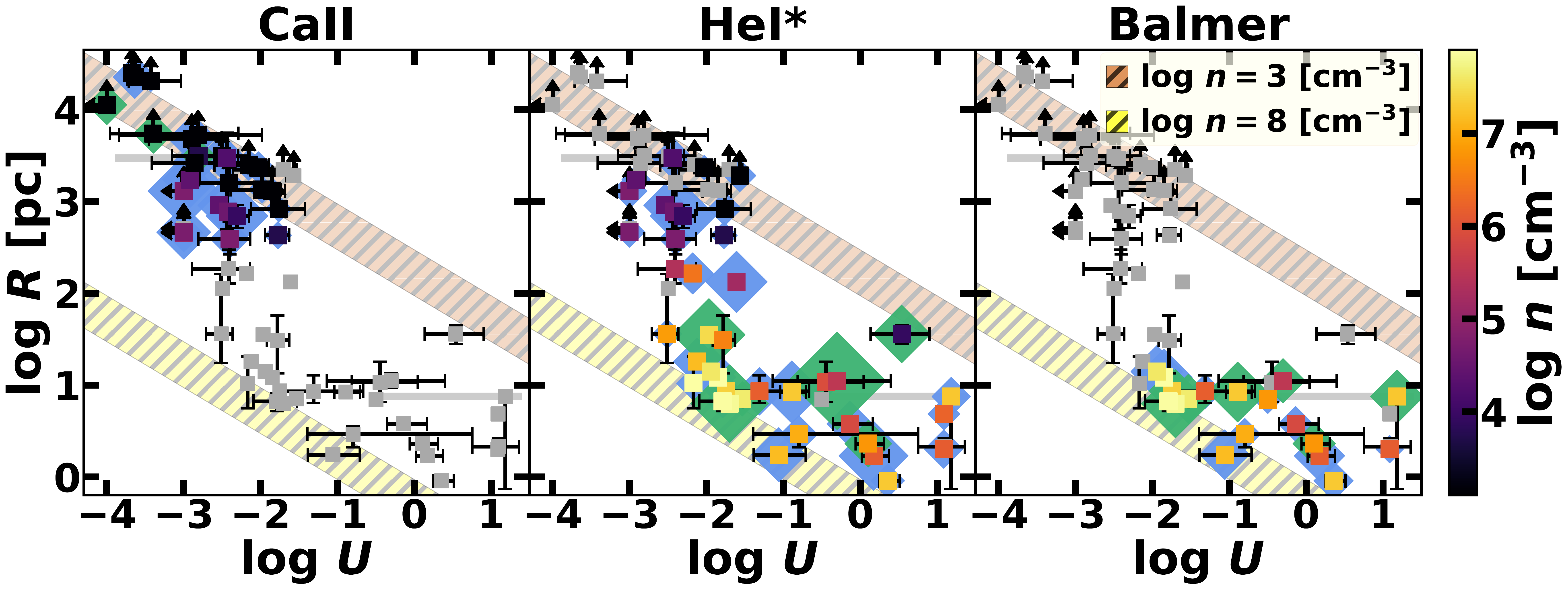}
\caption{
Figure~\ref{fig:fitpar_aR} modified to
show the distributions of absorption strengths (\S~\ref{subsubsec:gen_opacity_prof}) of select line transitions predicted using the $I/I_0$ models.
Our results point towards the presence of specific absorption lines as being key diagnostics for the location of the outflow ($R$).
The blue diamonds represent the absorption strengths calculated from best-fitting models.
The absorption strengths represented with green diamonds were calculated from the extrapolated {\it SimBAL} models (see text).
The grey squares represent the BAL components that are not predicted show any opacity from the line transition featured in each panel.
The size of the diamonds  is proportional to values of the absorption strength parameter (i.e., larger markers represent stronger absorption).
The \ion{Ca}{2} absorption lines are only found in distant BALs ($\log R\gtrsim3$ [pc]) and Balmer absorption lines are found in BALs with $\log R\lesssim1$ [pc].
\ion{He}{1}* absorption lines are predicted and found in nearly all FeLoBALs.
The error bars show 2$\sigma$ (95.45\%) uncertainties and the grey shaded bars represent the range of the values among the tophat model bins for each BAL.
\label{fig:ii0_three_lines}}
\end{figure}

Out of 60 BAL components analyzed, we found 26, 40, and 18 BAL components are predicted to show opacity from \ion{Ca}{2} H, K, \ion{He}{1}*$\lambda 3889$, and Balmer transitions, respectively.
Because our sample contains a wide range of redshifts ($0.66 < z < 1.63$), not all SDSS/BOSS spectra we analyzed included these transitions in the bandpass.
Therefore, we had to extrapolate the best-fitting models to longer wavelengths to calculate the absorption strengths for some of the BALs.
Figure~\ref{fig:ii0_three_lines} shows the distributions of absorption strength predicted by the models (\S~\ref{subsubsec:gen_opacity_prof}) along the $\log U$ and $\log R$ axes.

The opacity from Balmer transitions is only found in BALs with $\log R\lesssim1$ [pc].
In other words, we suggest that the presence of Balmer absorption lines can be used as an indicator for compact BAL winds.
In contrast, only the kiloparsec-scale BAL winds showed opacity from \ion{Ca}{2}.
This is expected given the distant BALs have low $\log U$ and low $\log n$ (Figure~\ref{fig:fitpar_ac_ad}), the physical conditions required to create Ca$^+$ ions.
However, in a few cases \ion{Ca}{2} BALs were observed with other high-$\log U$ or high-$\log n$ absorption lines at similar velocities (e.g., \ion{He}{1}*, excited state \ion{Fe}{2}), and therefore these \ion{Ca}{2} outflows likely lie at small radii ($\log R\ll$kpc) and have unusual physical conditions.
For instance, \citet{leighly14} inferred a density increase
at the hydrogen ionization front to explain \ion{Na}{1}D absorption in Mrk 231.
\citet{hall02} suggested a significant gas temperature change to explain the \ion{Ca}{2} absorption lines observed in SDSS~J0300$+$0048.
The opacity from \ion{He}{1}* was found in the majority of FeLoBALs over a wide range of $\log R$.
This result is consistent with what has been reported by \citet{liu15} where they also found a large fraction of \ion{Mg}{2} selected LoBALQs with \ion{He}{1}* absorption lines observed in the spectra.
The presence of \ion{He}{1}* absorption lines in (Fe)LoBALQs is a direct consequence of absorbing gas having high enough column density to produce observable low-ionization absorption lines such as \ion{Mg}{2} (\ion{Fe}{2}).

In summary, the absorption lines from \ion{Ca}{2}, \ion{He}{1}*, and Balmer transitions observed and predicted in rest-optical spectra of FeLoBALQs provide us with critical information about the physical properties.
In particular, the presence of \ion{Ca}{2} or Balmer absorption lines can be used to estimate the size scales of the BAL outflows and the \ion{He}{1}* absorption line can be used to identify (Fe)LoBALs.
This shows that even without a detailed photoionization modeling of rest-UV (FeLo)BALQ spectrum, one could potentially
predict the approximate outflow spectral properties
from the rest-optical spectrum.

\subsection{Overlapping Trough FeLoBALs}\label{subsec:ot_obj}
Overlapping troughs show magnificent absorption features in the rest-UV spectra where the continuum emission is often nearly completely absorbed between $\lambda\sim2000$ \AA\/ and $\lambda\sim2800$ \AA\/ by a multitude of \ion{Fe}{2} absorption lines.
\citet{hall02} introduced several objects with overlapping troughs and discussed their spectral features; however, in-depth analysis of overlapping trough BALs has not been possible with conventional methods (e.g., measuring ionic column densities from the individual line profiles) due to extreme line blending.
{\it SimBAL} can be used to analyze spectra with overlapping troughs, as demonstrated in SDSS~J1352$+$4239, a heavily-absorbed overlapping-trough object \citep{choi20}.

The term ``overlapping trough'' has been used differently by different authors \citep[e.g.,][]{hall02,meusinger16}.
In the literature, overlapping-trough objects generally refer to the FeLoBALQs with very broad absorption features reaching near-zero flux at the bottom, following the more stringent criteria introduced by \citet{hall02}.
We used a modified criterion to identify overlapping troughs that is based only on BAL morphology: such BALs
were identified based on whether the continuum emission at $\lambda\sim2500$ \AA\/ (outflow reference frame) where the absorption lines from the excited state \ion{Fe}{2} are expected to appear in the spectrum is heavily absorbed or not.
This method allowed us to focus only on the morphology of the troughs and identify all extremely wide troughs regardless of the amount of partial covering or the continuum shape.

\begin{figure*}
\epsscale{1.05}
\plotone{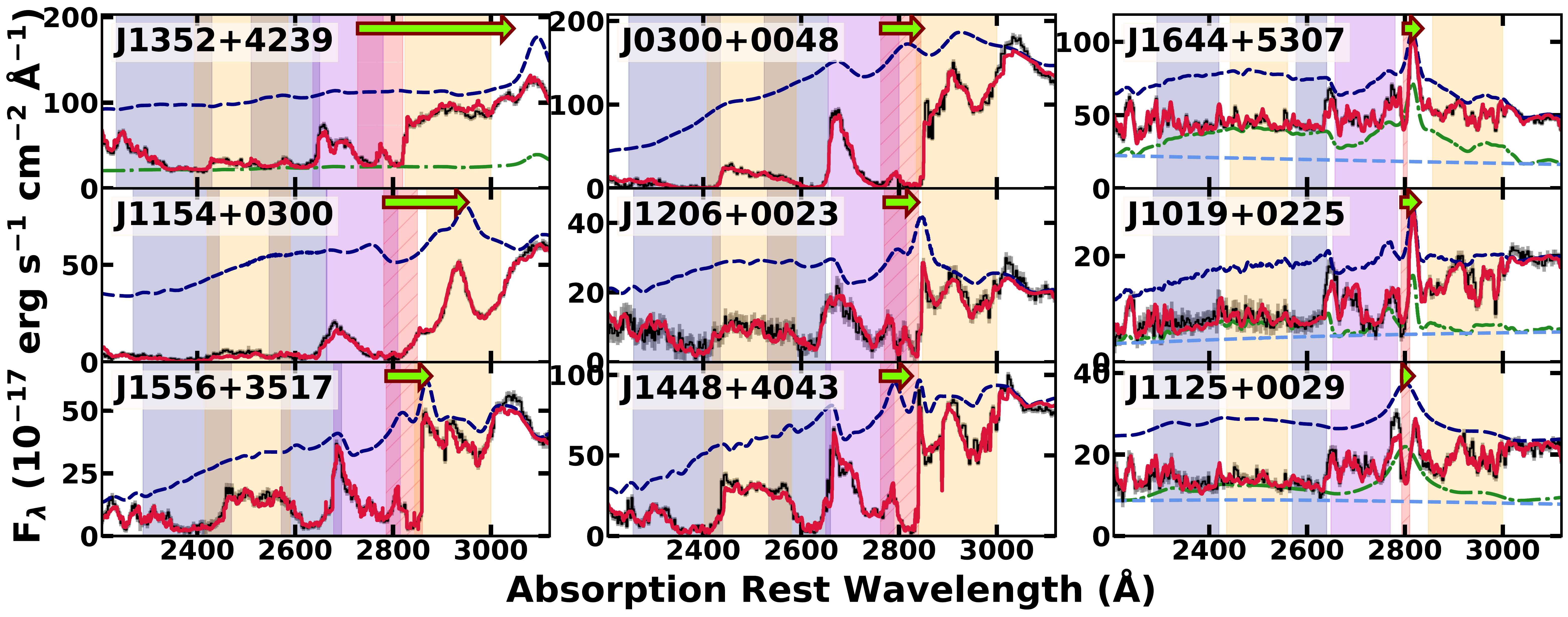}
\caption{The eight objects with overlapping troughs are plotted with SDSS~J1352$+$4239 \citep{choi20}.
All the spectra have been shifted to the reference frame of the main BAL trough.
The arrows extend from the estimated center of the \ion{Mg}{2} trough to the center of emission line to illustrate the BAL offset velocity.
The full spectral models and the continuum models are plotted in red and dashed blue lines, respectively.
The green dot-dashed lines show the unabsorbed line emission and continuum emission if present.
The additional blue dashed lines in the three objects on the right column represent the unabsorbed power-law continuum emission.
The shaded regions represent various absorption lines and absorption-line classes: red, \ion{Mg}{2}; blue, low-excitation \ion{Fe}{2}; pink, high-opacity moderate-excitation \ion{Fe}{2}; yellow, low-opacity moderate-excitation \ion{Fe}{2}.
Also see Figure~\ref{fig:felobal_anatomy}.
\label{fig:ot_objs}}
\end{figure*}

Figure~\ref{fig:ot_objs} shows the eight objects with overlapping troughs we found in our sample as well as SDSS~J1352$+$4239 \citep{choi20}.
We observe a great diversity in spectral morphology and gas kinematics as well as outflow gas properties.
Their broad troughs have been fit with tophat models and five objects (SDSS~J0300$+$0048, SDSS~J1154$+$0300, SDSS~J1556$+$3517, SDSS~J1206$+$0023, SDSS~J1448$+$4043) required the use of general reddening \citep{choi20} to model the anomalous reddening.
Three out of the eight objects (SDSS~J1019$+$0225, SDSS~J1125$+$0029, SDSS~J1644$+$5307) also required an additional unabsorbed components underneath the troughs (\S~\ref{subsubsec:special_covering_model}).
A similar plot can be found in Figure 12 of \citet{lucy14} where they show different spectral morphologies of FeLoBALs with narrower lines.

Although the significant absorption seen in overlapping troughs indicates high-opacity and high-column density gas, this feature does not mean necessarily mean the overlapping trough BALs have high kinetic luminosities ($L_{KE}$).
The wide range of outflow velocities
seen in the overlapping trough BALs (marked by the lengths of the arrows in Figure~\ref{fig:ot_objs}) shows that some of them have no significant outflow velocity and thus are not carrying any significant mass or energy in the wind.
As shown in \S~\ref{subsec:corr}, the most important factor in determining the kinetic luminosity of an outflow is the outflow velocity.
Thus, we found that five objects have overlapping trough BALs that are powerful outflows with $\vert v_{off}\vert>4000\rm\ km\ s^{-1}$ and $L_{KE}/L_{Bol}>0.5\%$.
These five objects also show blended troughs reaching near-zero flux at the bottom and can be called overlapping-trough objects based on the classification criteria by \citet{hall02}.
They are further distinguished by the presence of anomalous reddening \citep[e.g.,][]{choi20}.
The other three objects, represented in the third column in Figure~\ref{fig:ot_objs}, have outflows with low velocities ($\vert v_{off}\vert<1600\rm\ km\ s^{-1}$) thus do not have the high $L_{KE}$ required for the quasar feedback.

In contrast to the wide range of outflow velocities found in the overlapping trough BALs,
all of these outflows were found at similar distances from the central engine ($\log R\lesssim1$~[pc]).
Figure~\ref{fig:fitpar_aR} shows where the overlapping trough BALs are found in the distributions of $\log R$ and $\log U$ as well as $\log n$.
The BAL gas that creates overlapping troughs has higher densities ($\log n\ga6\rm\ [cm^{-3}]$), higher ionization parameters ($\log U\gtrsim-1.85$), and higher hydrogen column density (Figure~\ref{fig:fitpar_aNh}) compared to the BALs in the overall sample.
These conditions are required to produce the high-excitation transitions that yield the necessary opacity at $\lambda\sim2500$ \AA.
While the overlapping trough BALs have a range of $\log U$ that spans about two dex, they are all found to be located in the vicinity of the dusty torus ($\log R\lesssim1$~[pc]).
Based on the results from our sample, we conclude that the overlapping trough features in the FeLoBALQ spectra can be used to identify compact BAL winds.
Overlapping trough BALs could give us information about the inner regions of quasars where we expect most of the acceleration to occur for radiatively driven outflows \citep[e.g.,][]{arav94}.

We identified three main ways FeLoBAL outflows create overlapping troughs.
The most straightforward method is with large velocity widths.
Although the kinematic properties of the BALs showed a wide range among the overlapping troughs in our sample, the high velocity overlapping troughs plotted in the left two panels in Figure~\ref{fig:ot_objs} represent the FeLoBALs with the highest outflow velocities and widths in our sample (Figure~\ref{fig:fitpar_vel_a_width}).
For instance, SDSS~J1154$+$0300 has the largest BAL width with $\sim7400\rm\ km\ s^{-1}$ and the offset velocity of $\sim-15,400\rm\ km\ s^{-1}$.
Naturally, with larger widths the line blending is significant and the high opacity gas in these winds will be able to create a wide overlapping trough FeLoBAL features.

Secondly, overlapping troughs can be produced from moderate to narrow width BALs with large amount of opacity from rare transitions.
The width of the trough in SDSS~J1644$+$5307 is only $\sim900\rm\ km\ s^{-1}$, a value that is comparable or slightly smaller to the average value from other non-overlapping FeLoBAL troughs.
The overlapping trough feature in that object is caused by the large number of excited state \ion{Fe}{2} transitions and absorption lines from multiple iron-peak elements between $\sim$2000 \AA\/ and $\sim$3000 \AA.
If an outflowing gas has a high enough density, ionization parameter, and column density to have a significant population of highly excited state \ion{Fe}{2} ($\rm E_{lower-level}\gtrsim3\ eV$), then the gas will be able to create thousands of absorption lines (see Figure~\ref{fig:felobal_anatomy}).
Because these absorption lines are densely packed, they can easily form a wide trough by line blending even with a narrow absorption line velocity width.

Finally, objects that have an ordinary FeLoBAL component at lower velocity with an additional high-opacity component at higher velocity may show overlapping troughs in the spectra.
As discussed in \S~\ref{subsec:multiple_BAL}, some of the objects in the sample required more than one BAL component, and overlapping trough features were created by the higher velocity, higher $\log U$ components.
For example, in Figure~\ref{fig:multiple_bals} we see that most of the higher-velocity components in objects with blended multi-BALs features (red arrows) are also identified as overlapping trough BALs (pink circles).
Figure~\ref{fig:multiple_bals_model_2} shows how these higher-velocity components produce the majority of the opacity needed to complete the overlapping trough features near $\lambda\sim2500$~\AA\/.
The lower-velocity components (Figure~\ref{fig:multiple_bals_model_2}, plotted in green) resembled typical FeLoBALs, mainly showing BALs from the ground state \ion{Fe}{2} and \ion{Mg}{2} transitions with little to no highly excited state \ion{Fe}{2} transitions (see Figure~\ref{fig:felobal_anatomy}; \S~\ref{subsubsec:felobal_anatomy}).
Without the higher-velocity components, these objects would appear nearly indistinguishable from the non-overlapping trough FeLoBALQs.

Some compact HiBAL outflows show BAL variability possibly due to the transverse motion of the outflow clouds or change in photoionization state of the outflow gas \citep[e.g.,][]{capellupo13}.
For instance, the disappearance of overlapping troughs has been observed in FBQS~J140806.2$+$305448 \citep{hall11} and in SDSS~J123103.70$+$392903.6 \citep{rafiee16,mcgraw15}.
It is plausible that in some cases the variability might be coming from the disappearance of the higher-velocity component.
For example, while the overlapping trough features from \ion{Fe}{2} disappeared in these two objects, the strength of the deep absorption feature from \ion{Mg}{2} at the lower velocity end remained consistent.
The sample of radio-selected quasars analyzed by \citet{zhang15b} showed an enhanced spectral variability rate for the overlapping-trough objects.
However, three objects (SDSS~J0300$+$0048, SDSS~J1125$+$0029, SDSS~J1154$+$0300) have been analyzed by \citet{mcgraw15}
but no evidence for BAL variability was found.
\citet{shi16} found evidence for spectral variability in SDSS~J1125$+$0029, the variability they found was attributed to the unabsorbed line emission flux underneath the \ion{Fe}{2} trough and they did not find any significant change in the BAL troughs.

Some of the multi-component objects with overlapping troughs (SDSS~J1125$+$0029 and SDSS~J1206$+$0023) showed a lower-velocity and lower-$\log U$ component that only produces significant opacity from \ion{Mg}{2} with little \ion{Fe}{2} opacity (\S~\ref{subsec:multiple_BAL}).
The velocity of these components
extends across $\sim0\rm\ km\ s^{-1}$ and they are located at about an order of magnitude larger distances than the higher-velocity components (Figure~\ref{fig:multiple_bals}).
The origins of these BAL components are uncertain.
Discussions related to redshifted BALs that have similar kinematic characteristics have suggested rotationally dominated outflows or infalls as potential origins \citep{hall13}.

Three objects (SDSS~J1019$+$0225, SDSS~J1125$+$0029, SDSS~J1644$+$5307)
were modeled with a modified partial covering scheme (\S~\ref{subsubsec:special_covering_model}),
because the usual power-law partial covering that is used in {\it SimBAL} was not sufficient to model the observed non-black saturation.
Although these three objects show overlapping trough BALs in the spectra with no continuum emission recovery shortward of $\lambda\sim2800$ \AA\/, they do not meet the usual criteria for overlapping-trough object classification due to significant flux beneath the troughs.
\citet{choi20} discussed the unabsorbed component under the overlapping troughs in SDSS~J1352$+$4239 as scattered flux where $\sim29\%$ of the light from the accretion disk and the broad line region (BLR) is scattered directly into the line of sight.
In these three objects, the inferred scattered fraction would have to be
greater than $50\%$; this is unphysical.
All three of these objects have compact outflows with
$\log R\lesssim1$ [pc], and their close proximity to the BLR and the accretion disk
suggests that the BAL gas is physically only covering part of the continuum and line emission.
Thus the non-zero flux at the bottom of the troughs are due to strong partial covering effect.
In Paper III \citep{choi_prep_paper3}, we discuss the angular size scales of the accretion disk and broad line region seen from the locations of the BAL winds using the black hole masses and Eddington ratio estimates obtained in Paper II \citep{leighly_prep}.
We conclude that the large angular size scales of the accretion disk seen from the BAL gas at these small $\log R$ values can plausibly produce this partial covering scenario.

\subsection{``Loitering'' Outflows}\label{subsec:loiter}
\begin{figure}
\epsscale{.56}
\plotone{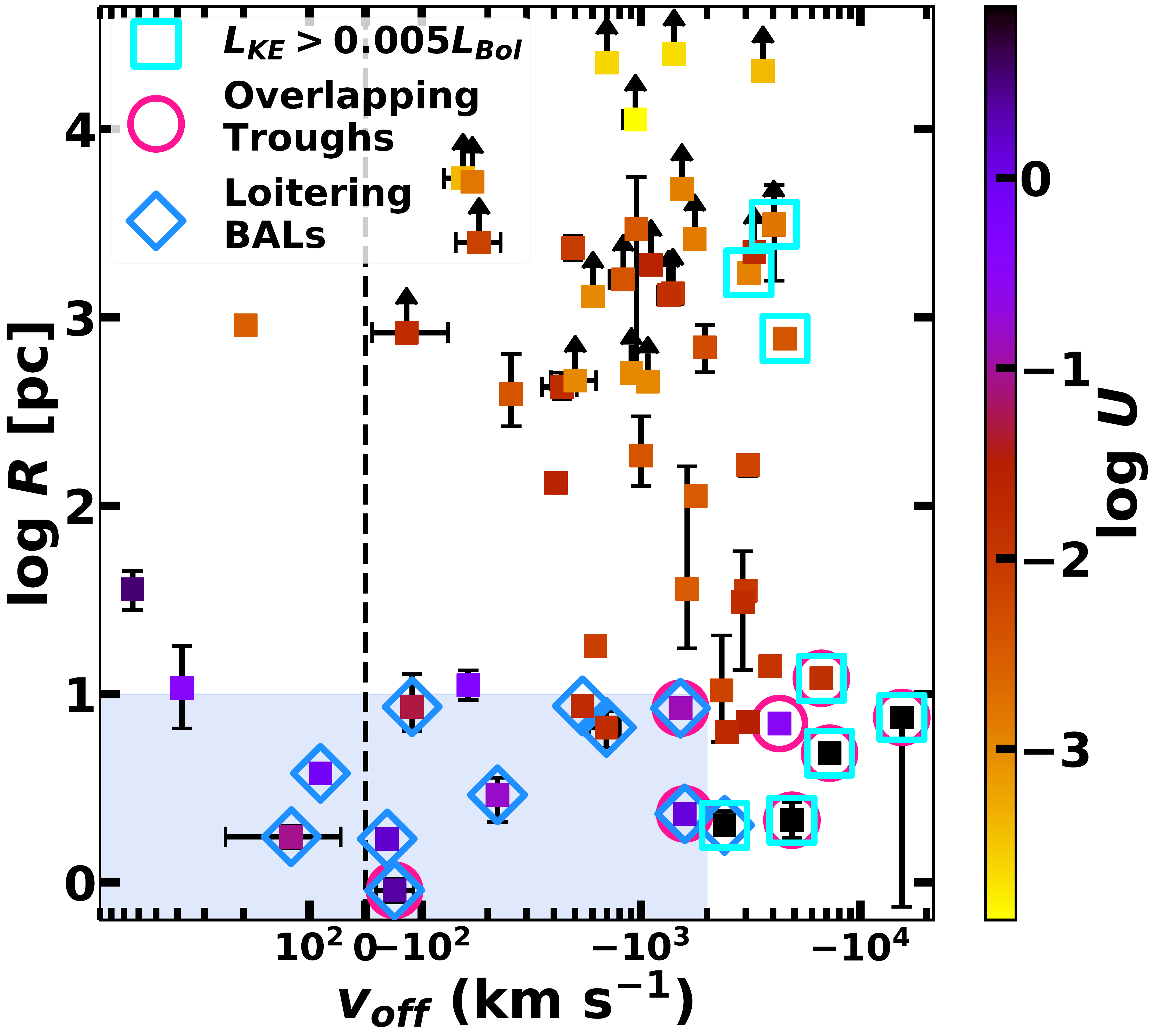}
\caption{The distributions of the offset velocities ($v_{off}$) and the distances of the BALs from the central SMBHs ($\log R$).
The blue shaded region represents the defining criteria for the loitering outflows ($\log R<1,\ \vert v_{off,\ \mathrm{FeII\ excited}}\vert<2000\rm\ km\ s^{-1}$, and $v_{width,\ \mathrm{FeII\ excited}}<2000\rm\ km\ s^{-1}$).
Excluding the loitering outflows and the inflows ($v_{off}>0$), we found robust statistical evidence ($p<0.05$) for a correlation between $\log R$ and $v_{off}$ where the high velocity flows
are found closer to the central engine.
The error bars show 2$\sigma$ (95.45\%) uncertainties.
\label{fig:loiter_vel_logR}}
\end{figure}
We identified a group of eleven compact FeLoBAL winds with small offset velocities and distinct properties, and classified them ``loitering'' outflows (we use the term ``outflow'' in the nomenclature, but note that some loitering outflows have $v_{off}>0\rm\ km\ s^{-1}$).
The loitering outflows are defined by the following properties: a) $\log R<1$ [pc]; b) $\vert v_{off,\ \mathrm{FeII\ excited}}\vert<2000\rm\ km\ s^{-1}$ and $v_{width,\ \mathrm{FeII\ excited}}<2000\rm\ km\ s^{-1}$.
In other words, the loitering outflows are relatively static gas clouds that are located close to the central SMBHs within the vicinity of the torus and thus they are appeared to be neither outflowing or inflowing.
Figure~\ref{fig:loiter_vel_logR} shows how the distribution of the physical properties of loitering outflows differs from the other more typical BAL winds and our defining criteria.

Instead of using $v_{off}$ and $v_{width}$ estimated from the \ion{Mg}{2} transition, we used the values measured from the excited-state \ion{Fe}{2}$^*\lambda 2757$ extracted from the $I/I_0$ models discussed in \S~\ref{subsubsec:gen_opacity_prof}.
This was done to avoid excluding any BALs with narrow \ion{Fe}{2} absorption features that may have larger $v_{off}$ or $v_{width}$ due to extended \ion{Mg}{2} opacity profiles.
Moreover, the opacity profile of the excited state \ion{Fe}{2} traces the high-density cores within the BAL gas structure that produce the majority of opacity and carry most of the mass and energy in the wind.
However as can be see in Figure~\ref{fig:loiter_vel_logR}, our classifications would not have been significantly affected if we had used the standard $v_{off}$ and $v_{width}$ from \ion{Mg}{2}.
Only one BAL, in SDSS~J1006$+$0513, showed modest differences between the values ($v_{off}\sim-2400\rm\ km\ s^{-1}$, $v_{width}\sim3500\rm\ km\ s^{-1}$; $v_{off,\ \mathrm{FeII\ excited}}\sim-1420\rm\ km\ s^{-1}$, $v_{width,\ \mathrm{FeII\ excited}}\sim830\rm\ km\ s^{-1}$).
We note that the selection criteria were chosen based on the visual inspection of the distribution of parameters obtained from our low redshift sample.
Future work with larger samples and high redshift objects may modify our selection criteria.

The loitering outflows have high $\log U$ with lower partial covering (high $\log a$)
and five out of eleven BALs required a modified partial-covering model.
Conversely, five out of 6 objects modeled with modified partial-covering were found to have loitering outflows.
These BALs show an extremely large number of absorption line transitions in the spectra because the high-$\log U$ FeLoBALs also have high column density and such a thick gas slab can produce a plethora of rare line transitions from various excited states \ion{Fe}{2} as well as rarer iron-peak elements such as Co and Zn (Figure~\ref{fig:felobal_anatomy}).
Three of the eleven (SDSS~J1019$+$0225, SDSS~J1125$+$0029, and SDSS~J1644$+$5307) loitering outflows are also classified as overlapping trough BALs.
These three were notable because they required step-function partial covering for the power-law continuum emission and unabsorbed line emission in the model.
In addition to these objects, SDSS~J1128$+$0113 and SDSS~J1321$+$5617 were modeled with modified partial-covering where the line emission was unabsorbed (Figure~\ref{fig:no_line_model}).
The remaining six were modeled using the standard power-law partial covering.
These objects were characterized by lower signal-to-noise ratios (median SNR$\sim3-6$), and it is possible that modified partial-covering would be required in higher signal-to-noise ratio spectra where the bottoms of the troughs would be better defined.

The locations of the loitering outflows in the quasar suggests the torus as a potential origin for the absorbing gas.
The dust sublimation radius for quasars with $\log L_{bol}\sim46.0-47.0\ \rm[erg\ s^{-1}]$ is $R_{sub}\sim0.2-0.6$ pc \citep{laor93}.
The outer radius of the torus was estimated to be $R_o\sim40-120$ [pc] using the equation $R_o<12L_{45}^{1/2}$ pc
\citep[$L_{45}=L_{bol}/10^{45}\ \rm erg\ s^{-1}$;][]{nenkova08}.
The loitering outflows are located at $R\sim1-10$ pc which is within the region where we expect the dusty torus to be.
A wind origin of the torus has been proposed by \citet{elitzur06}, and
recent studies using magnetohydrodynamic models of a dusty wind have been successful in finding potential connections between the outflowing winds and the windy torus structure \citep[e.g.,][]{keating12,gallagher15}.
Paper II \citep{leighly_prep} investigates the accretion properties of the FeLoBALQs that have loitering outflows (or so-called ``loitering outflow objects'').
We found that the loitering outflows all
had lower-than-average accretion rates. \citet{elitzur06}
predict that at low accretion rates, the wind forming the torus
fails. In Paper IV \citep{leighly_prep_paper4}, we conjecture that in the loitering
outflow objects, the torus wind is on the verge of failing, so that it
is not optically thick enough to reprocess continuum into the infrared
band, but is still optically thick enough to produce the observed
\ion{Fe}{2} absorption.

As expected from the similar size scales of the outflows ($\log R\lesssim1.0$~[pc]), the spectral morphology of the loitering outflows resembles that of FeLoBALQs with Balmer absorption lines (\S~\ref{subsubsec:he_h_profile}).
Using this spectral property, we can use the presence of the narrow Balmer absorption lines (as well as \ion{He}{1}* to differentiate BALs from galaxy contamination) in the optical spectra to search for quasars with loitering outflows.

We found a significant correlation between the $v_{off}$ of the BAL outflows and $\log R$ ($p=0.02$, Kendall $\tau$; Figure~\ref{fig:loiter_vel_logR}) when the loitering outflows and the inflows are removed ($N=46$).
The remaining compact BAL winds that are located at $\log R\lesssim1.0$ [pc] have the highest outflow velocities and most of them were also identified as overlapping trough BALs with powerful outflows ($L_{KE}/L_{bol}>0.5\%$; \S~\ref{subsec:ot_obj}).
Assuming the BAL clouds have not traveled significantly such that their current locations in the quasars represent where they were initially launched, simulations and statistical calculations predict that such a correlation should exist.
For the line-driven outflows the terminal velocities of the winds roughly correlate with the Keplerian circular velocity or the escape velocity at the launch radius \citep[e.g.,][]{proga04,giustini19}.
A simple equation of motion derivation using the radiative acceleration also predicts $v_{off}$ or $v_\infty\propto R_{in}^{-1/2}$ where $R_{in}$ is the inner wind radius or the launch radius \citep[e.g.,][]{hamann98,leighly09,choi20}.
Because the correlation was not apparent when the loitering outflows were included in the analysis,
we postulate that the loitering outflows may represent a different BAL phenomenon with potentially different acceleration mechanisms involved (see also Paper III, \citealt{choi_prep_paper3}).

\section{Discussion}\label{sec:disc}
\subsection{Location and Origin of FeLoBAL Winds}\label{subsec:loc_origin_felobal}
Our results show that the FeLoBAL winds span a large range of radii or distances from the central SMBH.
However, the number density of objects may not be constant with $R$.
There is an apparent gap near $\log R\sim2$ [pc]; see Figure~\ref{fig:fitpar_aR}.
While a larger sample may fill this gap, we can use the assumed break around $\log R\sim2$ [pc] to divide the FeLoBALs into two groups: compact outflows with special spectral morphologies, and the distant galactic-scale outflows that could have potentially formed in-situ \citet{fg12b}.
In other words, these two groups that are largely separated in physical size may represent two intrinsically different types of BAL outflows that have different origins and physical processes including their acceleration.

The winds observed in these objects are not continuous but are clumpy \citep[e.g.,][]{hamann11}.
Without an external confinement mechanism, the clumps should dissipate in order of sound crossing time $t_{sc}=l/c_s$ where $l$ is the characteristic cloud size and $c_s$ is the sound speed \citep{hamann01,schaye01,finn14}.
For $l\sim \Delta R\sim0.01$ pc (median from the sample) and T$\sim10^4$ K BAL gas, the cloud will survive for $\sim670$ yr; however, it is more likely that the BAL wind is comprised of many smaller clouds with $l\sim \Delta R/N$ where N is the number of clouds \citep{hamann13}.
In comparison, a characteristic flow time $t_f\sim R/v_{outflow}$ for a BAL wind with $v_{outflow}\sim1,000\rm\ km\ s^{-1}$ and $R\sim1$ pc is $t_f\sim1,000$ yr.
Based on these calculations and considering that the dissipation time-scale cannot exceed the flow time-scale, we can assume the FeLoBAL winds have not traveled far from their origin.

Theoretical disk wind models \citep[e.g.,][]{foltz87,arav94} suggest the location of disk winds at $R\sim0.01$ pc for luminous quasars \citep[e.g.,][]{proga00,proga04}.
However, none of the FeLoBAL outflows in our sample was found at such a compact scale.
Rather, the compact outflows in our sample ($R\sim1-10$ pc) suggest a torus wind model where the winds originate from the dusty torus \citep[e.g.,][]{gallagher15,chan16,chan17,vollmer18}.
It is possible that the polar dust discovered in spatially resolved mid-infrared observation \citep[e.g.,][]{honig13} has the same origin as these winds.
Similar to the dusty torus winds, polar dust models also predict dust at comparable size-scales $R\sim1-100$ pc with large dust masses, $M_{dust}\sim100$s M$_\odot$ \citep{honig17,stalevski19}.

The torus wind models may also provide intriguing explanations for some of the inflows we observed in the sample (\S~\ref{subsubsec:felobal_classification}; \S~\ref{subsec:derived_pars}).
Most of the inflowing FeLoBAL gas is found at $R<100$ pc (one at $R\sim1000$ pc; Figure~\ref{fig:loiter_vel_logR}). 
It is plausible that these compact inflows may represent some kind of disruption in the vicinity of the torus where the resulting material falls towards the central engine.
The distant inflow might be cause by some compaction of gas flows in the host galaxy.

Between $\sim$ 100 pc and $\sim$ 1000 pc from the center, photoionized gas can be observed in emission lines from the narrow line region (NLR).
Observational evidence for quasar driven winds at this size-scale is often observed in blueshifted [\ion{O}{3}] lines \citep[e.g.,][]{zakamska16,vayner21}.
These outflows have mass outflow rates comparable to the FeLoBAL outflows ($1\lesssim\log\dot M_{out}\lesssim3\rm \ [M_\odot\ yr^{-1}]$).
Although a direct connection between the [\ion{O}{3}] outflows and BAL outflows is yet inconclusive (Paper III, \citealt{choi_prep_paper3}), it is possible that outflowing gas in the NLR can manifest as either BAL winds or emission line outflows or both depending on the physical conditions of the gas and/or sightlines.

Lastly, the potential origin of the kiloparsec scale BAL winds can be explained by a model introduced by \citet{fg12b}, where a low radial filling factor results from in situ formation of outflows from the interaction between a dense interstellar medium and a quasar blast wave.
They concluded that the FeLoBAL outflowing gas may have properties comparable to massive molecular outflows that are generally found at similar distances from the central SMBHs (e.g., high momentum flux ratio; \S~\ref{subsec:other_outflows}).

\subsection{Acceleration Mechanisms and SED Properties}\label{subsec:redden}
We used {\it Cloudy} to calculate the force multiplier (FM) for a radiatively driven outflow using the physical parameters constrained from the best-fitting models to investigate the relationship between the photoionization properties of the gas and the wind acceleration (i.e., outflow velocity).
The force multiplier is defined as the ratio of the total cross section (line and continuum processes) to the Thompson cross section.
It represents how much radiative force the cloud can harness from the photons to power the outflow acceleration.

\begin{figure*}
\includegraphics[width=.49\linewidth]{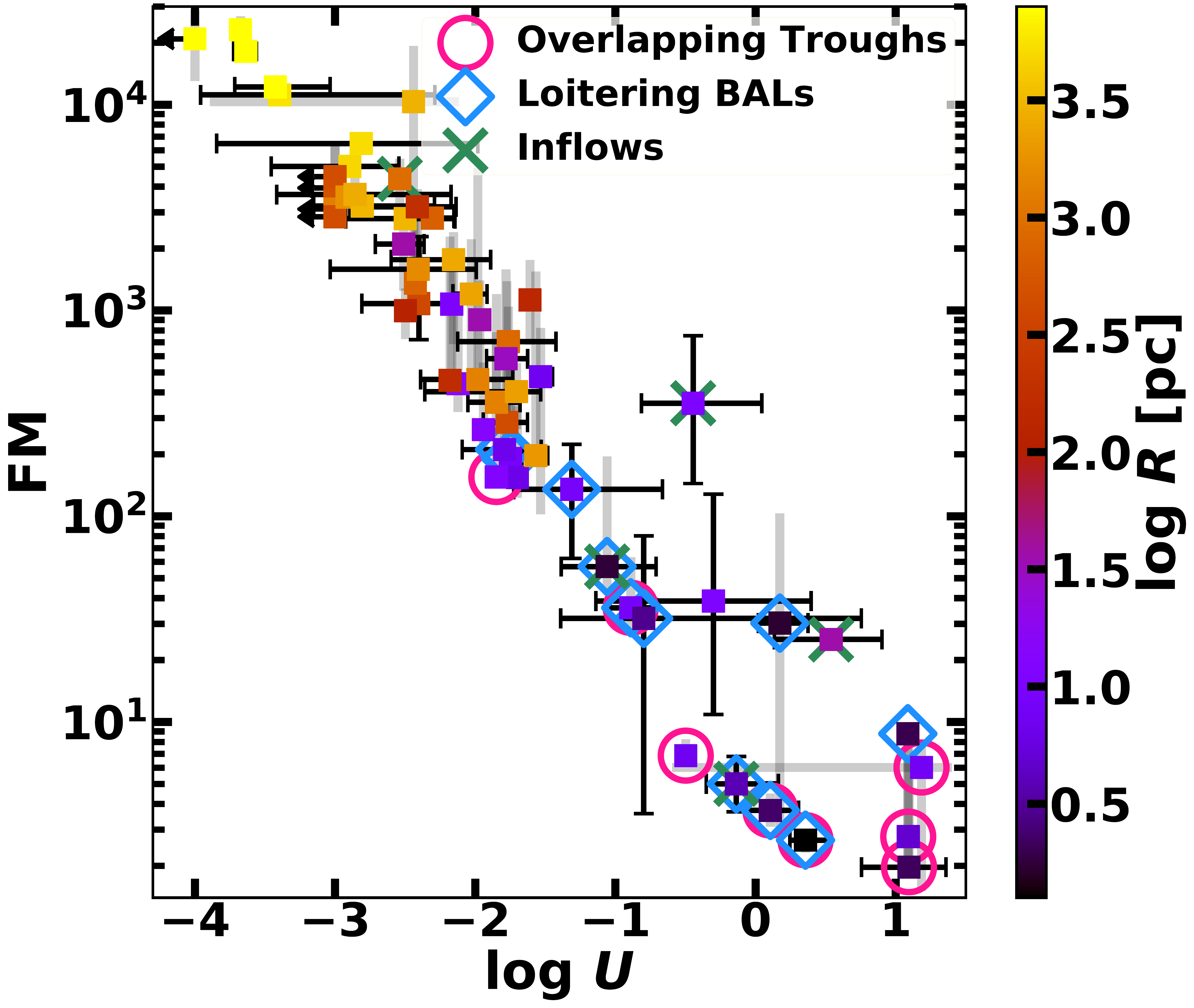}
\includegraphics[width=.5\linewidth]{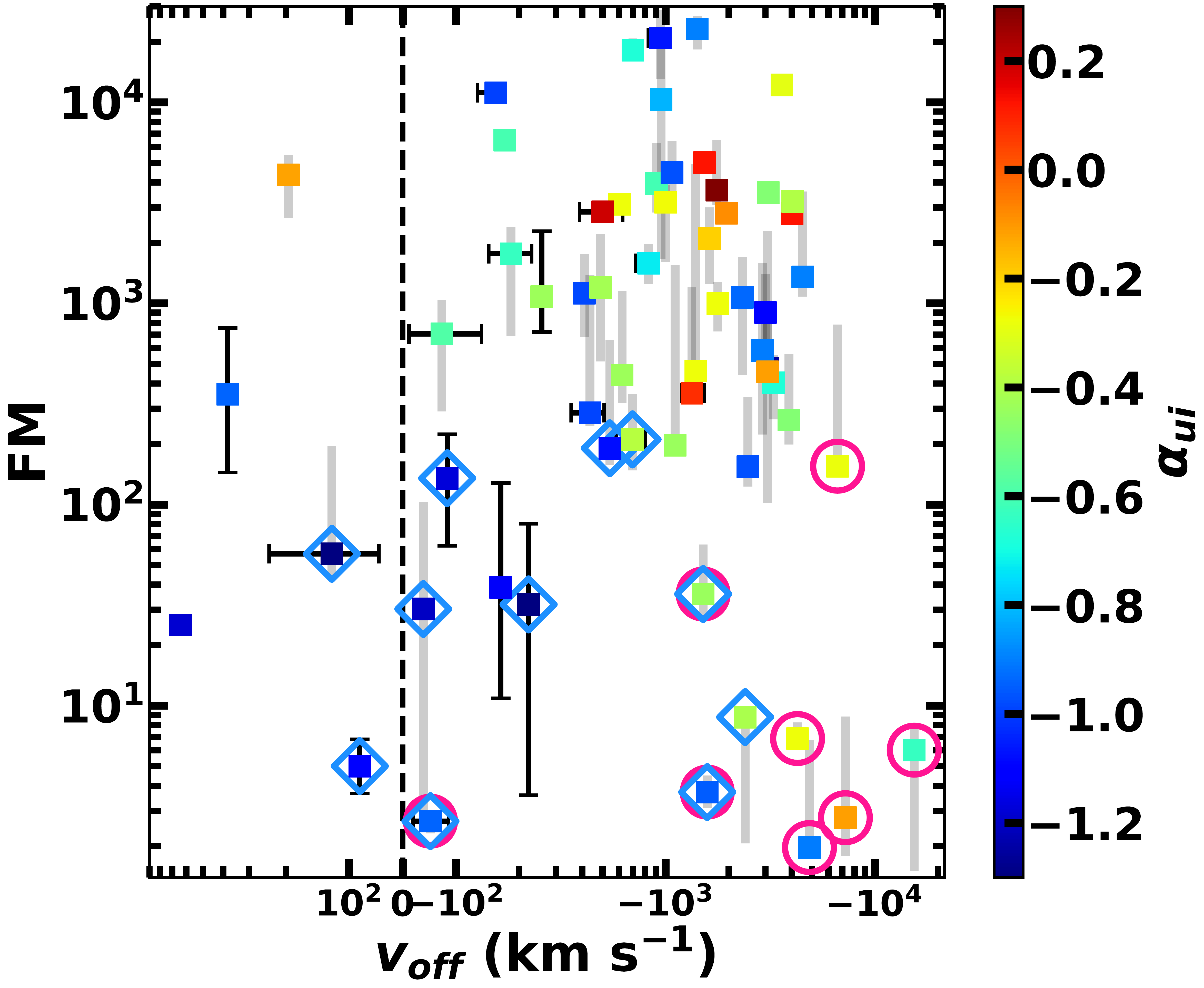}
\caption{The force multiplier (FM) calculated using {\it Cloudy} with physical parameters from the best-fitting {\it SimBAL} models.
{\it Left panel}: The force multiplier is strongly correlated with ionization parameter.
{\it Right panel}: We did not find robust correlation between the force multiplier and the outflow velocity.
The outflows with the highest outflow velocities have relatively smaller FM.
An extra source of opacity to capture the photon momentum, such as dust, or another acceleration mechanisms may be needed to explain the high velocity FeLoBALs.
The high values of $\alpha_{ui}$, flat or red SEDs, found in these objects could potentially indicate dusty outflows.
Markers and error bars as in Figure~\ref{fig:fitpar_ac_ad}.
\label{fig:fm}}
\end{figure*}

Figure~\ref{fig:fm} shows the distribution of FM as a function of $\log U$ and $v_{off}$.
FM decreases with $\log U$ which is consistent with analytical calculations that used the equation derived from the definition of FM \citep[e.g.,][]{castor75,arav94,arav94b}.
High $\log U$ gas is not only highly ionized but also has larger hydrogen column density ($\log N_H$) and more material since $\log N_H-\log U$ is nearly constant in our FeLoBAL sample (\S~\ref{subsubsec:best_fit_pars_photo}).
We found no correlation between FM and outflow velocity from our sample.
In fact, the outflows with the highest velocities have among the lowest values of FM.

The low FM values in BALs with extreme outflow velocities ($v_{off}\sim-10,000\rm\ km\ s^{-1}$) suggest that another acceleration mechanism is also playing a significant role.
For instance, radiation pressure on dust may play a significant role in accelerating the gas \citep[e.g.,][]{thompson15,murray05, ishibashi17}.
Some of these high outflow velocity BALs are overlapping trough BALs with anomalous reddening (\S~\ref{subsec:ot_obj}) and they have flatter SED slopes ($\alpha_{ui}$) than the rest (represented by the colors of the marker in the right panel of Figure~\ref{fig:fm}).
The flat SED slopes may indicate reddening in the quasar which may suggest the BALs in these objects are dusty.
That is not to say that the BALs themselves have high dust content; dust reddening tends to suppress photoionization.

\begin{figure}
\epsscale{.6}
\plotone{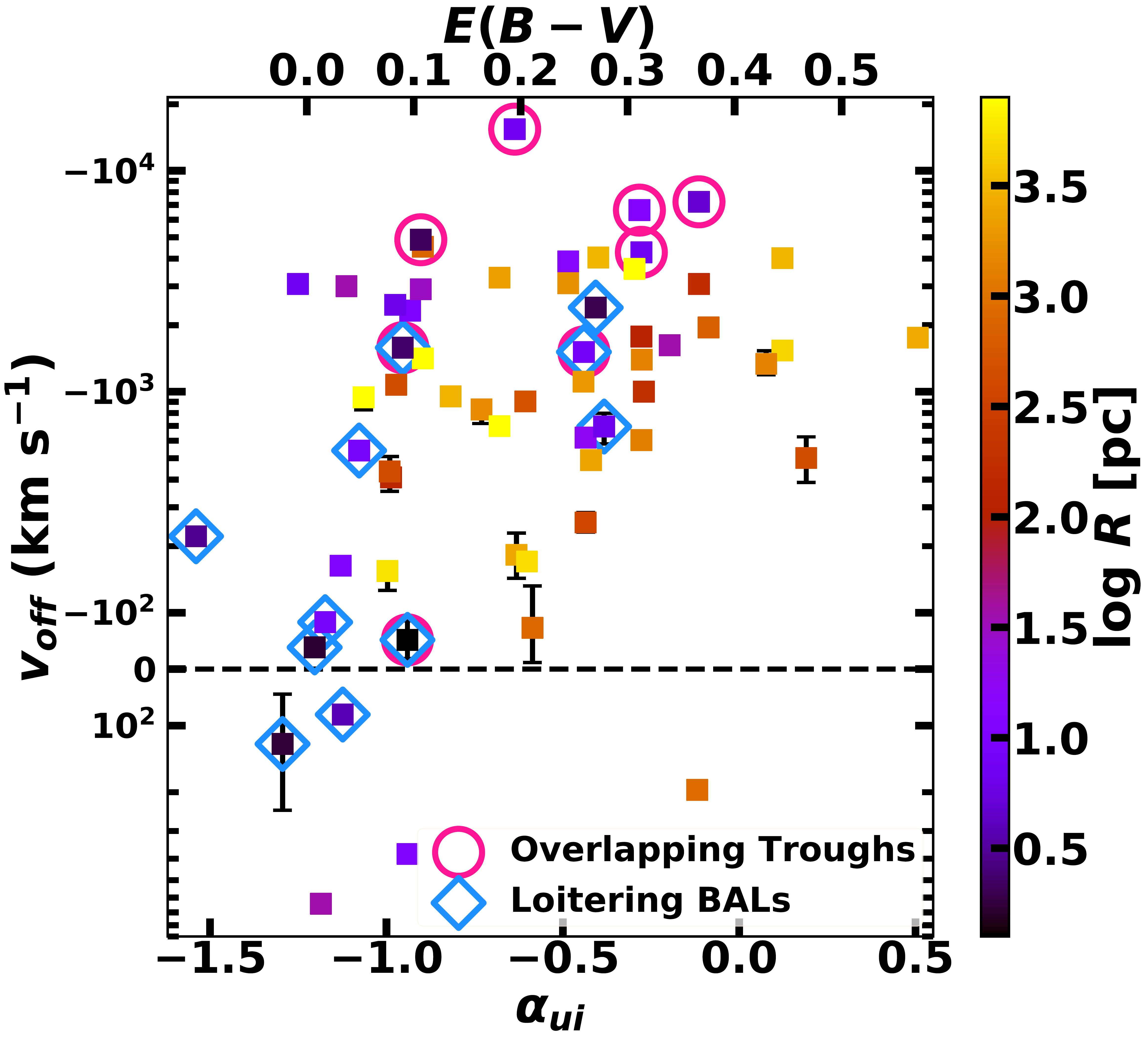}
\caption{The outflow velocity ($v_{off}$) is plotted against the SED slope parameter ($\alpha_{ui}$).
FeLoBALs with higher outflow velocities are found in objects with flatter or redder SEDs, where FeLoBALQs with steeper or bluer SEDs have compact outflows ($\log R\lesssim1$ [pc]).
Most of the compact outflows with steep SEDs are loitering outflows, and $\alpha_{ui}$ in these objects might be potentially affected by other SED properties than reddening such as the strength of IR emission from the torus \citep{leighly_prep_paper4}.
The top axis shows the values of $E(\bv)$ that correspond to the range of $\alpha_{ui}$ plotted on the bottom axis
(assuming a composite quasar SED \citep{richards06} and SMC reddening \citep{prevot84}).
The vertical error bars show 2$\sigma$ (95.45\%) uncertainties.
\label{fig:alpha_ui}}
\end{figure}

Figure~\ref{fig:alpha_ui} reveals that the outflow velocity is correlated with $\alpha_{ui}$ ($p=0.003$, Kendall $\tau$).
The top axis shows the inferred values of $E(\bv)$ calculated using the composite quasar SED from \citet{richards06} ($\alpha_{ui}=-1.23$) and SMC reddening \citep{prevot84}.
This strong correlation between the SED slope and outflow velocities in the FeLoBAL outflows is consistent with what has been found for extremely red quasars (ERQs).
\citet{hamann17} analyzed a unique sample of ERQs at $2.0<z<3.4$ and discovered a high BAL fraction ($\sim30-68$ \%) and frequently the presence of outflow features.
\citet{perrotta19} analyzed the [\ion{O}{3}] emission lines in a subsample of ERQs from \citet{hamann17}.
They found a correlation between $\rm i-W3$ color and outflow velocity where faster and more powerful outflows were found in redder quasars.
It is plausible that this trend can be explained by FeLoBALQs being at a similar evolutionary stage as dusty, red quasars in a transitional phase in quasar evolution where obscured quasars are expelling gas and dust via outflows to become normal quasars \citep[e.g.,][]{hopkins05,urrutia08,glikman17,glikman18}.
However, FeLoBALQs in our sample are not ERQs and the link between the red quasars and FeLoBALQs is still uncertain; LoBALQs are found among red quasars but no enhanced merger rates or star formation rate has been found for FeLoBALQs \citep[e.g.,][]{violino16,villforth19}.

A simpler explanation for the correlation between the two properties is that the outflow itself as the source of reddening.
First of all, a quasar outflow from the dusty torus can form a dusty wind \citep[e.g.,][]{gallagher15}.
The dust is then sublimated, and we see BAL.
Further downstream, the dust precipitates out of the gas \citep{elvis02} producing the reddening.
For instance, \citet{dunn15} suggested that the reddening in FeLoBALQs occurs at larger radial distances than the outflows.
A multiwavelength SED analysis of red quasars supports the idea that dust in the winds are responsible for the reddening in these objects \citep{calistrorivera21}.
Thus for our FeLoBALs, flatter or redder SEDs can be the result of high-velocity outflows that carry more mass and energy causing more dust reddening.
Similarly, the FeLoBALQs with the steepest, or the bluest, SEDs have the compact, low mass outflows with the lowest outflow velocities that are characteristics of loitering outflows (\S~\ref{subsec:loiter}).

In summary, FM analysis using {\it Cloudy} suggested that radiative line driving is insufficient for outflows with the highest outflow velocities located close to the central engine.
We found a compelling evidence that FeLoBALQs with redder SEDs have faster outflows.
Based on these results, we speculate that the additional acceleration mechanism is acceleration by dust for these FeLoBALs.
We note that the SED slope ($\alpha_{ui}$) calculated from specific fluxes at rest-frame 2000 \AA\/ and 3 $\rm\mu m$ used in this work can be affected not only by reddening but
by other properties of the quasar such as the strength of torus.
This point is investigated further in Paper IV \citep{leighly_prep_paper4}.

\subsection{Comparison with other forms of outflows}\label{subsec:other_outflows}
Observational evidence for quasar outflows can be found both in blueshifted absorption lines and blueshifted or broad emission lines \citep[e.g.,][]{fabian12,kp15}.
All together, AGN-driven winds are found at a wide range of distance scales, from the ultra fast outflows (UFOs) seen in X-ray band that are located in subparsec scales \citep[e.g.,][]{tombesi10} to molecular winds at kiloparsec scales \citep[e.g.,][]{cicone14}.
BAL outflows \citep[e.g.,][]{arav18,leighly18a,choi20} and the outflows seen as blueshifted [\ion{O}{3}]$\lambda\lambda 4959,5007$ emission lines \citep[e.g.,][]{harrison14,zakamska16} are often found at parsec to kiloparsec scales, i.e., between the size scales of the X-ray and molecular outflows.
It is conceivable that the different forms of outflows we observe are related and originate in the same AGN-driven outflow phenomenon.

\begin{figure*}
\includegraphics[width=.495\linewidth]{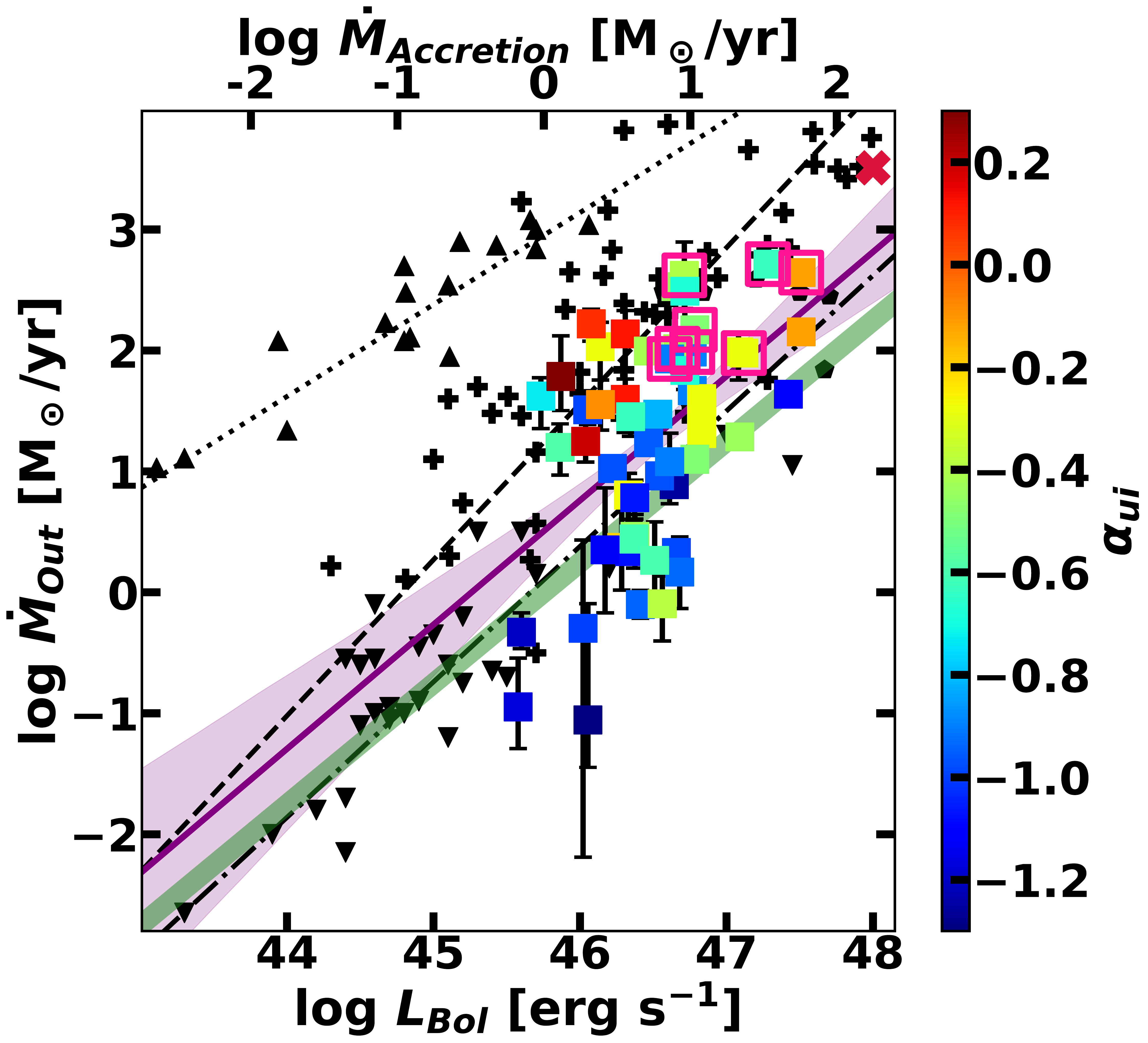}
\includegraphics[width=.495\linewidth]{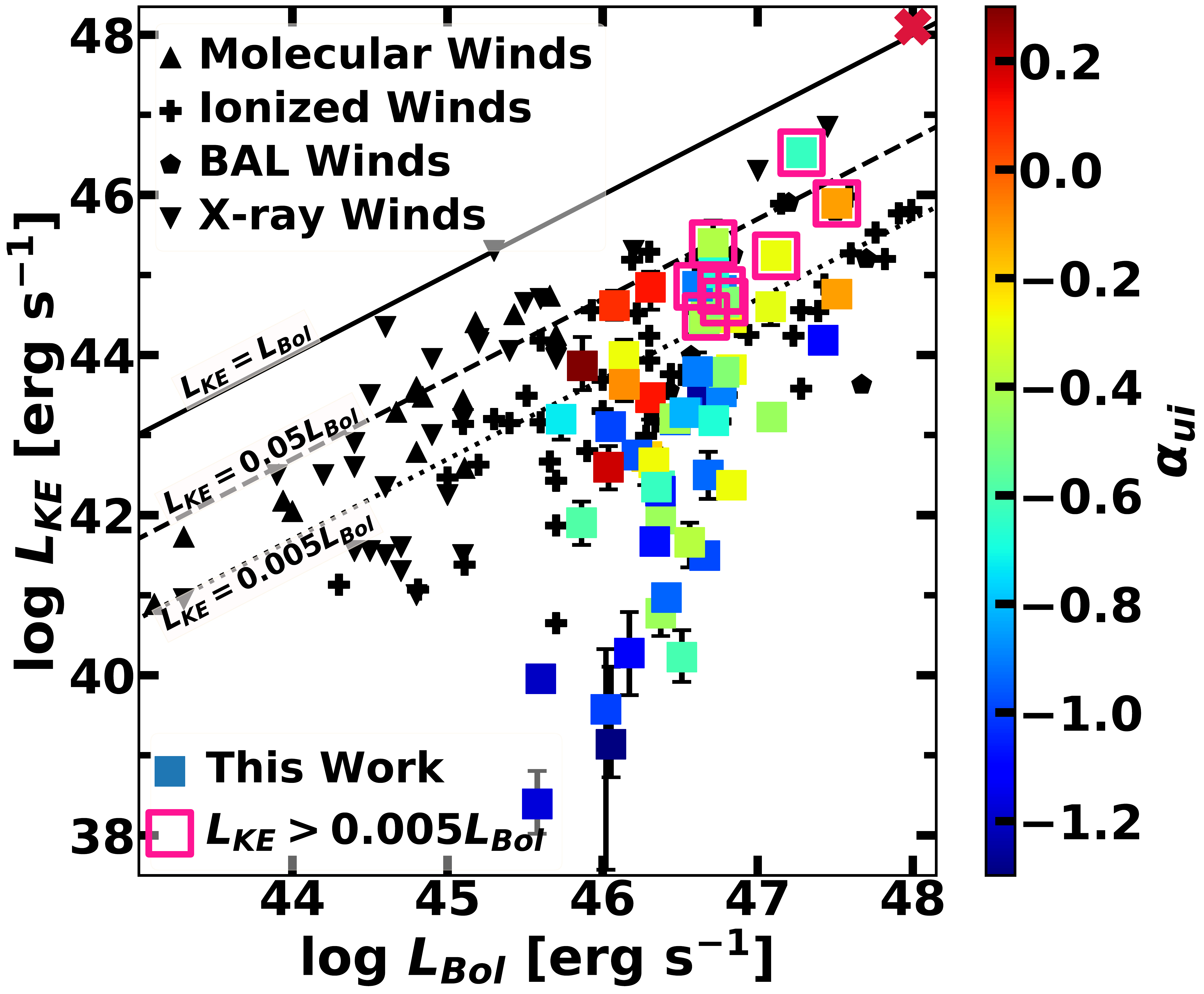}
\caption{Our sample of FeLoBAL outflows and the compilation of outflows from \citet{fiore17}.
{\it Left panel}: the regression slopes shown as the dotted, dashed, and dotted-dashed lines for the molecular outflows, ionized winds, and X-ray outflows, respectively, taken from \citet{fiore17}.
The green line has a slope of one that represents $\dot M_{Out}=\dot M_{Accretion}$ (assuming the energy conversion efficiency, $\eta=0.1$).
The purple line (shade) showing the regression for our sample of FeLoBALs has the slope of $1.02\pm0.25$ which is consistent with both ionized winds ($1.29\pm0.38$; dashed line) and X-ray outflows ($1.12\pm0.16$; dotted-dashed line).
This result is also consistent with $\dot M_{Out}/\dot M_{Accretion}\sim3$, although we observe a wide range of this ratio from $\sim$0.04 to $\sim$80 among the objects in our sample.
{\it Right panel}: the solid, dashed, and dotted lines show $L_{KE}=1.0,\ 0.05,\ 0.005\ L_{Bol}$, respectively.
Pink square outlines denote the powerful outflows ($L_{KE}/L_{Bol}>0.005$) in our sample that have  well-constrained physical parameters and outflow properties.
The red cross represents the outflow in SDSS~J1352$+$4239 \citep{choi20}.
The vertical error bars show 2$\sigma$ (95.45\%) uncertainties.
\label{fig:logLbol_Mdot_Lke}}
\end{figure*}

\citet{fiore17} compiled an extensive list of quasar outflows and their properties with more than one hundred wind measurements from the literature, limited to those that have robust estimates of the physical sizes of the outflows.
Only 7 BAL outflows were included in their sample, possibly because BAL outflows were not the main focus of their investigation.
Figure~\ref{fig:logLbol_Mdot_Lke} shows our FeLoBAL outflows ($v_{off}<0\rm\ km\ s^{-1}$) combined with the \citet{fiore17} sample, as well as the FeLoBAL outflow in SDSS~J135246.37$+$423923.5 \citep{choi20}.
The color bar illustrates our observation of a strong relationship between the outflow properties and the observed shape of the SED.
Outflows found in objects with flatter SED slope tend to be more massive and powerful, which is expected given there is a strong relationship between the outflow velocity and the slope of the SED (\S~\ref{subsec:corr}; \S~\ref{subsec:redden}).

Similar to the other forms of quasar outflows, $\dot M$ and $L_{KE}$ both increase with $L_{Bol}$ in our FeLoBAL sample.
We performed a Bayesian linear regression using {\tt linmix}\footnote{https://github.com/jmeyers314/linmix/}, a python implementation of \citet{kelly07}, to determine how the correlation we found among the FeLoBALs in our sample compares with the other outflow channels.
We took into account the uncertainties associated with $\dot M_{Out}$ for the regression analysis.
The log linear slope for our FeLoBALs is $1.02\pm0.25$ which is similar to the values \citet{fiore17} found for other types of quasar outflows, except for the molecular outflows which were found to have a flatter slope of $0.76\pm0.06$ and higher mass outflow rates.
Our slope is steeper than the values expected from the theoretical models \citep[e.g.,][$\dot M_{out}\propto L_{Bol}^{1/3}$]{kp15}; however, as \citet{fiore17} pointed out in their discussion, the discrepancy might be explained by the presence of multiphase winds (i.e., underestimated $\dot M_{out}$).
Our regression slope for FeLoBALs corresponds to the ratio between the mass outflow rate and the mass accretion rate of 3.
However, there is a large range in this ratio among the objects in our sample, from $\sim$0.04 to $\sim$80, more than 2 orders of magnitude.
This result strongly suggests that a simple prescription of a fixed ratio between the mass outflow rate and mass accretion (or inflow) rate used in the subgrid physics in cosmological simulations \citep[e.g.,][]{choi12} may not be adequate to reproduce a realistic quasar outflows and mechanical quasar feedback from the BAL winds.
The sample of FeLoBALQs we analyzed contains a number of BALs with low outflow velocities ($\vert v_{off}\vert<1000\rm km\ s^{-1}$, Figure~\ref{fig:fitpar_vel_a_width}), which are seen in the right panel of Figure~\ref{fig:logLbol_Mdot_Lke} in the comparatively lower values of $L_{KE}$.
Objects with weak or less massive outflows possibly have not been included in the \citet{fiore17} compilation or have not been analyzed in detail due to publication bias.
The objects in our sample, on the other hand, were chosen to either have low redshift \citep{leighly_prep} or from the objects analyzed by \citet{farrah12}, and are included in the sample regardless of the strength of the outflow.

\begin{figure*}
\includegraphics[width=.495\linewidth]{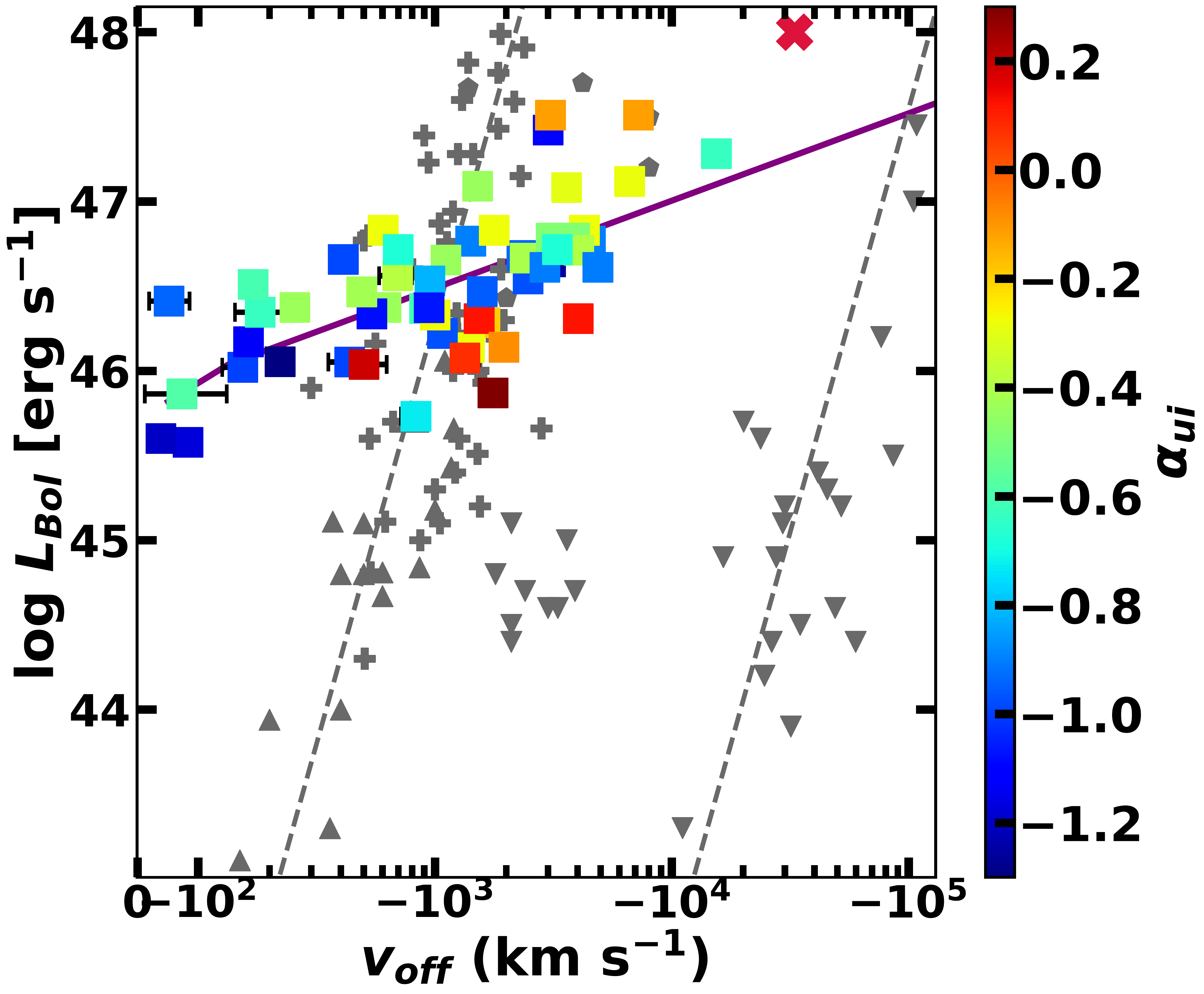}
\includegraphics[width=.495\linewidth]{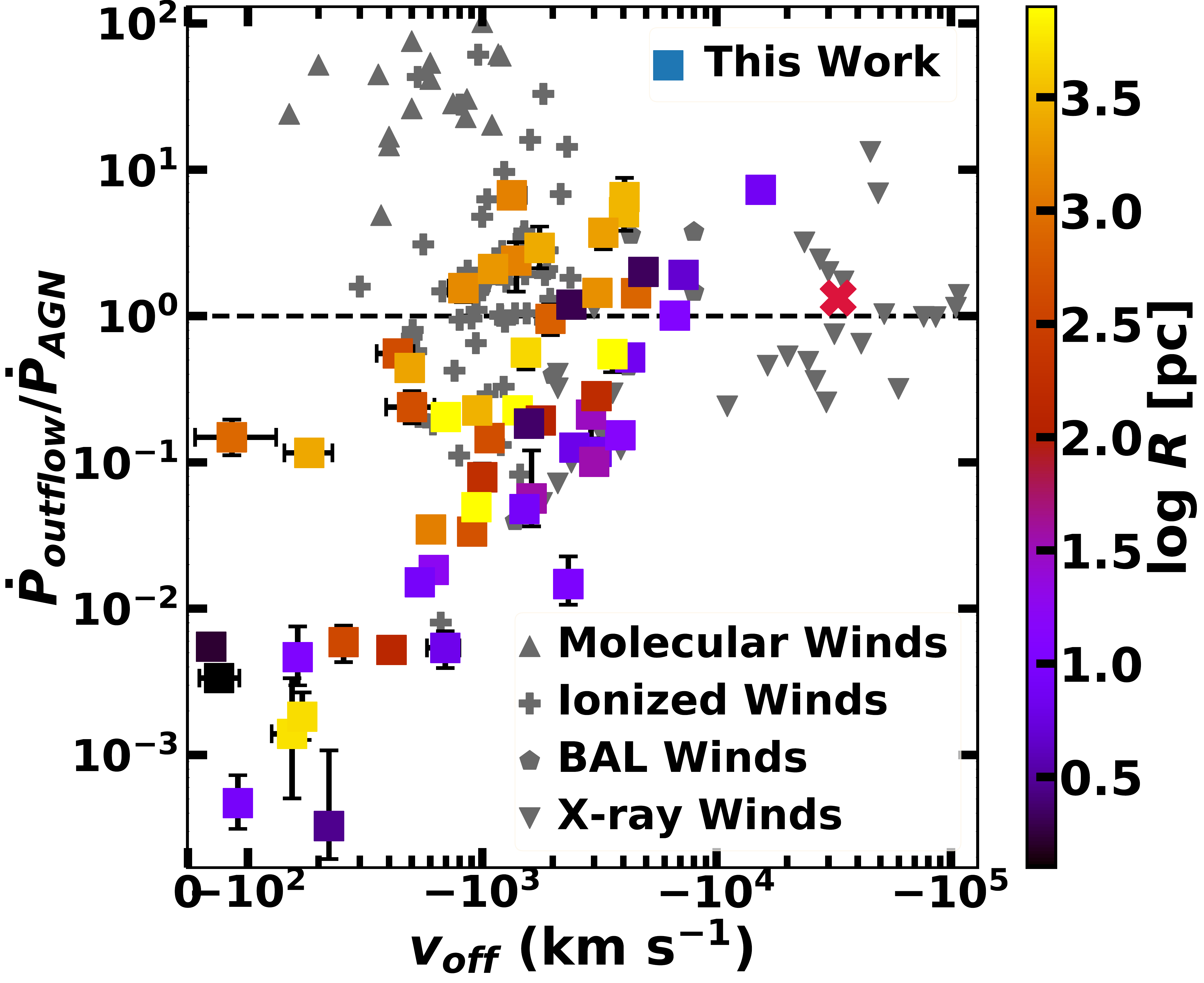}
\caption{{\it Left panel}: the bolometric luminosity as a function of outflow velocity.
The grey points represent the compilation of outflows from \citet{fiore17} and the red cross shows the FeLoBAL outflow in SDSS~J1352$+$4239 \citep{choi20}.
The purple line represents the regression slope, $L_{Bol}\propto v_{off}^{0.5}$, for our sample of FeLoBAL outflows which differs dramatically from the $L_{Bol}\propto v_{off}^5$ scaling (dashed grey lines) described in \citet{fiore17}.
{\it Right panel}: the momentum flux ratio ($\dot P_{outflow}/\dot P_{AGN}$) as a function of outflow velocity.
The dashed horizontal line marks the expected ratio for a momentum conserving outflow.
The error bars show 95\% uncertainties.
\label{fig:vel_Lbol_Prat}}
\end{figure*}

We found a positive correlation between $L_{Bol}$ and $v_{off}$ in our sample of FeLoBAL outflows (left panel in Figure~\ref{fig:vel_Lbol_Prat}) that is consistent with trends observed in quasar outflows in general \citep[e.g.,][]{laor02,ganguly07,spoon13,veilleux13b,fiore17}.
Also, BALs with higher outflow velocities are found in objects with flatter or redder SED for a given bolometric luminosity (\S~\ref{subsec:redden}).
This further propagates to the mass outflow rates and kinetic luminosity of the outflows (Figure~\ref{fig:logLbol_Mdot_Lke}) where for a given bolometric luminosity, objects that are redder have more massive and powerful outflows.

In a simple model for radiatively accelerated outflows where the wind is driven by the scattering of photons, and that process provides the momentum to accelerate the gas (i.e., a momentum-driven/conserving outflow), the maximum value for the momentum flux ratio ($\dot P_{outflow}/\dot P_{AGN}$; $\dot P_{outflow}=\dot M_{out}v_{outflow},\ \dot P_{AGN}=L_{Bol}/c$) is about 1 \citep[e.g.,][]{king03a,king03b}.
The right panel in Figure~\ref{fig:vel_Lbol_Prat} shows that the FeLoBALs from our sample mostly show $\dot P_{outflow}/\dot P_{AGN}\lesssim1$ and only a small fraction of outflows with high velocities have $1\lesssim\dot P_{outflow}/\dot P_{AGN}\lesssim10$.
The distribution in momentum flux ratios can be ascribed to the relationship that outflows at larger distances from the center ($\log R$) have larger momentum flux ratios due to the winds being more massive ($\dot M\propto R N_H v$).
However, it is also possible that the bolometric luminosity may have changed since the winds were launched thus potentially creating a large range of the momentum flux ratios depending on how the luminosities evolved in these objects \citep[e.g.,][]{ishibashi18,zubovas18}.
For instance, \citet{king11} showed that an outflow may persist for an order of magnitude longer than the duration of the AGN event that powered it.

The outflows that have $\dot P_{outflow}/\dot P_{AGN}\gg1$ may require different acceleration mechanisms to explain the large momentum load.
Dust in the outflows can increase the opacity of the gas and harness the momentum of the photons more effectively and this mechanism could potentially produce a momentum flux ratio above unity \citep[e.g.,][]{fabian08,fabian18}.
For the compact outflows with the highest outflow velocities, it seems likely the dust opacity has contributed to the momentum flux (\S~\ref{subsec:redden}).
Molecular winds located at large distances ($\log R\gtrsim2$ [pc]) also have high momentum flux ratios ($\dot P_{outflow}/\dot P_{AGN}>10$); these outflows are thought to be accelerated by an energy conserving outflow mechanism \citep[e.g.,][]{kp15}.
A large fraction of the FeLoBAL outflows in our sample with high $\dot P_{outflow}/\dot P_{AGN}$ are found at large distances ($\log R\sim3$ [pc]) which supports the energy conserving outflow scenario.
Conversely, not all distant FeLoBAL outflows have high momentum flux ratios and the in-situ wind formation model postulated by \citet{fg12b} may explain the properties of these outflows more suitably.

\subsection{Implications for AGN feedback}\label{subsec:feedback}
In order for cosmological models and theoretical calculations to successfully explain the co-evolution of the galaxies and the central black holes and to reproduce AGN feedback, the energy input from the AGNs to the host galaxies needs to be at least $0.5\sim5\%$ of the bolometric luminosity of the quasar \citep[e.g.,][]{scannapieco04,dimatteo05,hopkins10}.
From the 55 FeLoBAL outflows ($v_{off}<0\rm\ km\ s^{-1}$), we found that 8 BALs that have kinetic luminosities greater than the 0.5\% of the quasar bolometric luminosity.
Out of 50 FeLoBALQs from the sample, 9 objects with BAL signatures were identified with powerful BAL outflows (note, two high-velocity outflows in SDSS~J1448$+$4043 combined have $L_{KE}>0.005L_{Bol}$).
Five objects out of the 9 with energetic outflows showed overlapping trough features in the spectra (\S~\ref{subsec:ot_obj}).

We identify a couple of reasons why this number may be underestimating the feedback potential of FeLoBAL outflows.
First, we may not be finding many energetic outflows because our sample is dominated by objects with relatively low bolometric luminosities (median $\log L_{bol}\sim46.4\ \rm[ erg\ s^{-1}]$).
Energetic outflows are found in luminous quasars and the outflow strength is correlated with the bolometric luminosity (Figure~\ref{fig:logLbol_Mdot_Lke} right panel).
This is further highlighted by the fact that the 9 objects with energetic outflows are among the most luminous quasars in our sample ($\log L_{bol}>46.6\ \rm[erg\ s^{-1}]$).
For example, if we limit our sample to include only these high luminosity objects, then we find that 50\% of our FeLoBALQs have energetic outflows sufficient to power quasar feedback ($46.6<\log L_{Bol}<47.6\ \rm[erg\ s^{-1}]$; 18 objects).
This result is similar to what \citet{miller20} found in their sample of BALQs that have similar bolometric luminosity range as this high luminosity subset.
The flux-limited nature of survey such as the SDSS means that more luminous objects are found at higher redshifts, and we expect such objects to have more powerful outflows.

Another reason that the outflow energy may be underestimated for FeLoBALs analyzed in this work is that they were calculated from the BAL physical properties estimated from the low-ionization lines, mainly using \ion{Fe}{2} and \ion{Mg}{2} absorption lines.
(Fe)LoBALs also show absorption lines from the high-ionization species (e.g., \ion{C}{4}, \ion{Si}{4}) in the spectra and they often show larger widths with higher outflow velocities \citep[e.g.,][]{voit93, hamann19}.
Thus we are not including the $L_{KE}$ contributed by the higher velocity portion of the high-ionization lines.
In other words, the kinetic luminosity estimates based only on low-ionization lines may be taken as lower limits.
Once taking these two effects into consideration, the distributions of FeLoBAL outflows in our sample described in \S~\ref{subsec:other_outflows} may shift to higher velocities and may not appear relatively weaker or less massive compared to other forms of outflows included in the \citet{fiore17} compilation.

\section{Summary and Conclusions}\label{sec:summary}
\begin{figure*}
\epsscale{1.1}
\plotone{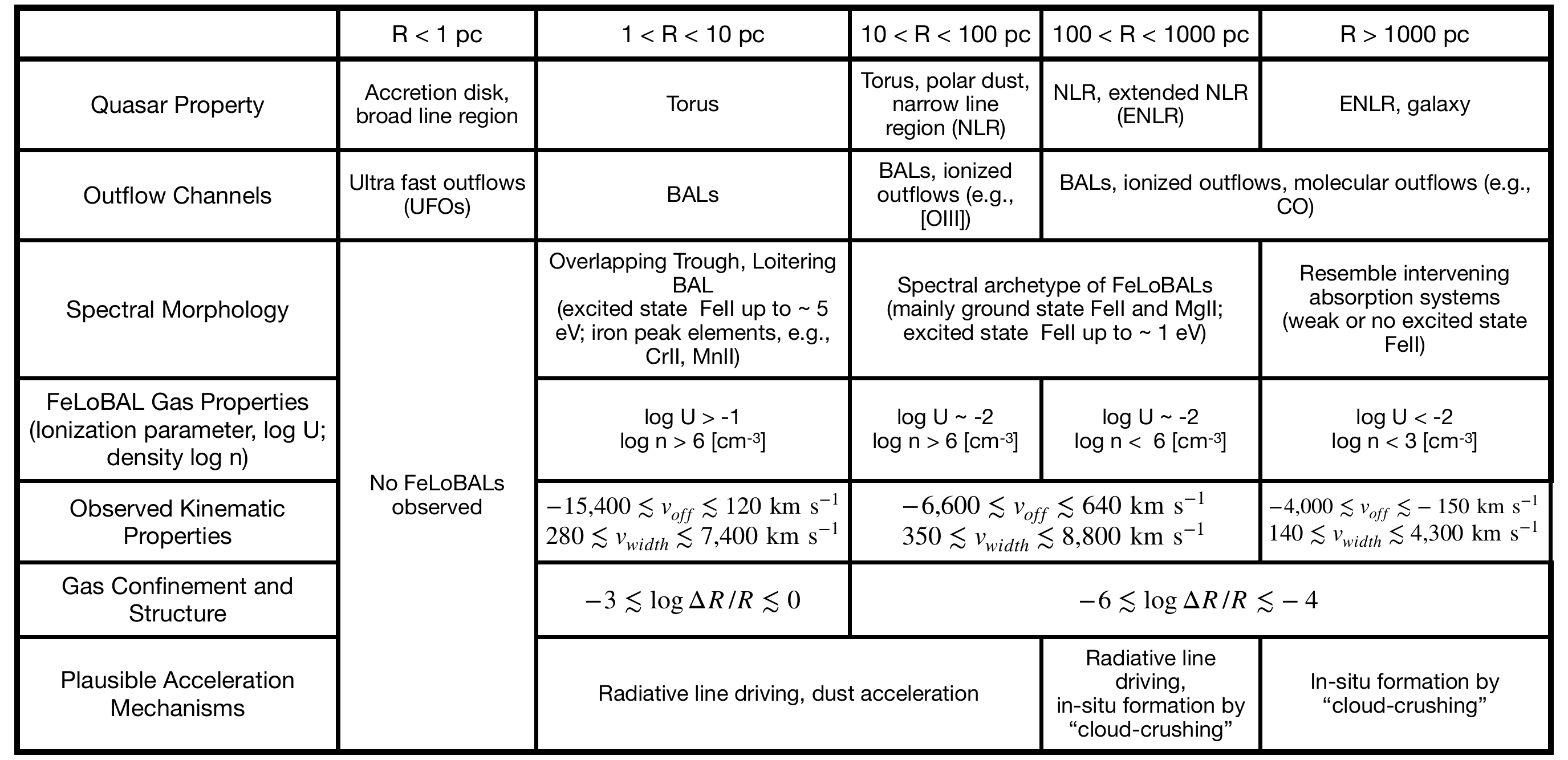}
\caption{The summary of FeLoBAL properties found in our sample.
\label{fig:summary_tbl}}
\end{figure*}
In this work, we presented the results and analysis from the first systematic study of a large sample of low redshift FeLoBALQs.
This work increases the number of well studied FeLoBALQs by a factor of five.
We were able to constrain the physical properties of the FeLoBAL outflows from the best-fitting {\it SimBAL} models and quantify the outflow properties.
The summary of FeLoBAL properties is shown in Figure~\ref{fig:summary_tbl}.
Our principal results are the following:
\begin{itemize}
\item We performed the first systematic study of a sample of 50 low redshift
($0.66 < z < 1.63$) FeLoBALQs using {\it SimBAL}.
From the best-fitting {\it SimBAL} models, we were able to identify 60 FeLoBAL components and constrain their physical properties as well as calculate their outflow properties (Figure~\ref{fig:fitfig1}; \S~\ref{subsec:best_fit_pars}).
\item We found a wide range of ionization parameters ($\log U\sim-4$ to 1.2) and densities ($\log n\sim2.8-8.0\ \rm[cm^{-3}]$) from our FeLoBALs, each spanning more than five orders of magnitude (Figure~\ref{fig:fitpar_ac_ad}).
The forward modeling technique used in {\it SimBAL} enabled us to analyze high-$\log U$ FeLoBALs from heavily absorbed FeLoBALQ spectra.
\item The outflow properties calculated using the physical properties extracted from best-fitting {\it SimBAL} models revealed a wide range of outflow locations ($\log R\sim0.0-4.4$ [pc]; Figure~\ref{fig:fitpar_aR}).
We found a significant correlation between outflow strength ($L_{KE}$) and outflow velocity ($v_{outflow}$) from our sample (Figure~\ref{fig:vel_ke}) and confirmed that outflow velocity is the principal factor in determining the outflow strength.
\item From the best-fitting {\it SimBAL} models, we identified multiple outflow components in $\sim18\%$ of the FeLoBALQs in our sample.
The higher-velocity components had higher $\log U$ and some of them played a role in creating the overlapping trough BALs (\S~\ref{subsec:multiple_BAL};~\ref{subsec:ot_obj}).
The line profiles extracted from the {\it SimBAL} models showed discrete outflow gas structures and demonstrated how the BALs from the rare line transitions (e.g., \ion{He}{1}*, Balmer series) that are found in the rest-optical band alone can be used to estimate the distances and physical properties of the FeLoBAL outflows (Figure~\ref{fig:ii0_three_lines}).
\item Eight FeLoBALQs in our sample showed overlapping trough features in the spectra (Figure~\ref{fig:ot_objs}).
All of the overlapping trough BALs were found close to the central engine  $\log R\lesssim1$~[pc].
Their kinematic properties showed a wide range ($v_{off}\sim-15,400$ to $-50\rm\ km\ s^{-1}$, $v_{width}\sim900-7,400\rm\ km\ s^{-1}$), a fact that suggests that a large width is not required to create an overlapping trough BAL.
The five objects ($\sim63\%$) that showed typical overlapping trough features (e.g., high-velocity troughs reaching near-zero flux at the bottom) and anomalous reddening in the spectra have powerful outflows with $L_{KE}$ exceeding 0.5\% of $L_{Bol}$.
\item We identified a new class of FeLoBALs dubbed loitering outflows (\S~\ref{subsec:loiter}).
They are characterized by compact outflows ($\log R<1$ [pc]) and low outflow velocities ($\vert v_{off}\vert\lesssim2,000\rm\ km\ s^{-1}$).
Loitering outflows have high $\log U$ and high $\log n$ gas with large opacity; however, $\sim50\%$ of them showed no absorption in the emission lines and $\sim27\%$ of loitering outflow objects required an additional step-function partial covering in the model because only a fraction of continuum emission was absorbed by the BAL (\S~\ref{subsubsec:special_covering_model}).
The FeLoBALQs with loitering outflows can be identified by predicted Balmer absorption lines and some of them show overlapping troughs as well.
Their outflow property distributions (Figure~\ref{fig:loiter_vel_logR}) suggest that these objects may represent a distinct sub-population within FeLoBALQs.
\item We found that the compact outflows are located within the vicinity of a dusty torus where a dusty wind scenario can be used to explain the origin and the acceleration mechanism of these FeLoBALs (\S~\ref{subsec:redden}).
A force multiplier analysis showed that radiative line driving alone may not be sufficient to accelerate the compact outflows at $R\lesssim100$ pc to extreme velocities ($v_{off}\sim-10,000\rm\ km\ s^{-1}$).
An additional mode of acceleration (e.g., dust opacity; Figure~\ref{fig:fm}) may be needed to explain these high velocity outflows.
In-situ formation of FeLoBALs from the ISM \citep{fg12b} is a plausible model to describe the outflow properties of the kiloparsec-scale winds.
\item We found a significant correlation between SED slope ($\alpha_{ui}$) and outflow velocity (Figure~\ref{fig:alpha_ui}).
The objects that have flatter SED slopes have faster and more powerful FeLoBAL outflows.
A flatter SED slope may indicate strong dust reddening; noting that the SED slope also depends on intrinsic AGN properties such as slope of the rest-optical/UV power-law and the strength of torus emission.
\item We found that more luminous quasars have more powerful outflows (Figure~\ref{fig:corr_matrix};~\ref{fig:logLbol_Mdot_Lke}), consistent with trends observed in other forms of quasar outflows.
The FeLoBAL outflows in our sample showed a wide range of the ratio between the mass outflow rate and the mass accretion rate, from $\sim$0.04 to $\sim$80, more than 2 orders of magnitude.
This wide distribution suggests that a simple fixed ratio prescription used in some cosmological simulations may be insufficient for accurate depiction of feedback by BAL outflows 
\item Only nine objects out of 50 FeLoBALQs have sufficiently powerful outflows to produce quasar feedback.
We suspect that a low fraction of powerful outflows is a consequence of the low redshift and therefore lower luminosity of our sample (median $\log L_{bol}\sim46.4\ \rm[ erg\ s^{-1}]$).
In addition, our analysis relied exclusively on the information extracted from the low-ionization lines, and it is known that the high-ionization lines from the same outflow tend to show extended structures to higher velocities.
Therefore, our kinetic luminosity estimates may be considered to be lower limits.
\end{itemize}

We have expanded the number of FeLoBALQs that are analyzed in detail by a factor of five.
More importantly, the detailed analysis made possible by {\it SimBAL} has fleshed out our picture of quasar outflows, allowing us to study trends as a function of location and velocity.
But there are still many questions left unanswered that we plan to investigate in the future.
Do the outflow properties show correlations with emission line properties in these FeLoBALQs (Paper III, \citealt{choi_prep_paper3})?
Do outflows in high redshift FeLoBALQs have similar physical properties as the low redshift objects presented in this paper?
We expect that FeLoBAL quasars with higher redshifts and therefore higher luminosities to have more powerful outflows than the low redshift objects.
However, it is uncertain whether the outflows in high redshift FeLoBALQs have comparable physical properties such as ionization parameter and density.
Preliminary results \citep{voelker21} found a higher fraction of objects with powerful outflows in a sample of higher redshift objects.
They also found intriguing evidence suggesting that the distribution of physical properties of the FeLoBALs may change as a function of luminosity.
Furthermore, we are developing a convolutional neural net to identify FeLoBALQs from the SDSS archive with which we will be able to build larger samples of FeLoBALQs to perform follow up studies \citep[FeLoNET;][]{dabbieri20, dabbieri_prep}.
Future studies using spectroscopic time series may provide additional information about the origin and acceleration mechanism of FeLoBAL winds.
The outflow properties of our newly discovered class of FeLoBALs, loitering BALs (e.g.,small $R$), potentially suggest that they might experience higher probability of spectral variability \citep[e.g.,][]{zhang15b}.
New data may allow us to investigate the variability in these objects that could further constrain the geometry of the BAL gas around the accretion disk and the BLR to help us understand the partial-covering we see in these BALQs.

\begin{acknowledgments}
The author thanks the current and past {\it SimBAL} group members and Jens-Kristian Krogager for useful discussions and comments on drafts.
The work is funded by NSF grant AST-1518382 and AST-2006771 to the University of Oklahoma.
This research has made use of the data from the SDSS/BOSS archive.
Some of the computing for this project was performed at the OU Supercomputing Center for Education \& Research (OSCER) at the University of Oklahoma.

Long before the University of Oklahoma was established, the land on which the University now resides was the traditional home of the ``Hasinais'' Caddo Nation and ``Kirikiris'' Wichita \& Affiliated Tribes.
This land was also once part of the Muscogee Creek and Seminole nations.
We acknowledge this territory once also served as a hunting ground, trade exchange point, and migration route for the Apache, Comanche, Kiowa and Osage nations.
Today, 39 federally-recognized Tribal nations dwell in what is now the State of Oklahoma as a result of settler colonial policies designed to assimilate Indigenous peoples.
The University of Oklahoma recognizes the historical connection our university has with its Indigenous community. We acknowledge, honor and respect the diverse Indigenous peoples connected to this land. We fully recognize, support and advocate for the sovereign rights of all of Oklahoma's 39 tribal nations.
This acknowledgement is aligned with our university's core value of creating a diverse and inclusive community. It is our institutional responsibility to recognize and acknowledge the people, culture and history that make up our entire university community.

Funding for the Sloan Digital Sky Survey IV has been provided by the Alfred P. Sloan Foundation, the U.S. Department of Energy Office of Science, and the Participating Institutions.
SDSS-IV acknowledges support and resources from the Center for High Performance Computing  at the University of Utah. The SDSS website is www.sdss.org.
SDSS-IV is managed by the Astrophysical Research Consortium for the Participating Institutions of the SDSS Collaboration including the Brazilian Participation Group, the Carnegie Institution for Science, Carnegie Mellon University, Center for Astrophysics | Harvard \& Smithsonian, the Chilean Participation Group, the French Participation Group, Instituto de Astrof\'isica de Canarias, The Johns Hopkins 
University, Kavli Institute for the Physics and Mathematics of the Universe (IPMU) / University of 
Tokyo, the Korean Participation Group, Lawrence Berkeley National Laboratory, Leibniz Institut f\"ur Astrophysik Potsdam (AIP),  Max-Planck-Institut f\"ur Astronomie (MPIA Heidelberg), 
Max-Planck-Institut f\"ur Astrophysik (MPA Garching), Max-Planck-Institut f\"ur Extraterrestrische Physik (MPE), National Astronomical Observatories of China, New Mexico State University, New York University, University of Notre Dame, Observat\'ario Nacional / MCTI, The Ohio State University, Pennsylvania State University, Shanghai 
Astronomical Observatory, United Kingdom Participation Group, Universidad Nacional Aut\'onoma de M\'exico, University of Arizona, University of Colorado Boulder, University of Oxford, University of Portsmouth, University of Utah, University of Virginia, University of Washington, University of Wisconsin, Vanderbilt University, and Yale University.
\end{acknowledgments}
\vspace{5mm}

\software{emcee \citep{emcee}, Sherpa \citep{freeman01}, SimBAL \citep{leighly18a}, 
          Cloudy \citep{ferland17}}
          
\appendix

\section{Modeling the Continuum}\label{app:model_cont}

The absorption line optical depths depend on the level of the
continuum.  When there is significant blending, it is important to
have a reasonably robust method for modeling the continuum.  In the
first paper reporting {\it SimBAL} results, \citet{leighly18a}, we
modeled the continuum, 
and then divided the spectrum by the model to isolate the absorption
lines.  In \citet{leighly18b}, we kept the shape of the continuum
fixed, but allowed the normalization to vary.  In \citet{choi20}, we
developed  emission-line models from an {\it HST} observation of
Mrk~493 and modeled the continuum with those and a power law.  

All of these methods were stop-gap approximations that served while we
developed a robust and reliable method that could routinely be used in
{\it SimBAL}.  Here we describe the method based on principal components
analysis (PCA) that we found to work the best. Previous work on this
topic has been reported by \citet{leighly_aas17}, \citet{marrs17}, and 
\citet{wagner17}.  We note that we have also tried a variational
autoencoder method \citep{mcleod20}, but we found it that it did not
offer any advantage over the PCA method that we describe below.

Spectral principal components analysis has a long history in AGN and
quasar astronomy, both as a method to study quasar line emission
\citep[e.g.,][]{francis92,shang03,yip04}, or as a tool to model quasar spectra
\citep[e.g.,][]{suzuki06,paris11}.  Our methodology differs from
typical analysis in several ways. First, we subtract the continuum, and
only perform analysis on the emission lines.  Second, we use the {\tt
  EMPCA} algorithm \citep{bailey12}, which weights the contribution of a
spectrum to the PCA according to the uncertainty.  Third, we split the
results into groups of objects with similar spectra using k-means
clustering on the eigenvector coefficients, and then we compute the
eigenvectors for each group.
This grouping method is useful because we can explain more
of the variance with fewer eigenvectors and therefore require fewer
parameters for continuum modeling in the {\it SimBAL} modeling procedure.
Finally,
we use the PCA model coefficients of the training set to set priors on 
the relative normalization of the eigenvectors compared with mean
spectrum. These methods are described further below. 

\subsection{The Sample and Continuum Modeling}\label{pcasample}

The parent sample was drawn from SDSS DR4 quasars with redshifts
between 1.2 and 1.8.  These were examined by eye and the $\sim 12000$
objects with relatively narrow \ion{Mg}{2} lines were selected.  All
of these objects were modeled between 2200 and 3050\AA\/ using the
method described in \citet{lm06}.  The spectral fitting was done using
IRAF {\tt SpecFit} using a model consisting of a iron template developed
by us from the I~Zw~1 {\it HST} observation following \citet{vb01}, a
power law continuum, and additional gaussians modeling
\ion{C}{2}]$\lambda 2325$,   \ion{Fe}{2}$\lambda\lambda 2419.3,
2438.9$, and \ion{Mg}{2}$\lambda\lambda 2795.5, 2802.7$.  The
redshifts of the spectra were fine tuned based on 
\ion{Mg}{2} fit wavelengths.
Objects with FWHM of \ion{Mg}{2} greater than 3500 $\rm km\, s^{-1}$
were excluded, and, of those, the objects with signal-to-noise ratios
less than the median value between 2200 and 2600\AA\/ were excluded.
The final sample consisted of 2626 objects.  These were cleaned of bad
points and absorption lines from intervening absorbers.  

We chose to remove the continuum from the spectra before performing
the PCA analysis.  The continuum of quasars has a large range of
slopes \citep[e.g.,][]{krawczyk15}, and those differences in continuum
shape adds variance to the spectra.  PCA measures the variance among
spectra.  Therefore, if the continuum is included in the
quasar spectrum, one or more of the first several principal components
will be dominated by continuum variance.  For example, see the second 
eigenvector from \citet{shang03} (their Fig.\ 3).  In addition, using
such eigenvectors complicates continuum modeling.  To avoid these
complications, the continuum was modeled and subtracted.  We used an
empirical model (a broken power law) between relatively line-free
areas near 1675\AA\/, 2200\AA\/, and 3050\AA\/.  The spectra were then
normalized near 2200\AA\/, and the scaled continuum was subtracted.   

\subsection{The Weighted Expectation Maximization Principal
  Components Analysis Method}\label{empca}

Traditionally, principal components analysis is performed using
singular value decomposition \citep[e.g.,][]{suzuki06, paris11}.  While
straightforward, this is not the only way to compute PCA:
\citet{bailey12} presented a method to compute PCA that is done
iteratively  using an expectation-maximization algorithm ({\tt EMPCA}).  

One of the difficulties with the singular value decomposition method
is that there is a range of signal-to-noise ratios among the spectra.
Noise adds variance, and because PCA detects variance, the resulting
eigenvectors may be noisy despite a large training data set
\citep[e.g.,][their Fig.\ 5]{bailey12}.  \citet{bailey12}'s method 
weights the contribution of each spectrum according to the inverse
variance.
Weighting can be done spectrum by spectrum, but there is an
additional advantage of using the point-by-point version: if a point
is weighted to zero, it will not be used for computing the PCA.
Thus missing data can be accounted for without interpolation.  Although we
use a common  bandpass in this application, in principle this
capability allows the construction of PCA eigenvectors with a range of
redshifts and bandpasses.   

The weighted mean spectrum was subtracted from each spectrum before
{\tt EMPCA} was run.  
The output of {\tt EMPCA} are the specified number of eigenvectors.
Because the method does not solve the eigen problem, the eigenvalues
are not output.  Instead, the fractional variance modeled is made
available; the fractional variance can be used in a scree plot as usual.  Also output are
the reconstruction coefficients of the training sample.  Note that
because of the weighting, these do not have the same value as the dot
product of an eigenvector and a spectrum as in the singular value
decomposition method.

\subsection{K-means Grouping of Spectra}\label{kmeans}

How many eigenvectors are needed to model a spectrum?  Often a scree
plot is used to determine this number.  But that number may be large;
for example, both \citet{suzuki06} and  \citet{paris11} produce 10
eigenvectors, although not that many may be necessary for any
particular application.  However, for application in {\it SimBAL}, we
want to use as few eigenvectors as possible, since each one requires a
model fit parameter that has to be constrained by the MCMC procedure.
One way to reduce the number of eigenvectors that need to be used is
to group like spectra together and make several sets of eigenvectors.
Spectra are conveniently grouped using the PCA coefficients for the
whole sample \citep[e.g.,][Fig.\ 4]{suzuki06}, since objects with
similar coefficients will have similar reconstructions.  

\begin{figure*}[!t]
\epsscale{1.1}
\begin{center}
\includegraphics[width=4.5truein]{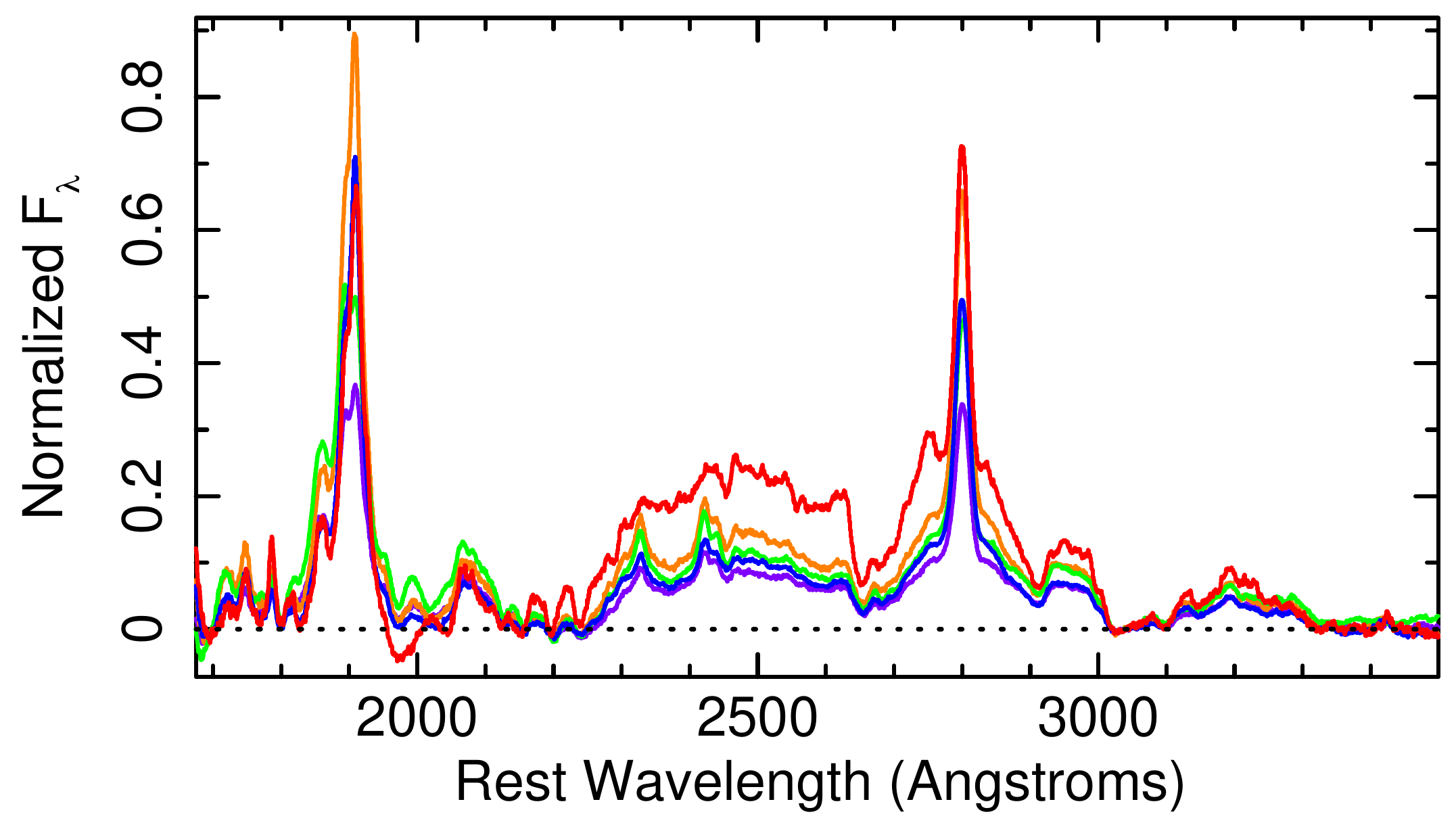}
\caption{The mean spectra corresponding to our five principal sets of
  eigenvectors for near-UV model fitting.  The five subsamples were
  constructed using k-means clustering on the first four {\tt EMPCA}
  eigenvector reconstruction coefficients from analysis of the whole
  sample.  The five sets of eigenvectors (plus an extreme
  \ion{Fe}{2}/\ion{Mg}{2} set) are sufficient to model most  near-UV line
  emission   morphologies.  \label{mean_spectra}}
\end{center}
\end{figure*}

We used the SciPy\footnote{\href{https://www.SciPy.org}{SciPy.org}} implementation of K-means
applied to the first four coefficients of the EMPCA result for the
2626 spectra.  We arbitrarily specified five clusters.  The resulting
clusters contained from 179 to 829 spectra. 
The weighted mean spectrum from each cluster can be seen in
Fig.~\ref{mean_spectra}. The means differ in the relative contribution
of \ion{Fe}{2} emission, \ion{Mg}{2} emission, and \ion{C}{3}
 emission.  For example, one particular cluster may be used when the 
 near-UV \ion{Fe}{2} emission is relatively strong, while another
 cluster is more useful when \ion{C}{3}] is very  strong.   
One additional set of spectra was considered: 75 objects with the
highest values of \ion{Fe}{2}/\ion{Mg}{2} ratio.

We choose the appropriate set of eigenvectors to use by examination of
the spectra.  In cases of extreme absorption, the best choice is
sometimes not clear \citep{macinnis18}.
However, because of the $e^{-\tau}$ nature of
absorption, uncertainties in continuum placement become less important
for larger absorption columns.  

\subsection{PCA Coefficients as Priors}\label{pcapriors}

One of the challenges in using PCA eigenvectors to model BAL quasar
spectra is that, if unconstrained, the eigenvectors can model a
portion of the absorption.  The solution in the MCMC framework is to
constrain the eigenvector amplitudes relative to the mean using
priors. Specifically, a power-law-plus-emission-line continuum model
may be expressed as:  
$$C(\lambda)=N_{PL} (\lambda/\lambda_0)^\alpha+ N_{L} \times
(L_{mean}+C_1*E1+C_2*E2+C_3*E3+C_4*E4)$$ 
where $N_{PL}$ is the power law normalization at wavelength
$\lambda_0$, $\alpha$ is the slope of power law, $N_L$ is the
normalization of the line emission, $L_{mean}$ is the weighted mean
emission-line spectrum, $C_1$--$C_4$ are the fit coefficients for the
eigenvectors, and $E1$--$E4$ are PCA eigenvectors.  Thus this model
has seven fit parameters - the two model component normalizations, the 
power law slope, and the four eigenvector coefficients.  

This problem of the PCA components modeling absorption can be ameliorated if suitable priors are set on the
eigenvector coefficients $C1$--$C4$.  Initially, we fit a
multidimensional Gaussian to the coefficients generated by training
sample.  Because eigenvectors are orthogonal, the result was a
nearly-diagonal matrix.  Then, a Gaussian prior was used for each of
the coefficients.    

In some cases, the Gaussian prior was not sufficient
to control the eigenvector coefficients basically because a Gaussian profile has
long tails.  To more strongly disallow deviations very far from the
center of the distribution, we also fit the coefficients with
generalized Gaussian models, which have more sharply decreasing tails.
In particular, a 8th order generalized Gaussian was fit and the
priors for this distribution were provided as an option to the user.  

\subsection{The Long-Wavelength Extension}\label{app:long_wavelength}

Our sets of eigenvectors model the data satisfactorily between 1675
and 3500\AA\/.  Some of the spectra of our lower-redshift objects
extend to longer wavelengths, and include interesting absorption lines  
such as \ion{Ca}{2}$\lambda\lambda 3994,3968$ and
\ion{He}{2}*$\lambda 3889$.   Ideally, we would have a set of
eigenspectra that extended to longer wavelengths; that infrastructure
is planned for a future version of {\it SimBAL}.  In the meantime, we
use a single template spectrum that models the emission between
2870\AA\/ and 4750\AA\/.  The construction of this template spectrum
is described in \citet{leighly11}.  The template is normalized to the
eigenvectors using a lookup table matching the strong \ion{Fe}{2}
feature between 2820 and 3020\AA\/ that appears in both the template
and the mean spectrum from the EMPCA analysis.  Thus, no additional normalization
parameter is needed to use the long-wavelength extension.

\section{FeLoBAL Outflow Properties\label{app:data_table}}
The best-fit parameters from the {\it SimBAL} models and the derived outflow properties are presented in Table~\ref{tbl:simbal_fit} and \ref{tbl:outflow_prop}, respectively.
95\% uncertainties estimated from the posterior probability distributions are reported.
For the BAL components that were fit using tophat accordion models, the range of values among the bins are reported.
We report the opacity (column density) weighted velocity as the representative BAL velocity ($v_{off}$) and the widths of BALs have been measured from the continuum emission normalized spectra ($I/I_0$) using the \ion{Mg}{2}$\lambda2796$ line transition (\S~\ref{subsubsec:kinematics_measure}).
The covering-fraction-corrected hydrogen column density ($\log N_H$; \S~\ref{subsec:derived_pars}) for each individual BAL is reported in Table~\ref{tbl:outflow_prop}.

\startlongtable
\begin{deluxetable*}{lCCCCCCr}
\tabletypesize{\scriptsize}
\tablecaption{\it{SimBAL} Fit Results\label{tbl:simbal_fit}}
\tablehead{\\
\colhead{Name} &\colhead{$\log U$} &\colhead{$\log n$} &\colhead{$\log N_H-\log U$} &\colhead{$v_{off}$} &\colhead{$v_{width}$} &\colhead{$\log a$} &\colhead{Type\tablenotemark{a}} \\ \colhead{} &\colhead{} &\colhead{[cm$^{-3}$]} &\colhead{[cm$^{-2}$]} &\colhead{(km s$^{-1}$)} &\colhead{(km s$^{-1}$)} &\colhead{} &\colhead{}}
\startdata
J0111+1426&-3.43^{+0.39}_{-0.29} & < 2.80 & 22.53-23.15 & -3600\pm30 & 760^{+0}_{-30} & -0.53\ \mathrm{to}\ 0.48 & \\ 
J0158$-$0046&-1.06^{+0.35}_{-0.33} & 7.15^{+0.39}_{-0.38} & 23.06-23.64 & 120^{+120}_{-80} & 1600^{+210}_{-80} & 0.70-1.13 & LB, IF\\ 
J0242$-$0722&-3.89\ \mathrm{to}\ -2.12 & 4.24^{+0.17}_{-0.18} & 22.96-23.44 & -950^{+60}_{-80} & 1300^{+70}_{-80} & 0.13-0.51 & \\ 
J0258$-$0028a&-2.12^{+0.08}_{-0.07} & 7.15^{+0.08}_{-0.12} & 22.21-23.73 & -620\pm20 & 2300^{+0}_{-20} & -0.99\ \mathrm{to}\ 0.89 & \\ 
J0258$-$0028b&-2.40^{+0.27}_{-0.41} & 4.74^{+0.49}_{-0.40} & 23.69^{+0.08}_{-0.18} & -260^{+20}_{-30} & 620^{+70}_{-30} & 0.44^{+0.13}_{-0.23} & \\ 
J0300+0048&-1.85\pm0.001 & 7.96^{+0.01}_{-0.02} & 23.09-23.72 & -6600\pm30 & 8800^{+0}_{-30} & 0.08-0.60 & OT\\ 
J0802+5513&-2.54^{+0.04}_{-0.05} & 4.40\pm0.04 & 22.99-23.10 & 200^{+9}_{-10} & 1400^{+0}_{-10} & -1.45\ \mathrm{to}\ 0.36 & IF\\ 
J0809+1818&-1.53^{+0.09}_{-0.07} & 7.63^{+0.07}_{-0.08} & 22.22-23.68 & -3100^{+50}_{-60} & 4600^{+0}_{-60} & -1.19\ \mathrm{to}\ 1.41 & \\ 
J0813+4326&< -4.00 & < 2.80 & 22.08-22.92 & -940^{+120}_{-60} & 550^{+70}_{-60} & -0.54\ \mathrm{to}\ 0.11 & \\ 
J0835+4242&-1.61\pm0.06 & 5.18\pm0.05 & 23.00-23.03 & -410\pm10 & 1200^{+70}_{-10} & 0.58-1.26 & \\ 
J0840+3633&-1.97\pm0.03 & 7.45^{+0.03}_{-0.01} & 23.09-23.22 & -3000\pm8 & 3400^{+0}_{-8} & -0.68\ \mathrm{to}\ 0.99 & \\ 
J0916+4534&-0.14^{+0.30}_{-0.21} & 5.87^{+0.22}_{-0.28} & 23.67^{+0.08}_{-0.21} & 80\pm10 & 550^{+70}_{-10} & 1.20\pm0.05 & LB, IF\\ 
J0918+5833&0.54^{+0.36}_{-0.41} & 3.92^{+0.27}_{-0.21} & 23.01\pm0.07 & 640\pm10 & 830^{+70}_{-10} & 0.88^{+0.06}_{-0.07} & IF\\ 
J0944+5000&-2.51^{+0.15}_{-0.20} & 6.85^{+0.67}_{-1.29} & 22.50-23.18 & -1600\pm90 & 3800^{+140}_{-90} & 0.15-1.78 & \\ 
J1006+0513&1.09^{+0.01}_{-0.07} & 6.13^{+0.06}_{-0.10} & 22.95-23.71 & -2400\pm110 & 3500^{+140}_{-110} & 0.86-2.43 & LB\\ 
J1019+0225&-0.89^{+0.18}_{-0.15} & 7.29^{+0.15}_{-0.18} & 23.20-23.68 & -1500\pm20 & 1900^{+0}_{-20} & 0.39-1.42 & OT, LB\\ 
J1020+6023a&-1.94^{+0.05}_{-0.01} & 7.60^{+0.03}_{-0.05} & 23.16-23.73 & -3900^{+10}_{-20} & 1300^{+70}_{-20} & 0.16-0.38 & \\ 
J1020+6023b&-2.92^{+0.12}_{-0.08} & 4.40^{+0.10}_{-0.13} & 22.55-23.42 & -3100^{+20}_{-30} & 1700^{+70}_{-30} & -1.18\ \mathrm{to}\ 0.80 & \\ 
J1022+3542&-1.98^{+0.25}_{-0.41} & < 2.80 & 22.10-22.93 & -1400^{+50}_{-60} & 2200^{+0}_{-60} & -0.25\ \mathrm{to}\ 2.08 & \\ 
J1023+0152&-0.80^{+1.55}_{-0.59} & 7.04^{+0.64}_{-1.26} & 23.42^{+0.26}_{-0.40} & -220\pm20 & 550^{+70}_{-20} & 1.27\pm0.10 & LB\\ 
J1030+3120&-2.17\pm0.04 & 7.97^{+0.57}_{-0.59} & 22.73-23.10 & -2300\pm20 & 3000^{+410}_{-20} & -0.21\ \mathrm{to}\ 1.55 & \\ 
J1039+3954a&-2.50^{+0.35}_{-0.42} & < 2.80 & 23.10^{+0.19}_{-0.12} & -4000\pm10 & 410^{+70}_{-10} & 0.07^{+0.20}_{-0.24} & \\ 
J1039+3954b&-2.89^{+0.35}_{-0.56} & < 2.80 & 23.22\pm0.15 & -1500\pm10 & 550^{+70}_{-10} & 0.10^{+0.18}_{-0.32} & \\ 
J1044+3656a&-2.43\pm0.01 & 4.58^{+0.04}_{-0.05} & 21.98-23.02 & -4500^{+6}_{-8} & 1600^{+0}_{-8} & -1.04\ \mathrm{to}\ 0.32 & \\ 
J1044+3656b&-3.67^{+0.04}_{-0.05} & < 2.80 & 21.57-23.40 & -1400^{+10}_{-9} & 1400^{+0}_{-9} & -0.96\ \mathrm{to}\ 2.29 & \\ 
J1125+0029a&0.36^{+0.16}_{-0.11} & 7.30^{+0.13}_{-0.15} & 23.21-23.74 & -50\pm30 & 830^{+0}_{-30} & 1.24-2.04 & OT, LB\\ 
J1125+0029b&-0.45^{+0.49}_{-0.37} & 5.93^{+0.41}_{-0.35} & 23.06\pm0.02 & 380\pm30 & 2800^{+140}_{-30} & 0.94\pm0.07 & IF\\ 
J1128+0113&-1.75^{+0.007}_{-0.001} & 7.37^{+0.02}_{-0.01} & 22.16-23.71 & -540^{+10}_{-20} & 3000^{+0}_{-20} & -1.00\ \mathrm{to}\ 2.08 & LB\\ 
J1129+0506&-2.81^{+0.51}_{-0.35} & 3.72^{+0.34}_{-0.44} & 21.91-23.42 & -4000^{+280}_{-270} & 2600^{+0}_{-270} & -0.41\ \mathrm{to}\ 0.57 & \\ 
J1145+1100&-0.30^{+0.70}_{-0.84} & 5.54^{+0.76}_{-0.58} & 23.11^{+0.22}_{-0.09} & -160\pm4 & 350^{+0}_{-4} & 0.31^{+0.13}_{-0.21} & \\ 
J1154+0300&-0.60\ \mathrm{to}\ 1.40 & 7.27\pm0.04 & 22.87-23.72 & -15400\pm50 & 7400^{+0}_{-50} & 0.83-1.43 & OT\\ 
J1158$-$0043&-2.15^{+0.27}_{-0.45} & < 2.80 & 22.08-23.03 & -180^{+40}_{-50} & 1500^{+0}_{-50} & -0.36\ \mathrm{to}\ 0.37 & \\ 
J1200+6322&-1.77^{+0.15}_{-0.17} & 3.73^{+0.13}_{-0.16} & 21.76-23.05 & -440^{+80}_{-70} & 3200\pm70 & -0.35\ \mathrm{to}\ 1.51 & \\ 
J1206+0023a&1.09^{+0.26}_{-0.34} & 6.03^{+0.16}_{-0.09} & 22.94-23.36 & -4900^{+160}_{-130} & 3700^{+1200}_{-130} & 1.42-2.08 & OT\\ 
J1206+0023b&-1.78^{+0.15}_{-0.14} & 6.59^{+0.70}_{-0.52} & 23.01-23.07 & -2900^{+160}_{-150} & 5400^{+70}_{-150} & 0.12-0.59 & \\ 
J1208+6240&-1.31^{+0.65}_{-0.41} & 6.17^{+0.49}_{-0.66} & 23.56^{+0.13}_{-0.27} & -80^{+6}_{-4} & 280^{+0}_{-4} & 0.74^{+0.14}_{-0.13} & LB\\ 
J1212+2514&-2.31^{+0.15}_{-0.17} & 3.93\pm0.29 & 23.02^{+0.06}_{-0.04} & -2000^{+9}_{-10} & 480^{+0}_{-10} & -0.92^{+0.50}_{-0.55} & \\ 
J1214$-$0001&-2.81^{+0.83}_{-1.03} & < 2.80 & 23.07^{+0.33}_{-0.13} & -170\pm10 & 280^{+0}_{-10} & 0.62^{+0.19}_{-0.31} & \\ 
J1214+2803&-1.70^{+0.004}_{-0.001} & 7.81\pm0.01 & 22.15-23.72 & -2500^{+6}_{-5} & 2000^{+0}_{-5} & -1.22\ \mathrm{to}\ 0.53 & \\ 
J1235+0132&-1.79^{+0.26}_{-0.30} & 7.88^{+0.44}_{-0.32} & 22.55-23.69 & -700^{+110}_{-100} & 2100^{+70}_{-100} & 0.25-1.37 & LB\\ 
J1240+4443&-1.76^{+0.34}_{-0.36} & < 2.80 & 22.20-23.38 & -70\pm60 & 1000^{+70}_{-60} & 0.33-1.26 & \\ 
J1321+5617&0.17^{+0.20}_{-0.15} & 6.13^{+0.13}_{-0.17} & 23.07-23.75 & -40^{+10}_{-20} & 620^{+0}_{-20} & 0.91-1.14 & LB\\ 
J1324+0320&< -3.00 & 5.04^{+0.35}_{-0.25} & 22.71-23.53 & -900^{+70}_{-60} & 2200^{+70}_{-60} & 0.30-2.19 & \\ 
J1336+0830&< -3.00 & 5.06^{+0.03}_{-0.04} & 21.95-23.79 & -1100^{+20}_{-7} & 1900^{+0}_{-7} & -1.07\ \mathrm{to}\ 0.70 & \\ 
J1355+5753&-2.41^{+0.41}_{-0.63} & < 2.80 & 22.11-23.42 & -830^{+110}_{-90} & 1600^{+0}_{-90} & -0.22\ \mathrm{to}\ 1.86 & \\ 
J1356+4527&-1.85^{+0.17}_{-0.20} & < 2.80 & 21.93-23.14 & -1300^{+140}_{-200} & 4300^{+70}_{-200} & -0.16\ \mathrm{to}\ 1.07 & \\ 
J1427+2709a&-1.71^{+0.17}_{-0.65} & < 2.80 & 22.08-23.22 & -3300^{+30}_{-60} & 970^{+0}_{-60} & -0.54\ \mathrm{to}\ 1.09 & \\ 
J1427+2709b&-3.64^{+0.07}_{-0.09} & < 2.80 & 22.30-23.40 & -700\pm10 & 1200^{+0}_{-10} & -1.05\ \mathrm{to}\ 1.31 & \\ 
J1448+4043a&-0.50^{+0.008}_{-0.002} & 6.80\pm0.01 & 23.49-23.74 & -4300\pm30 & 4500^{+0}_{-30} & 1.24-1.67 & OT\\ 
J1448+4043b&-2.50^{+0.01}_{-0.02} & 6.38^{+0.01}_{-0.02} & 22.73-23.80 & -1800\pm20 & 4300^{+0}_{-20} & -1.48\ \mathrm{to}\ 0.74 & \\ 
J1448+4043c&< -3.00 & 4.76^{+0.05}_{-0.06} & 23.72\pm0.03 & -600\pm2 & 550^{+0}_{-2} & -0.82^{+0.11}_{-0.13} & \\ 
J1517+2328&< -3.00 & 4.73^{+0.20}_{-0.25} & 23.03-23.60 & -500^{+110}_{-120} & 1700^{+280}_{-120} & -0.23\ \mathrm{to}\ 0.62 & \\ 
J1527+5912&-1.57^{+0.08}_{-0.10} & < 2.80 & 22.04-22.94 & -1100\pm30 & 2600^{+0}_{-30} & -0.86\ \mathrm{to}\ 0.87 & \\ 
J1531+4852&-3.40^{+1.11}_{-0.56} & < 2.80 & 22.51^{+0.34}_{-0.44} & -150^{+30}_{-7} & 140^{+70}_{-7} & -0.89^{+0.84}_{-0.59} & \\ 
J1556+3517a&1.09^{+0.005}_{-0.013} & 6.21^{+0.06}_{-0.03} & 22.95-23.28 & -7200^{+50}_{-60} & 3500^{+0}_{-60} & 1.03-2.06 & OT\\ 
J1556+3517b&-2.18^{+0.06}_{-0.04} & 6.42^{+0.12}_{-0.08} & 22.71-23.72 & -3100^{+40}_{-50} & 4300^{+0}_{-50} & -1.33\ \mathrm{to}\ 0.67 & \\ 
J1644+5307&0.11^{+0.20}_{-0.17} & 6.79^{+0.14}_{-0.16} & 23.23-23.61 & -1600\pm30 & 900^{+70}_{-30} & 0.93-2.10 & OT, LB\\ 
J2107+0054&-2.86^{+0.69}_{-0.56} & < 2.80 & 22.05-23.13 & -1800^{+90}_{-110} & 1900^{+550}_{-110} & -0.37\ \mathrm{to}\ 1.14 & \\ 
J2135$-$0320&-2.03^{+0.11}_{-0.13} & 2.92^{+0.15}_{-0.11} & 22.28-23.11 & -490\pm20 & 1200^{+0}_{-20} & -0.93\ \mathrm{to}\ 0.71 & \\ 
J2307+1119&-2.41^{+0.28}_{-0.48} & 5.38^{+0.28}_{-0.23} & 22.98-23.05 & -1000^{+60}_{-50} & 2500^{+70}_{-50} & 0.30-1.01 & \\ 
\enddata
\tablenotetext{a}{OT: Overlapping Trough BAL; LB: Loitering BAL; IF: Inflow}
\end{deluxetable*}

\startlongtable
\begin{deluxetable*}{lCCCCCCCC}
\tabletypesize{\scriptsize}
\tablecaption{Derived Outflow Properties\label{tbl:outflow_prop}}
\tablehead{\\
\colhead{Name} &\colhead{$\log N_H$} &\colhead{$\log R$} &\colhead{$\log \dot M_{out}$} &\colhead{$\log \dot M_{in}$} &\colhead{$\log \dot P$} &\colhead{$\log L_{KE}$} &\colhead{$\log L_{Bol}$} &\colhead{$\alpha_{ui}$} \\ \colhead{} &\colhead{[cm$^{-2}$]} &\colhead{[pc]} &\colhead{[M$_\odot$ yr$^{-1}$]} &\colhead{[M$_\odot$ yr$^{-1}$]} &\colhead{[dyne]} &\colhead{[erg s$^{-1}$]} &\colhead{[erg s$^{-1}$]} &\colhead{}}
\startdata
J0111+1426&19.35^{+0.28}_{-0.17} & > 4.31 & 1.99^{+0.15}_{-0.23} & \nodata & 36.35^{+0.14}_{-0.23} & 44.60^{+0.14}_{-0.23} & 47.08 & -0.30\\ 
J0158$-$0046&21.92^{+0.28}_{-0.26} & 0.24\pm0.06 & -1.31^{+0.42}_{-0.51} & -0.82^{+0.42}_{-0.35} & 32.06^{+0.43}_{-0.56} & 39.33^{+0.44}_{-0.63} & 45.42 & -1.29\\ 
J0242$-$0722&20.56^{+0.16}_{-0.15} & 3.47^{+0.28}_{-0.77} & 1.47^{+0.10}_{-0.09} & \nodata & 35.41\pm0.09 & 43.28\pm0.09 & 46.53 & -0.82\\ 
J0258$-$0028a&21.61\pm0.04 & 1.26\pm0.04 & 0.41^{+0.07}_{-0.05} & \nodata & 34.16^{+0.08}_{-0.05} & 41.95^{+0.09}_{-0.06} & 46.37 & -0.44\\ 
J0258$-$0028b&20.70^{+0.24}_{-0.33} & 2.59^{+0.21}_{-0.17} & 0.46^{+0.18}_{-0.26} & \nodata & 33.67^{+0.20}_{-0.27} & 40.77^{+0.22}_{-0.28} & 46.37 & -0.44\\ 
J0300+0048&22.31\pm0.01 & 1.08\pm0.01 & 1.98\pm0.01 & \nodata & 36.64\pm0.01 & 45.24\pm0.01 & 47.12 & -0.28\\ 
J0802+5513&20.92^{+0.03}_{-0.04} & 2.96^{+0.03}_{-0.02} & 0.33\pm0.05 & 1.03\pm0.03 & 33.55\pm0.06 & 40.73\pm0.06 & 46.62 & -0.12\\ 
J0809+1818&21.79^{+0.06}_{-0.05} & 0.85^{+0.02}_{-0.01} & 0.89\pm0.06 & \nodata & 35.24\pm0.06 & 43.54\pm0.07 & 46.64 & -1.25\\ 
J0813+4326&18.94^{+0.35}_{-0.17} & > 4.05 & 0.78^{+0.14}_{-0.18} & \nodata & 34.59^{+0.14}_{-0.18} & 42.30^{+0.15}_{-0.18} & 46.37 & -1.06\\ 
J0835+4242&20.83\pm0.05 & 2.12\pm0.03 & 0.33\pm0.04 & \nodata & 33.90\pm0.05 & 41.50\pm0.06 & 46.66 & -0.99\\ 
J0840+3633&21.85\pm0.02 & 1.55^{+0.005}_{-0.008} & 1.64^{+0.01}_{-0.02} & \nodata & 35.95^{+0.01}_{-0.02} & 44.18^{+0.01}_{-0.02} & 47.42 & -1.11\\ 
J0916+4534a&22.29^{+0.18}_{-0.16} & 0.58\pm0.05 & \nodata & -0.47^{+0.19}_{-0.18} & \nodata & \nodata & 45.73 & -1.12\\ 
J0918+5833&22.62^{+0.28}_{-0.33} & 1.56^{+0.10}_{-0.11} & \nodata & 1.74^{+0.20}_{-0.26} & \nodata & \nodata & 46.41 & -1.19\\ 
J0944+5000&20.83^{+0.13}_{-0.15} & 1.56^{+0.65}_{-0.32} & 0.37^{+0.59}_{-0.35} & \nodata & 34.56^{+0.58}_{-0.36} & 42.73^{+0.58}_{-0.36} & 46.28 & -0.20\\ 
J1006+0513&23.57^{+0.04}_{-0.08} & 0.30^{+0.07}_{-0.03} & 2.01^{+0.06}_{-0.05} & \nodata & 36.27^{+0.07}_{-0.06} & 44.48\pm0.08 & 46.67 & -0.41\\ 
J1019+0225&22.42^{+0.10}_{-0.07} & 0.92\pm0.03 & 1.29^{+0.10}_{-0.08} & \nodata & 35.29^{+0.10}_{-0.09} & 43.23^{+0.10}_{-0.09} & 47.09 & -0.44\\ 
J1020+6023a&21.60^{+0.03}_{-0.02} & 1.15^{+0.01}_{-0.02} & 1.10\pm0.03 & \nodata & 35.49\pm0.03 & 43.79\pm0.03 & 46.78 & -0.48\\ 
J1020+6023b&20.68\pm0.07 & 3.24\pm0.05 & 2.17^{+0.07}_{-0.06} & \nodata & 36.46^{+0.07}_{-0.06} & 44.66^{+0.07}_{-0.06} & 46.78 & -0.48\\ 
J1022+3542&20.98^{+0.26}_{-0.40} & > 3.13 & 2.03^{+0.20}_{-0.45} & \nodata & 36.04^{+0.20}_{-0.45} & 43.98^{+0.21}_{-0.45} & 46.14 & -0.28\\ 
J1023+0152&21.35^{+1.11}_{-0.35} & 0.46^{+0.09}_{-0.14} & -1.05^{+0.96}_{-0.39} & \nodata & 32.09^{+0.97}_{-0.40} & 39.14^{+0.97}_{-0.41} & 46.05 & -1.54\\ 
J1030+3120&21.01^{+0.04}_{-0.05} & 1.02^{+0.29}_{-0.27} & 0.17^{+0.29}_{-0.30} & \nodata & 34.37^{+0.29}_{-0.30} & 42.50^{+0.29}_{-0.30} & 46.68 & -0.93\\ 
J1039+3954a&20.27^{+0.29}_{-0.33} & > 3.49 & 2.14^{+0.19}_{-0.28} & \nodata & 36.54^{+0.19}_{-0.28} & 44.85^{+0.19}_{-0.28} & 46.31 & 0.12\\ 
J1039+3954b&19.96^{+0.27}_{-0.34} & > 3.68 & 1.60^{+0.17}_{-0.27} & \nodata & 35.58^{+0.17}_{-0.27} & 43.47^{+0.17}_{-0.27} & 46.31 & 0.12\\ 
J1044+3656a&20.68\pm0.01 & 2.89\pm0.02 & 1.99^{+0.03}_{-0.02} & \nodata & 36.45^{+0.03}_{-0.02} & 44.81^{+0.03}_{-0.02} & 46.77 & -0.90\\ 
J1044+3656b&19.36\pm0.03 & > 4.40 & 1.67\pm0.01 & \nodata & 35.65^{+0.01}_{-0.02} & 43.54\pm0.02 & 46.77 & -0.90\\ 
J1125+0029a&23.05\pm0.09 & -0.04\pm0.06 & -0.10^{+0.12}_{-0.11} & -0.31^{+0.14}_{-0.12} & 33.46\pm0.13 & 40.97^{+0.14}_{-0.15} & 46.41 & -0.94\\ 
J1125+0029b&21.63^{+0.43}_{-0.33} & 1.03^{+0.22}_{-0.21} & \nodata & 0.01^{+0.28}_{-0.27} & \nodata & \nodata & 46.41 & -0.94\\ 
J1128+0113&21.84\pm0.01 & 0.94\pm0.01 & 0.34\pm0.01 & -0.50\pm0.04 & 34.04\pm0.01 & 41.67^{+0.01}_{-0.02} & 46.34 & -1.08\\ 
J1129+0506&20.78^{+0.29}_{-0.27} & 3.49^{+0.21}_{-0.30} & 2.62^{+0.28}_{-0.27} & \nodata & 37.05\pm0.28 & 45.39^{+0.28}_{-0.29} & 46.71 & -0.40\\ 
J1145+1100&22.33^{+0.57}_{-0.55} & 1.05\pm0.08 & 0.35^{+0.51}_{-0.52} & \nodata & 33.36^{+0.51}_{-0.52} & 40.28^{+0.51}_{-0.53} & 46.17 & -1.13\\ 
J1154+0300&23.62\pm0.02 & 0.88^{+0.03}_{-1.01} & 2.71^{+0.03}_{-0.02} & \nodata & 37.67^{+0.03}_{-0.02} & 46.53^{+0.03}_{-0.02} & 47.28 & -0.64\\ 
J1158$-$0043&20.90^{+0.25}_{-0.36} & > 3.40 & 1.45^{+0.13}_{-0.16} & 0.87^{+0.22}_{-0.30} & 34.93^{+0.14}_{-0.16} & 42.35^{+0.14}_{-0.17} & 46.35 & -0.63\\ 
J1200+6322&21.31^{+0.14}_{-0.15} & 2.63^{+0.08}_{-0.07} & 1.51\pm0.12 & 1.01\pm0.17 & 35.32\pm0.13 & 43.11^{+0.14}_{-0.13} & 46.05 & -0.99\\ 
J1206+0023a&23.15^{+0.12}_{-0.16} & 0.33^{+0.10}_{-0.09} & 1.93^{+0.07}_{-0.09} & \nodata & 36.44^{+0.07}_{-0.09} & 44.86^{+0.08}_{-0.10} & 46.61 & -0.90\\ 
J1206+0023b&21.37^{+0.14}_{-0.13} & 1.49^{+0.27}_{-0.36} & 1.08^{+0.24}_{-0.35} & \nodata & 35.46^{+0.24}_{-0.35} & 43.79^{+0.24}_{-0.35} & 46.61 & -0.90\\ 
J1208+6240&21.43^{+0.39}_{-0.29} & 0.93^{+0.17}_{-0.13} & -0.95^{+0.40}_{-0.34} & \nodata & 31.77^{+0.41}_{-0.36} & 38.39^{+0.41}_{-0.37} & 45.58 & -1.17\\ 
J1212+2514&20.66^{+0.14}_{-0.15} & 2.84^{+0.12}_{-0.13} & 1.55^{+0.20}_{-0.21} & \nodata & 35.64^{+0.20}_{-0.21} & 43.63^{+0.20}_{-0.21} & 46.14 & -0.09\\ 
J1214$-$0001b&19.55^{+0.66}_{-0.67} & > 3.72 & 0.27^{+0.32}_{-0.30} & \nodata & 33.29^{+0.33}_{-0.30} & 40.23^{+0.34}_{-0.31} & 46.51 & -0.60\\ 
J1214+2803&22.01\pm0.01 & 0.80^{+0.007}_{-0.005} & 0.96\pm0.01 & \nodata & 35.17\pm0.01 & 43.28\pm0.01 & 46.54 & -0.97\\ 
J1235+0132&21.47^{+0.22}_{-0.25} & 0.82^{+0.09}_{-0.11} & -0.09^{+0.23}_{-0.31} & -1.45^{+0.29}_{-0.37} & 33.82^{+0.23}_{-0.31} & 41.66^{+0.25}_{-0.31} & 46.56 & -0.38\\ 
J1240+4443&21.33\pm0.28 & > 2.92 & 1.20^{+0.19}_{-0.23} & 0.92^{+0.22}_{-0.25} & 34.56^{+0.23}_{-0.25} & 41.91^{+0.26}_{-0.28} & 45.86 & -0.59\\ 
J1321+5617&22.89^{+0.13}_{-0.12} & 0.23^{+0.03}_{-0.04} & -0.33^{+0.16}_{-0.14} & -0.74^{+0.11}_{-0.15} & 32.86^{+0.16}_{-0.14} & 39.96^{+0.15}_{-0.14} & 45.60 & -1.20\\ 
J1324+0320a&20.01^{+0.16}_{-0.06} & > 2.71 & 0.44^{+0.15}_{-0.18} & \nodata & 34.42^{+0.16}_{-0.18} & 42.37^{+0.15}_{-0.17} & 46.37 & -0.61\\ 
J1336+0830&20.57\pm0.02 & > 2.66 & 1.02\pm0.02 & \nodata & 34.91\pm0.02 & 42.75^{+0.02}_{-0.03} & 46.22 & -0.97\\ 
J1355+5753&20.74^{+0.33}_{-0.51} & > 3.20 & 1.62^{+0.15}_{-0.27} & \nodata & 35.45^{+0.15}_{-0.26} & 43.20^{+0.15}_{-0.26} & 45.73 & -0.73\\ 
J1356+4527&21.21^{+0.16}_{-0.18} & > 3.12 & 2.22^{+0.12}_{-0.14} & \nodata & 36.42\pm0.16 & 44.62\pm0.18 & 46.08 & 0.08\\ 
J1427+2709a&20.85^{+0.15}_{-0.53} & > 3.35 & 2.49^{+0.08}_{-0.81} & \nodata & 36.81^{+0.08}_{-0.81} & 45.03^{+0.08}_{-0.80} & 46.72 & -0.68\\ 
J1427+2709b&19.88\pm0.04 & > 4.35 & 1.83\pm0.02 & \nodata & 35.55\pm0.02 & 43.18^{+0.02}_{-0.03} & 46.72 & -0.68\\ 
J1448+4043a&22.37\pm0.01 & 0.84^{+0.005}_{-0.006} & 1.60^{+0.01}_{-0.02} & \nodata & 36.07^{+0.01}_{-0.02} & 44.47^{+0.01}_{-0.02} & 46.83 & -0.28\\ 
J1448+4043b&21.41\pm0.01 & 2.05\pm0.01 & 1.47\pm0.01 & \nodata & 35.64\pm0.02 & 43.82\pm0.02 & 46.83 & -0.28\\ 
J1448+4043c&20.66\pm0.02 & > 3.11 & 1.31\pm0.02 & \nodata & 34.89\pm0.02 & 42.37\pm0.02 & 46.83 & -0.28\\ 
J1517+2328&21.12^{+0.17}_{-0.12} & > 2.66 & 1.25^{+0.18}_{-0.17} & -1.80^{+1.49}& 34.94\pm0.22 & 42.60^{+0.26}_{-0.28} & 46.04 & 0.19\\
J1527+5912&21.44^{+0.09}_{-0.10} & > 3.28 & 2.53^{+0.06}_{-0.07} & \nodata & 36.50^{+0.06}_{-0.08} & 44.43^{+0.07}_{-0.08} & 46.65 & -0.44\\ 
J1531+4852b&18.98^{+1.23}_{-0.64} & > 3.74 & -0.30^{+0.73}_{-1.89} & \nodata & 32.69^{+0.74}_{-1.95} & 39.57^{+0.75}_{-2.01} & 46.02 & -1.00\\ 
J1556+3517a&23.34\pm0.02 & 0.69^{+0.02}_{-0.03} & 2.64\pm0.03 & \nodata & 37.31\pm0.03 & 45.89\pm0.03 & 47.51 & -0.11\\ 
J1556+3517b&21.69\pm0.03 & 2.22^{+0.04}_{-0.06} & 2.15^{+0.06}_{-0.08} & \nodata & 36.49^{+0.06}_{-0.08} & 44.76^{+0.06}_{-0.08} & 47.51 & -0.11\\ 
J1644+5307a&22.91^{+0.12}_{-0.10} & 0.36\pm0.03 & 1.24^{+0.10}_{-0.09} & \nodata & 35.26^{+0.10}_{-0.09} & 43.19^{+0.10}_{-0.09} & 46.47 & -0.95\\ 
J2107+0054&20.36^{+0.64}_{-0.40} & > 3.42 & 1.79^{+0.34}_{-0.28} & \nodata & 35.86^{+0.35}_{-0.29} & 43.86^{+0.36}_{-0.30} & 45.87 & 0.51\\ 
J2135$-$0320&21.17^{+0.10}_{-0.11} & 3.37\pm0.06 & 2.00^{+0.09}_{-0.10} & \nodata & 35.63^{+0.09}_{-0.10} & 43.20^{+0.09}_{-0.11} & 46.47 & -0.42\\ 
J2307+1119&20.78^{+0.23}_{-0.36} & 2.27^{+0.21}_{-0.16} & 0.80^{+0.18}_{-0.20} & \nodata & 34.75^{+0.18}_{-0.20} & 42.65^{+0.18}_{-0.20} & 46.33 & -0.27\\ 
\enddata
\end{deluxetable*}

\section{Notes on Individual Objects\label{app:model_detail}}

{\bf 011117.36+142653.6}  This object is included in the
\citet{farrah12} sample, and has been observed twice by SDSS.  It was
originally classified as an 
FeLoBAL quasar by \citet{trump06}.  It was reported have been detected
by {\it ROSAT} \citep{scott07}, and was detected in the near-UV by
{\it GALEX} \citep{trammell07}.  It was observed using {\it SCUBA-2}
but was not detected \citep{violino16}.
Two Gaussian absorption profiles were used to model the absorption features.
They were constrained to have the same ionization parameter and density while other parameters were allowed to vary.
However, we combined the two Gaussians together for the analysis as a single BAL outflow component for determining the outflow mass rates and the hydrogen column densities.
The break in the continuum near 3000 \AA\/ required the use of a general reddening curve \citep{choi20}
\vskip 1pc

{\bf 015813.56-004635.5}  This object was observed once by SDSS and is
included in the DR14 SDSS quasar catalog \citep{paris18}.
A four-bin tophat model (single ionization parameter and single density) was used.
\vskip 1pc

{\bf 024254.66-072205.6} This object is included in the
\citet{farrah12} sample, and has been observed once by SDSS. It was
originally classified as an FeLoBAL quasar by \citet{trump06}.    It
was observed using {\it SCUBA-2} but was not detected \citep{violino16}.
A six-bin tophat with two-covering model was used.
All bins were constrained to have a single density parameter while other parameters, including the ionization parameter, were allowed to vary.
The SMC reddening for this object was fixed to a value of $E(\bv)=0.075$ obtained from a fit to the photometry.
\vskip 1pc

{\bf 025858.17-002827.0} This object has been observed three times
using SDSS, and is included in the DR14 SDSS quasar catalog
\citep{paris18}.
It was identified as a BALQ in the DR10 SDSS quasar catalog \citep{paris14}.
A combination of an eight-bin tophat (single ionization parameter and single density) and a Gaussian opacity profile were used to model the two BAL components found in the spectrum.
The general reddening law was used in the continuum model.
\vskip 1pc

{\bf 030000.57+004828.0}  This bright overlapping-trough BALQ has been
observed three times using SDSS, and is included in the
\citet{farrah12} sample.  It was first classified as an FeLoBAL quasar
by \citet{hall02}, and \citet{hall03} presented an analysis of the
spectacular narrow \ion{Ca}{2} absorption lines.  Despite the heavy
absorption, near-UV emission was detected by {\it GALEX}
\citep{trammell07}.  \citet{dipompeo11} reported spectropolarimetry
observations; the continuum is modestly polarized (2\%), but the
emission and absorption lines do not show any different polarization.
\citet{vivek12} obtained two additional spectra; they found no
variability in the optically-thickest portions of the outflow, but
reported variability that may be associated 
with the optically thinner portion or the underlying continuum or line
emission.  \citet{mcgraw15} also investigated the variability in this
object; they concluded that there is no variability in the BALs but
tentative variability in the associated absorption lines.   It was
observed using {\it SCUBA-2} but was not detected
\citep{violino16}.  \citet{villforth19} presented near-IR imaging
observations obtained using {\it HST}; the image was dominated by the
PSF.    \citet{lawther18} used {\it HST} to image the
object in the UV, within the BAL troughs, and in the near-IR.  They
found that that the host galaxy properties are consistent with those
of non-BAL quasars.
\citet{rogerson11} observed this object with {\it Chandra} to study the X-ray absorption and constrained the lower limit for the column density of the X-ray absorbing gas to be $\log N_H\geq24.3\ [\rm cm^{-2}]$ with density $\log n\sim6\ [\rm cm^{-3}]$.
A twelve-bin tophat (single ionization parameter and single density) model with two covering factors was used.
We only fit the main \ion{Fe}{2} trough and ignored the much narrower \ion{Ca}{2} absorption lines that were observed at the lower velocity end of the main overlapping trough.
The physical constraints on the main \ion{Fe}{2} trough from the best-fitting model are unaffected by this exclusion.
\citet{hall03} concluded that the \ion{Ca}{2} must have formed in a different region within the same BAL outflow gas with a significant temperature difference.
\vskip 1pc

{\bf 033810.84+005617.6}  This object is included in the
\citet{farrah12} sample, and has been observed five times by SDSS. It
was first classified as an FeLoBAL quasar by \citet{hall02}.  It was
detected in the near-UV by {\it GALEX} \citep{trammell07}.
Spectropolarimetry observations \citep{dipompeo11} revealed that it is
unpolarized.
A model with a nine-bin tophat bins (single ionization parameter and single density) and the general reddening law was used.
The best-fitting {\it SimBAL} model found a \ion{Mg}{2} trough spanning from $-43,300\rm\ km\ s^{-1}$ to $-26,400\rm\ km\ s^{-1}$,
exceeding that discovered in GQ 1309$+$2904, the previously discovered fastest LoBAL \citep{fynbo20}.
We excluded this object
from the analysis because we could not constrain the physical properties of the outflow gas since no other absorption line was found in the spectrum (also makes this object a LoBALQ instead of FeLoBALQ) and the BAL identification was uncertain due to the location of the trough in the region where strong \ion{Fe}{2} emission is generally found.
\vskip 1pc

{\bf 080248.18+551328.8} This object was observed twice by SDSS, and
was first classified as a BALQ by \citet{gibson09}. \citet{liu15}
reported \ion{He}{1}* absorption in the SDSS spectrum.  \citet{yi19}
reported variability in the \ion{Mg}{2} absorption line equivalent
width with  $3.4\sigma$ confidence.
A five-bin tophat model (single ionization parameter and single density) was used.
\vskip 1pc

{\bf 080957.39+181804.4} This object was observed twice by SDSS, and
is included in the DR14 SDSS quasar catalog \citep{paris18}.
\citet{liu15} reported \ion{He}{1}* absorption in the SDSS spectrum.
\citet{villforth19} presented near-IR imaging observations obtained
using {\it HST}; the image is dominated by the PSF.
A nine-bin tophat (single ionization parameter and single density) with two-covering model was used.
\vskip 1pc

{\bf 081312.61+432640.1} This object is included in the 
\citet{farrah12} sample, and has been observed three times by SDSS.  It
was first classified as an FeLoBAL quasar by \citet{trump06}.
Two Gaussian profiles were used to model the absorption feature, mainly following the shapes of the \ion{Mg}{2}$\lambda\lambda 2796,2803$ doublet lines.
They were combined and analyzed as a single BAL outflow.
\vskip 1pc

{\bf 083522.77+424258.3}  This object was discovered in the First
Bright Quasar Survey \citep[FBQS~J083522.7+424258][]{white00}, and is
included in the \citet{farrah12} sample.  It has been observed three
times by SDSS. The \ion{He}{1}* absorption lines were noted by
\citet{liu15}.  \citet{vivek12} reported no BAL variability over 5 years
in the object's rest frame.
Two Gaussian profiles were used based on the shape of the \ion{Mg}{2} doublet lines.
Only the lower-velocity Gaussian component of the two was used in the analysis because the other Gaussian component
presented no \ion{Fe}{2} lines.
The \ion{Fe}{2} and \ion{Mg}{2} absorption lines as well as the \ion{He}{1}$^*\lambda 3188$ and \ion{He}{1}$^*\lambda 3889$ transitions in this object were well modeled with {\it SimBAL}.
However, the model underpredicts the opacity for \ion{Ca}{2}$\lambda\lambda 3934,3969$ doublet lines as well as the \ion{Mg}{1}$\lambda 2853$ line.
\vskip 1pc

{\bf 084044.41+363327.8}  This object was identified as a radio-loud
BALQ in the FIRST Survey \citep[FBQS J0840+3633;][]{becker97}, and is included in the \citet{farrah12}
sample.  It was observed by SDSS three times.  It is highly polarized,
with complex polarization 
structure across the troughs \citep{brotherton97}.  It was detected by
{\it Chandra}, with inferred $\alpha_{ox}=2.11$ \citep{green01}.
\citet{dekool02f1} performed a heroic in-depth analysis of a Keck
echelle spectrum.  They found evidence for two absorption systems,
with the higher-excitation one located $\sim 230$ pc from the
nucleus.  \citet{lewis03} reported no detection in
sub-millimeter band from a {\it SCUBA} observation. Both \citet{vivek14} and
\citet{mcgraw15} found that the broad absorption lines were not
variable.
A nine-bin tophat (single ionization parameter and single density) model with two covering factors was used.
Even though the absorber velocity profile from our model encompasses both of the BAL components analyzed in \citet{dekool02f1}, we do not find compelling evidence for two separate outflows from our low-resolution data.
\vskip 1pc

{\bf 091658.43+453441.1}  This object was observed once 
using SDSS, and is included in the DR14 SDSS quasar catalog
\citep{paris18}.
Two Gaussian profiles were used based on the shape of the \ion{Mg}{2} doublet lines.
One of the Gaussian profile components was added to model the weak absorption feature seen only in \ion{Mg}{2} lines, thus this component was not included in the analysis.
\vskip 1pc

{\bf 091854.48+583339.6}  This object is included in the 
\citet{farrah12} sample, and has been observed once by SDSS.  It
was first classified as an FeLoBAL quasar by \citet{trump06}. It was
detected in the near-UV by {\it GALEX} \citep{trammell07}.
A single Gaussian model was used.
\vskip 1pc

{\bf 094404.25+500050.3}  This object was observed once
using SDSS, and is included in the DR14 SDSS quasar catalog
\citep{paris18}.   It was identified  as an FeLoBALQ using the
convolutional neural network FeLoNET \citep{dabbieri_prep}.
The model included an eleven-bin tophat, the general reddening law and a template spectrum for the long-wavelength region ($\sim4700$ \AA).
\vskip 1pc

{\bf 100605.66+051349.0} This object is included in the
\citet{farrah12} sample, and has been observed once by SDSS.  It
was first classified as an FeLoBAL quasar by \citet{trump06}. It was
detected in the FIRST survey and has an unresolved core morphology
\citep{kimball11}.
An eight-bin tophat model (single ionization parameter and single density) was used.
\vskip 1pc

{\bf 101927.37+022521.4}  This object is included in the
\citet{farrah12} sample, and has been observed once by SDSS.  It
was first classified as an FeLoBAL quasar by \citet{trump06}. It was
detected in the FIRST survey and has a resolved core morphology
\citep{kimball11}.  \citet{schulze17} presented near-infrared
observations of this object, determining that the
rest-frame optical-band-based redshift is 1.364.  The spectrum also
shows substantial Balmer absorption from the BAL outflow
\citep{schulze18}.
A four-bin tophat model (single ionization parameter and single density) with a modified partial covering was used.
This object showed significant non-zero offset flux at the bottoms of the troughs and we used a modified partial covering model in which the emission lines and a fraction of continuum emission were not absorbed by the outflow.
The width of absorption lines is not significantly larger than other objects ($\sim 3000\rm\ km\ s^{-1}$); nonetheless J1019$+$0225 may be classified as an overlapping trough FeLoBALQ as we see no continuum recovery around $\lambda\sim2500$ \AA\/ due to the large opacity from the rare excited-state \ion{Fe}{2} produced in the high-ionization and high-density gas.
The kinematic properties (narrow width and small $v_{off}$) combined with the compactness of the outflow ($\log R<1$ [pc]) classify the BAL found in this object as a loitering outflow.
\vskip 1pc

{\bf 102036.10+602339.0} This object is included in the
\citet{farrah12} sample, and has been observed twice by SDSS.  It
was first classified as an FeLoBAL quasar by \citet{trump06}. Despite
the heavy absorption, near-UV emission was detected by {\it GALEX} 
\citep{trammell07}.  \citet{liu15}
reported \ion{He}{1}* absorption in the SDSS spectrum. \citet{yi19}
found that the  \ion{Mg}{2} absorption line equivalent
width did not vary between the two observations.  \citet{villforth19} 
found that the host galaxy showed signs of disturbance in an  {\it
  HST} near-IR image.
A two-covering model with two sets of tophat bins with each group having a single ionization parameter and density was used.
The higher-velocity group ($v_{off}\sim-3900\rm\ km\ s^{-1}$) with 3 tophat bins produced most of the \ion{Fe}{2} opacity needed to create the iron troughs found in the spectrum with $\log U\sim-1.9$ and $\log n\sim7.6\ [\rm cm^{-3}]$.
A lower-velocity group ($v_{off}\sim-3100\rm\ km\ s^{-1}$) with 5 bins was needed to model the wide \ion{Mg}{2} trough and deep Ca H$+$K and \ion{Mg}{1} lines that were not sufficiently modeled with the higher-velocity group.
Also, some of the deeper dips in the troughs ($\sim$2300--2600 \AA\/) required the opacity from the bins from the lower-velocity group.
The lower-velocity group shows substantially lower ionization parameter and density compared to the higher-velocity group ($\log U\sim-2.9$ and $\log n\sim4.4\ [\rm cm^{-3}]$).
From the above values, one can come to a conclusion that the two outflow components are separated by more than 2 dex in distance ($R\propto (1/nU)^{1/2}$).
The FWHM of the emission model was fixed to a value determined from the long-wavelength part of the spectrum where the absorption is not severe ($\lambda\ga3000$ \AA\/, FWHM $\sim11000\rm\ km\ s^{-1}$).
\vskip 1pc

{\bf 102226.70+354234.8} This object was observed twice by SDSS and is
included in the DR14 SDSS quasar catalog \citep{paris18}.
A nine-bin tophat model (single ionization parameter and single density) was used.
\vskip 1pc

{\bf 102358.97+015255.8} This object is included in the
\citet{farrah12} sample, and has been observed once by SDSS.  It
was first classified as an FeLoBAL quasar by \citet{trump06}.
A single Gaussian model was used.
\vskip 1pc

{\bf 103036.92+312028.8}  This object was observed four times by SDSS
and is included in the DR14 SDSS quasar catalog \citep{paris18}.
A model with a six-bin tophat (single ionization parameter and single density) and the general reddening law was used.
We used a long-wavelength template spectrum to model this object to $\sim4000$ \AA.
\vskip 1pc

{\bf 103903.03+395445.8} This object was observed once by SDSS
and is included in the DR14 SDSS quasar catalog \citep{paris18}. It
was identified  as an FeLoBALQ using the convolutional neural
network FeLoNET \citep{dabbieri_prep}.
The model included two Gaussian opacity profiles, the general reddening law and a template spectrum for the long-wavelength region ($\sim4700$ \AA).
Each Gaussian profile modeled one of the two BAL components that had a large velocity separation.
\vskip 1pc

{\bf 104459.60+365605.1} This object was discovered in the First
Bright Quasar Survey \citep[FBQS~J104459.5$+$365605][]{becker00,
  white00}; \citet{kimball11} reported that the radio emission is
unresolved.  It has been observed three times by SDSS.   A Keck HIRES
spectrum was analyzed in detail by \citet{dekool01}; they found that
the absorber is located $\sim 700\rm \, pc$ from the continuum source.
\citet{dipompeo10} reported detection of significant polarization that
increases toward shorter wavelengths.  \citet{runnoe13} observed the
rest-frame optical spectrum; they report that for a H$\beta$ FWHM of
3615 $\rm km\, s^{-1}$ and a log bolometric luminosity of
46.57 [$\rm erg\, s^{-1}$], the log black hole mass is 8.87
[M$_\odot$] and the Eddington ratio is 0.33.  While the H$\beta$ FWHM
and bolometric luminosity are similar, the other values are different
than those obtained by \citet{leighly_prep} from the BOSS spectrum
($\log M_{BH}=7.82$ [$\rm M_\odot$] and $\log L_{Bol}/L_{Edd}=0.8$).
\citet{mcgraw15} reported a tentative detection of absorption
variability in the high-velocity LoBAL feature, but not in the FeLoBAL
troughs modeled in this paper. \citet{yi19} also found evidence for
variability.
A thirteen-bin tophat was divided into two groups with each group having a single ionization parameter and single density.
The model also included a long-wavelength template spectrum to model this object to $\sim4000$ \AA.
The higher-velocity group ($-5500$ to $-4000\rm\ km\ s^{-1}$) was composed of 6 tophat bins and the lower-velocity group group ($-2600$ to $-640\rm\ km\ s^{-1}$) used the remaining 7 tophat bins.
\citet{dekool01} also found two velocity structures in this object; however, they did not treat them as two separate outflow components.
Although not included in the best-fitting {\it SimBAL} model, the shallow absorption feature observed at $\lambda\sim2650$ \AA\/ may be the \ion{Mg}{2} absorption lines from a high-velocity LoBAL outflow.
\vskip 1pc

{\bf 105748.63+610910.8}  This object was observed once using SDSS and
is included in the \citet{farrah12} sample.  It was first identified
as an FeLoBALQ by \citet{trump06}.
This object was modeled with a single Gaussian opacity profile.
The narrow absorption lines (FWHM $\sim150\rm km\ s^{-1}$) show full covering of the emission source which is uncommon for quasar outflows, and an unusually high outflow velocity $v_{Off}\sim-9500\rm km\ s^{-1}$.
The key spectral signatures of BAL outflows include the broad width of the absorption lines and partial coverage \citep[e.g.,][]{ganguly08}.
The absorption lines in this object, on the other hand, closely resembles the properties of an associated quasar absorber originating from a gas that is potentially located within the quasar host galaxy or its surrounding medium (Appendix~\ref{app:interv_absorp}).
In addition, this absorption system has been previously classified as an associated absorber in several catalogues \citep[e.g.,][]{quider11,chen18}; however, it is difficult to strictly differentiate between an outflow signature and an associated absorber therefore we proceed with caution in analyzing this outflow.
Although this object was included in \citet{farrah12}, we excluded this BAL from the sample due to its ambiguous nature.
\vskip 1pc

{\bf 112526.12+002901.3} This object was featured in \citet{hall02}
where they commented on the \ion{He}{1}*$\lambda 3889$ and \ion{Ca}{2}
H\&K absorption lines, as well as the low covering fraction and high
excitation \ion{Fe}{2} lines. \citet{hall13}
discussed the redshifted absorption found in this object and a handful
of other BAL outflows. \citet{shi16} compared the SDSS and BOSS 
spectra, and determined that variability between the two arises from a
change in broad line emission, which is not absorbed by the compact
outflow in this object.  \citet{zhang17} reported the X-ray
detection in a 3.8~ks {\it Chandra} observation; the object is 34.1
times X-ray weaker than an unabsorbed quasar.   Near-IR 
observations obtained using {\it HST} revealed that the image is
dominated by the  PSF \citep{villforth19}.
A four-bin tophat (single ionization parameter and single density) and a Gaussian opacity profile model were used.
We also used a modified partial covering model in which the emission lines and a fraction of continuum emission were not absorbed by the outflow.
The principal \ion{Fe}{2} opacity was modeled by the tophat component and the Gaussian component provided the extra opacity needed to create the deep \ion{Mg}{2} absorption trough.
This additional Gaussian component was allowed to freely absorb the emission lines and the continuum emission, unlike the main tophat component.
\vskip 1pc

{\bf 112828.31+011337.9} This object identified as an FeLoBALQ by
\citet{hall02}.   \citet{hall13} discussed the redshifted absorption
also found in this object.  \citet{yi19} found evidence for
variability among the three SDSS observations.  \citet{zhang17} reported
the results of a 5.4~ks {\it Chandra} observation; the object was not
detected, implying that it is more than 50 times weaker than a
comparable unabsorbed object.   Near-IR 
observations obtained using {\it HST} revealed a resolved
target with a somewhat disturbed morphology \citep{villforth19}.
An eight-bin tophat (single ionization parameter and single density) model was used.
The line emission is unabsorbed by the BAL in this model and we used a long-wavelength template spectrum to model this object to $\sim4000$~\AA.
\vskip 1pc

{\bf 112901.71+050617.0} This object was first classified as an
FeLoBALQ by \citet{trump06}.   It was
detected by the FIRST survey and has an unresolved core morphology
\citep{kimball11}.  No variability was detected between two SDSS
observations \citep{yi19}.
An eight-bin tophat (single ionization parameter and single density) model with two covering factors was used.
\vskip 1pc

{\bf 114556.25+110018.4}   This object is included in the
\citet{farrah12} sample, and was observed once by SDSS. It was first
classified as an FeLoBALQ by \citet{trump06}.
A single Gaussian opacity model was used.
The line emission is unabsorbed by the BAL and no reddening component was included in the model.
\vskip 1pc

{\bf 115436.60+030006.3}  This spectacular overlapping trough FeLoBALQ
was first identified by \citet{hall02}.  It has been observed twice by
SDSS and was included in the \citet{farrah12} sample.
\citet{mcgraw15} did not observe variability among three observations,
including one taken at MDM observatory.
A model with a seven-bin tophat and the general reddening law was used.
All bins were constrained to have a single density parameter but the three highest-velocity bins were group together and allowed to have a different ionization parameter than the rest of the four lower-velocity bins.
It has one of the highest outflow velocities ($v_{off}\sim-15,400\rm\ km\ s^{-1}$; $v_{max}\sim-17,000\rm\ km\ s^{-1}$) in the sample and also shows anomalous reddening.
\vskip 1pc

{\bf 115852.86-004301.9} This object was first classified as an
FeLoBAL quasar by \citet{trump06}.  It has been observed twice by
SDSS.  It was detected in the FIRST Survey; \citet{kimball11} reported a
jet with a recognizable morphology with a 21 cm flux density of
87~mJy.
A five-bin tophat (single ionization parameter and single density) model was used.
\vskip 1pc

{\bf 120049.54+632211.8} This object was first classified as an
FeLoBAL quasar by \citet{trump06}, and it is included in the
\citet{farrah12} sample.  It has been observed four times by
SDSS, but no variability among was found \citep{yi19}.
A twelve-bin tophat (single ionization parameter and single density) model with two covering factors was used.
\vskip 1pc

{\bf 120627.62+002335.4} This object was first classified as an
FeLoBAL quasar by \citet{trump06}, and is included in the
\citet{farrah12} sample.  It has been observed once by SDSS.
\citet{hutsemekers17} reported that the object is significantly
polarized at $1.7 \pm 0.36$\%.
The model included an eight-bin tophat that is divided into two groups (single ionization parameter and single density per group), the general reddening law and a template spectrum for the long-wavelength region ($\sim4400$ \AA).
The principal \ion{Fe}{2} opacity was modeled by the higher-velocity group ($-6800$ to $-1900\rm\ km\ s^{-1}$) with 4 tophat bins and the lower-velocity group group ($-5600$ to $-300\rm\ km\ s^{-1}$) used the remaining 4 tophat bins.
\vskip 1pc

{\bf 120815.03+624046.4} This object has been observed once by SDSS,
and is included in the DR14 SDSS quasar catalog 
\citep{paris18}.  It was identified  as an FeLoBALQ using the
convolutional neural network FeLoNET \citep{dabbieri_prep}.
A single Gaussian opacity model was used.
A long-wavelength template spectrum was included to model this object to $\sim4400$ \AA.
\vskip 1pc

{\bf 121231.47+251429.1} This object has been observed once by SDSS,
and is included in the DR14 SDSS quasar catalog
\citep{paris18}.  It was identified  as an FeLoBALQ using the
convolutional neural network FeLoNET \citep{dabbieri_prep}.
A single Gaussian opacity model was used.
A long-wavelength template spectrum was included to model this object to $\sim4700$ \AA.
\vskip 1pc

{\bf 121441.42$-$000137.8} A nine-bin tophat (single ionization parameter and single density) and a Gaussian opacity profile model were used.
A long-wavelength template spectrum was included to model this object to $\sim4500$ \AA.
The extremely broad tophat component seen in the \ion{Mg}{2} trough that extends from $-18000$ to $-6200\rm\ km\ s^{-1}$ was excluded in the analysis.
No substantial \ion{Fe}{2} opacity or other absorption lines were observed from this component and we could not extract robust physical constraints from only the \ion{Mg}{2} trough.
Moreover, the continuum placement near the assumed wavelength region where the troughs from the \ion{Fe}{2} are expected was highly uncertain due to the \ion{Fe}{2} emission lines.
\citet{pitchford19} observed this object in the mid-infrared to study the star formation property of the quasar and found that this object has the highest star formation rate among FeLoBALQs ($\sim2000\rm\ M_{\odot}yr^{-1}$).

\vskip 1pc

{\bf 121442.30+280329.1}  This object was discovered in the First
Bright Quasar Survey \citep[FBQS~J1214$+$2803;][]{becker00,
  white00}. It has been observed twice by SDSS.   A Keck HIRES
spectrum was analyzed in detail by \citet{dekool02f2}.  They found that
the absorber is located between 1 and 30 parsecs from the continuum
source.  \citet{branch02} and \citet{casebeer08} presented an
alternative resonance-scattering interpretation of the spectrum.  
\citet{dipompeo10} found a low continuum polarization ($0.4\pm
0.13$\%).  \citet{zhang17} found no evidence for variability between
the two SDSS spectra, but \citet{mcgraw15} found significant
variability among several MDM observations.
An eight-bin tophat (single ionization parameter and single density) model with two covering factors was used.
The outflow properties measured by \citet{dekool02f2} ($-2.0<\log U<-0.7$, $7.5<\log n<9.5\rm\ [cm^{-3}]$, and $21.4<\log N_H<22.2\rm\ [cm^{-2}]$) were consistent with the values we got from our best-fitting {\it SimBAL} model ($\log U=-1.7^{+0.002}_{-0.001}$, $\log n=7.8^{+0.005}_{-0.006}\rm\ [cm^{-3}]$, $\log N_H=22.0\pm0.003\rm\ [cm^{-2}]$).
\vskip 1pc

{\bf 123549.95+013252.6} This object was first classified as an
FeLoBAL quasar by \citet{trump06}, and is included in the
\citet{farrah12} sample.  It has been observed once by SDSS.  It was
observed using {\it SCUBA-2} but was not detected \citep{violino16}.
An eight-bin tophat (single ionization parameter and single density) model with was used.
A long-wavelength template spectrum was included to model this object to $\sim4000$ \AA.
\vskip 1pc

{\bf 124014.04+444353.4} This object was originally classified as a
type 2 quasar \citep{yuan16}.  It is actually a Seyfert 1.8, as a 
faint and broad H$\beta$ line can be seen in the sole SDSS spectrum \citep{leighly_prep}.
It was identified  as an FeLoBALQ using
the convolutional neural network FeLoNET \citep{dabbieri_prep}.
A six-bin tophat (single ionization parameter and single density) model with was used.
A long-wavelength template spectrum was included to model this object to $\sim4700$ \AA.
\vskip 1pc

{\bf 132117.24+561724.5}  This object was observed once by SDSS and is
included in the DR14 SDSS quasar catalog \citep{paris18}. It was
identified  as an FeLoBALQ using the convolutional neural network
FeLoNET \citep{dabbieri_prep}.
A two-bin tophat (single ionization parameter and single density) model was used.
The line emission is unabsorbed by the BAL in this model and we used a long-wavelength template spectrum to model this object to $\sim4400$ \AA.
\vskip 1pc

{\bf 132401.53+032020.5}  This object was first classified as an
FeLoBAL quasar by \citet{trump06}, and is included in the
\citet{farrah12} sample. \citet{liu15} observed \ion{He}{1}*
absorption in the SDSS spectrum.  \citet{young09} reported a
serendipitous X-ray detection with signal-to-noise ratio of 5.8 that
yields an $\alpha_{ox}=-1.85$.  \citet{yi19} found no variability
between the two SDSS observations.
An eight-bin tophat (single ionization parameter and single density) and a Gaussian opacity profile model were used.
A long-wavelength template spectrum was included to model this object to $\sim4400$ \AA.
An additional Gaussian component was included to model the narrow intervening absorber at $-21000\rm\ km\ s^{-1}$ relative to quasar rest frame; this component was excluded from the analysis.
\vskip 1pc

{\bf 133632.45+083059.9}  This object was observed twice by SDSS and is
included in the DR14 SDSS quasar catalog \citep{paris18}.  This object
was classified as an unusual BAL quasar by \citet{meusinger12}.  It 
was identified as an FeLoBALQ using the convolutional neural network
FeLoNET \citep{dabbieri_prep}.
A nine-bin tophat (single ionization parameter and single density) model with was used.
A long-wavelength template spectrum was included to model this object to $\sim4700$ \AA.
A strong degeneracy between the power-law slope and the reddening was observed in the {\it SimBAL} model, so we used the composite SED from \citet{richards06} with a normalization parameter in place of the power-law continuum model to eliminate the slope parameter.
\vskip 1pc

{\bf 135525.24+575312.7}  This object was observed once by SDSS and is
included in the DR14 SDSS quasar catalog \citep{paris18}. This object
was identified as an FeLoBALQ by visual examination.
An eight-bin tophat (single ionization parameter and single density) model with was used.
A long-wavelength template spectrum was included to model this object to $\sim4400$ \AA.
\vskip 1pc

{\bf 135640.34+452727.2} This object was observed once by SDSS and is
included in the DR14 SDSS quasar catalog \citep{paris18}.
A nine-bin tophat and a Gaussian opacity profile model were used.
The tophat bins and the Gaussian component were modeling a single BAL component together so they were given a single ionization parameter and single density.
A long-wavelength template spectrum was included to model this object to $\sim4700$ \AA.
\vskip 1pc

{\bf 142703.62+270940.4} This object was discovered in the First
Bright Quasar Survey \citep[FBQS~J142703.6$+$270940;][]{becker00,
  white00}.  \citet{dipompeo10} found significant continuum
polarization of about 2\% that rises toward short wavelengths.
The model included a nine-bin tophat that is divided into two groups (single ionization parameter and single density per group), the general reddening law and a template spectrum for the long-wavelength region ($\sim4700$ \AA).
The higher-velocity group ($-3700$ to $-2800\rm\ km\ s^{-1}$) was composed of 3 tophat bins and the lower-velocity group group ($-1100$ to $90\rm\ km\ s^{-1}$) used the remaining 6 tophat bins.
The FWHM of the emission model was fixed to a value determined from the long-wavelength part of the spectrum where the absorption is not severe ($\lambda\ga3000$ \AA\/, FWHM $\sim5000\rm\ km\ s^{-1}$).
\vskip 1pc

{\bf 144800.15+404311.7}  This spectacular overlapping trough object
was first classified as a broad absorption line quasar by
\citet{gibson09}.    \citet{villforth19} presented near-IR imaging
observations obtained using {\it HST}; the image was dominated by the
PSF.  It has been observed three times by SDSS.
The model included an eight-bin tophat that is divided into two groups (single ionization parameter and single density per group) and an extra Gaussian component with both using two covering factors, the general reddening law and a template spectrum for the long-wavelength region ($\sim4700$ \AA).
The higher-velocity group ($-6900$ to $-2400\rm\ km\ s^{-1}$) was composed of 4 tophat bins and the lower-velocity group group ($-5000$ to $-710\rm\ km\ s^{-1}$) used the remaining 4 tophat bins.
The additional Gaussian component at $\sim-600\rm\ km\ s^{-1}$ was required for the narrow outflow component that was identified by the narrow \ion{Mg}{1} and \ion{Ca}{2} H and K absorption lines.
\vskip 1pc

{\bf 151708.94+232857.5}  This object was observed once by SDSS and is
included in the DR14 SDSS quasar catalog \citep{paris18}.  It was
identified  as an FeLoBALQ using the convolutional neural network
FeLoNET \citep{dabbieri_prep}.
A seven-bin tophat (single ionization parameter and single density) with two-covering model was used.
A long-wavelength template spectrum was included to model this object to $\sim4700$ \AA.
\vskip 1pc

{\bf 152737.17+591210.1} This object was observed once by SDSS and is
included in the DR14 SDSS quasar catalog \citep{paris18}.
The model included a seven-bin tophat, the general reddening law and a template spectrum for the long-wavelength region ($\sim4700$ \AA).
\vskip 1pc

{\bf 153145.01+485257.2} It was identified  as an FeLoBALQ using the convolutional neural network FeLoNET \citep{dabbieri_prep}.
Three Gaussian profiles were used in the model.
Only one of the Gaussian components was included in the analysis because the other two Gaussian components were included to model a narrow intervening system at $\sim-23000\rm\ km\ s^{-1}$ relative to quasar rest frame and for a weak \ion{Mg}{2} opacity structure near rest velocity.
\vskip 1pc

{\bf 155633.77+351757.3}  This object was identified as a radio-loud BALQ
in the FIRST Survey \citep[FIRST~J155633.8$+$351758;][]{becker97}, 
and is included in the \citet{farrah12} sample.  It was observed by
SDSS three times.  It is highly polarized, up to $\sim 10$\% at short
wavelengths, with lower polarization and complex structure across the
deepest troughs \citep{brotherton97}.  \citet{najita00} presented a
near-infrared spectrum; they found strong Balmer lines and \ion{Fe}{2}
emission, and a Balmer-line-based redshift of $z=1.5008\pm 0.0007$.
 \citet{lewis03} found that the object was not detected in
sub-millimeter band using {\it SCUBA}.  \citet{brotherton05} reported
results of a {\it Chandra} observation; 40 photons were detected, and
they suggest that the X-ray emission was suppressed by a absorption by
a factor of 49.  A second longer {\it Chandra} observation netted 531
photons \citep{berrington13}; a heavily absorbed spectrum could be
ruled out, and a partial covering model was favored.  \citet{jiang03}
performed European VLBI Network observations at 1.6~GHz; they found
that the object is unresolved at 20~mas, and inferred a flat
spectrum.  \citet{yi19} reported no variability between the first two
SDSS observations.
The model included a nine-bin tophat that is divided into two groups (single ionization parameter and single density per group) with two-covering, the general reddening law and a template spectrum for the long-wavelength region ($\sim4100$ \AA).
The higher-velocity group ($-9000$ to $-5600\rm\ km\ s^{-1}$) was composed of 4 tophat bins and the lower-velocity group group ($-5500$ to $-1400\rm\ km\ s^{-1}$) used the remaining 5 tophat bins.
\vskip 1pc

{\bf 164419.75+530750.4} This object was first categorized as a QSO by
\citet{popescu96}.  It was observed once by SDSS and is included in
the DR14 SDSS quasar catalog \citep{paris18}.  It was identified as an
FeLoBAL quasar by visual inspection.
A six-bin tophat (single ionization parameter and single density) and a Gaussian opacity profile model were used.
We also used a modified partial covering model in which the emission lines and a fraction of continuum emission were not absorbed by the outflow.
The Gaussian component was excluded from the sample because it only produced \ion{Mg}{2} opacity
and thus we could not constrain the physical properties of that outflow component.

\vskip 1pc
{\bf 173753.97+553604.9} This object was first identified as an
FeLoBALQ by \citet{trump06}, and it is included in the
\citet{farrah12} sample. It was observed once by SDSS.
We did not find any absorption features in this object, and
therefore we excluded this object from the analysis.

\vskip 1pc
{\bf 210712.77+005439.4} This object was first identified as an
FeLoBALQ by \citet{trump06}, and it is included in the
\citet{farrah12} sample.  It was observed once by SDSS.  It was
observed using {\it SCUBA-2}
but was not detected \citep{violino16}.
An eight-bin tophat (single ionization parameter and single density) model with was used.
A long-wavelength template spectrum was included to model this object to $\sim4700$ \AA.

\vskip 1pc
{\bf 213537.44-032054.8}  This object was observed once by SDSS and is
included in the DR14 SDSS quasar catalog \citep{paris18}.
It was identified as a BALQ in the DR10 SDSS quasar catalog \citep{paris14}.
A six-bin tophat (single ionization parameter and single density) model with was used.
A long-wavelength template spectrum was included to model this object to $\sim4400$ \AA.

\vskip 1pc
{\bf 230730.69+111908.5}  This object was observed once by SDSS and is
included in the DR14 SDSS quasar catalog \citep{paris18}.
A five-bin tophat and a Gaussian opacity profile model were used.
The tophat bins and the Gaussian component were modeling a single BAL component together so they were given a single ionization parameter and single density.
A long-wavelength template spectrum was included to model this object to $\sim4100$ \AA.

\section{Intervening Absorbers\label{app:interv_absorp}}
\begin{figure*}
\includegraphics[width=.34\linewidth]{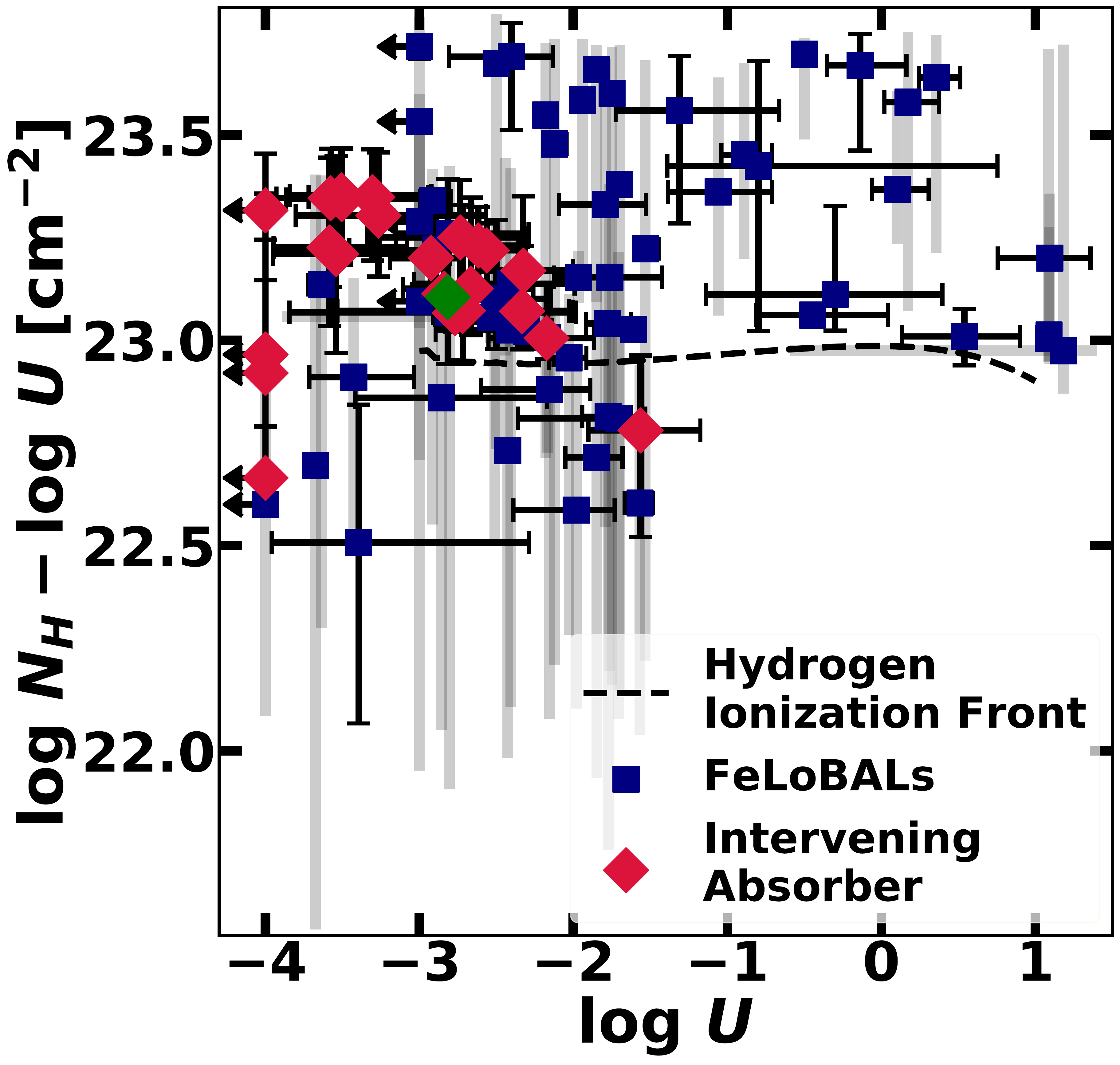}
\includegraphics[width=.31\linewidth]{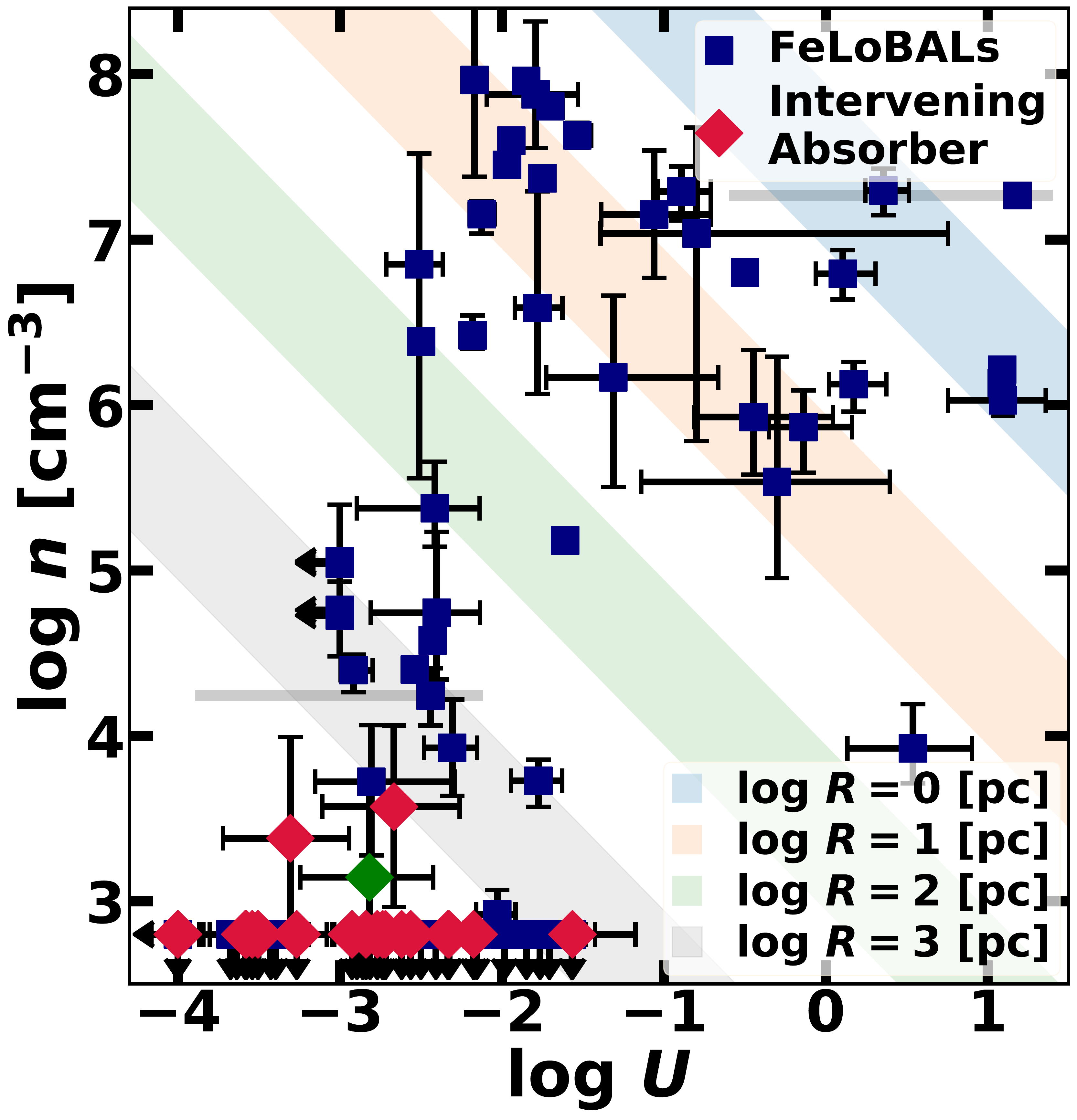}
\includegraphics[width=.33\linewidth]{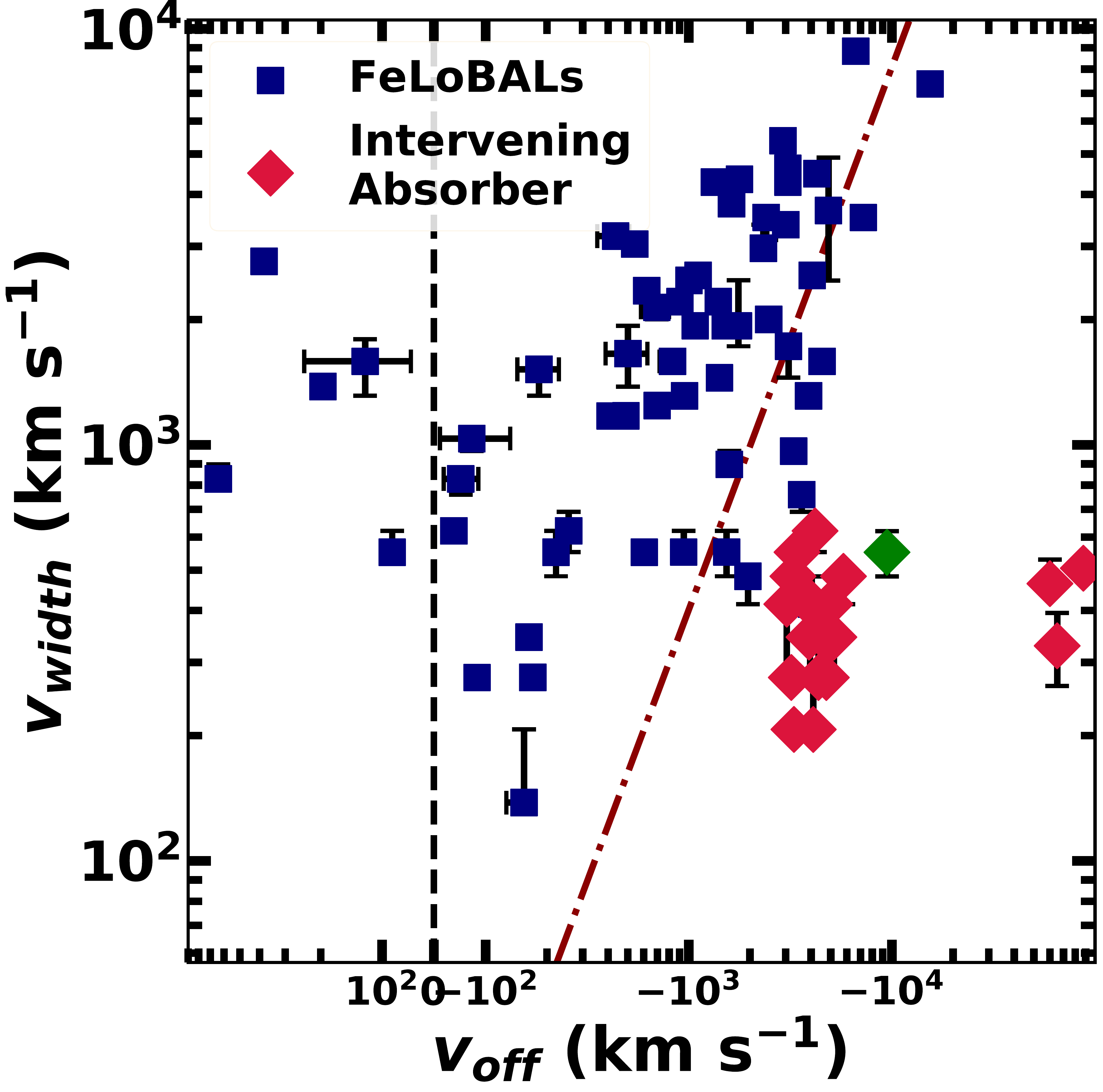}
\caption{{\it Left panel}:
The physical parameter distribution of intervening absorbers cannot be easily differentiated from that of FeLoBAL.
{\it Middle panel}:
Intervening absorbers have low densities and we were able to constrain only the density upper limits ($\log n\lesssim2.8\rm\ [cm^{-3}]$) for the majority of them.
Based on the ionization parameter and the density, we found the inferred distances from the central engine to the intervening absorber gas to be $\log R\gtrsim3$ [pc], if we assume the radiation from AGN as the ionizing source for these absorbers.
{\it Right panel}:
The kinematic properties show the clearest distinction between the intervening absorbers and FeLoBALs.
The intervening absorbers have very narrow width ($v_{width}\lesssim200\rm\ km\ s^{-1}$) and disproportionately large offset velocities ($v_{off}\ll-3,000\rm\ km\ s^{-1}$), relative to quasar rest frame, compared to FeLoBALs.
The brown dotted-dashed line shows a one-to-one ratio.
We rejected SDSS~J1057$+$6109 (green diamonds) from the analysis because the properties of the absorption feature seen in the spectra resembled more the intervening absorbers than the quasar driven outflows (\S~\ref{subsubsec:bal_exclude}).
A linear scale was used in the region $\vert v_{off}\vert<100\rm\ km\ s^{-1}$ and log scale was used elsewhere in the x-axis.
The error bars show 95\% uncertainties and the grey shaded bars represent the range of the values among the tophat model bins for each BAL.
}
\label{fig:interv}
\end{figure*}
The intrinsic quasar absorption lines (e.g., BAL) have five main spectroscopic characteristic that are not seen in intervening absorbers:
(1) wider absorption lines,
(2) partial coverage of the emission source
(4) time variability of the troughs
(3) higher ionization parameter
(4) higher metallicity
\citep[e.g.,][]{barlow97,ganguly08}.
Therefore, BALs that show narrow and deep features could easily be confused with intervening absorbers.
In order to systematically check whether we could differentiate the intervening absorbers from BALs using {\it SimBAL}, we fit 20 quasar spectra that were identified to have intervening absorption features.
The objects were drawn from the intervening \ion{Mg}{2} quasar absorption line catalogue by \citet{quider11} and we chose objects that have the similar redshift range ($1<z<1.5$).
We then selected the ones that have \ion{Mg}{2} absorbers located close to the background quasar ($z_{qso}\sim z_{abs}$) because these objects have the spectral features that most closely resembles BAL spectra.
The spectra of these objects show \ion{Mg}{2} absorption lines near $\lambda_{rest}\sim2750$ \AA\/ that could be confused with quasar-driven BAL or AAL \citep[e.g.,][]{hamann11}.

Figure~\ref{fig:interv} shows the distributions of physical and kinematic parameters of the intervening absorption lines and the FeLoBALs, both obtained from {\it SimBAL} modeling.
The distributions of physical parameters (e.g., ionization parameter and density) do not show a clear distinction between the intervening absorbers and FeLoBALs.
However, most of the intervening absorbers were constrained to have extremely low density ($\log n\lesssim3\rm\ [cm^{-3}]$).
The kinematic properties of the absorption line systems showed the clearest difference between the intervening \ion{Mg}{2} absorbers and FeLoBALs.
The intervening absorption lines have narrower widths and higher offset velocities than the majority of FeLoBALs in the sample.
We note that accurate modeling of intervening absorbers requires photoionization calculations using the correct photoionzing SED.
For this experiment we used the same ionic column density grid used with FeLoBALQs that was generated with quasar SED which may have a different shape than the photoionizing SEDs required for the intervening absorption line gas.
Nevertheless, we conclude that the intervening absorption systems could be identified by examining the kinematic properties of the absorption lines, in particular the width of the lines.
This result implies that intervening absorbers will be easily excluded in data from upcoming missions that will provide better spectral resolution, such as 4MOST \citep{dejong19}.

\end{CJK}
\end{document}